\documentclass[12pt,subeqn]{article}

\usepackage{amsmath}
\renewcommand{\theequation}{\arabic{equation}}
\newcommand{\tr}{\operatorname{Tr}}

\newcommand{\EQ}{\begin{equation}}
\newcommand{\EN}{\end{equation}}

\newcommand{\ket}[1]{\left|#1\right\rangle}      

\newcommand{\bear}{\begin{eqnarray}}
\newcommand{\ear}{\end{eqnarray}}
\newcommand{\bt} { \begin{tabular} }
\newcommand{\et}{ \end{tabular} }
\newcommand{\bc} { \begin{center} }
\newcommand{\ec}{ \end{center} }

\newcommand{\btb} { \begin{table} }
\newcommand{\etb}{ \end{table} }

\begin{document}

\topmargin 0pt
\oddsidemargin 5mm
\newcommand{\NP}[1]{Nucl.\ Phys.\ {\bf #1}}
\newcommand{\PL}[1]{Phys.\ Lett.\ {\bf #1}}
\newcommand{\NC}[1]{Nuovo Cimento {\bf #1}}
\newcommand{\CMP}[1]{Comm.\ Math.\ Phys.\ {\bf #1}}
\newcommand{\PR}[1]{Phys.\ Rev.\ {\bf #1}}
\newcommand{\PRL}[1]{Phys.\ Rev.\ Lett.\ {\bf #1}}
\newcommand{\MPL}[1]{Mod.\ Phys.\ Lett.\ {\bf #1}}
\newcommand{\JETP}[1]{Sov.\ Phys.\ JETP {\bf #1}}
\newcommand{\TMP}[1]{Teor.\ Mat.\ Fiz.\ {\bf #1}}

\renewcommand{\thefootnote}{\fnsymbol{footnote}}

\newpage
\setcounter{page}{0}
\begin{titlepage}
\begin{flushright}
UFSCARF-TH-09-10
\end{flushright}
\vspace{0.5cm}
\begin{center}
{\large Bethe ansatz for the XXX-$S$ chain with non-diagonal open boundaries}\\
\vspace{1cm}
{\large C.S. Melo, G.A.P. Ribeiro and M.J. Martins} \\
\vspace{1cm}
{\em Universidade Federal de S\~ao Carlos\\
Departamento de F\'{\i}sica \\
C.P. 676, 13565-905~~S\~ao Carlos(SP), Brasil}\\
\end{center}
\vspace{0.5cm}

\begin{abstract}
We consider the algebraic Bethe ansatz solution of the
integrable and isotropic XXX-$S$ Heisenberg 
chain with non-diagonal open boundaries. We show that the corresponding
$K$-matrices are similar to diagonal matrices with the help of suitable
transformations independent of the spectral parameter. When the boundary parameters satisfy certain
constraints we are able to formulate the diagonalization of the associated double-row transfer
matrix by means of the quantum inverse scattering method. This allows us to derive
explicit expressions for the
eigenvalues and the corresponding Bethe ansatz equations.  We also present evidences
that the eigenvectors can be build up
in terms of multiparticle states
for arbitrary $S$.
\end{abstract}

\vspace{.15cm}
\centerline{PACS numbers:  05.50+q, 02.30.IK}
\vspace{.1cm}
\centerline{Keywords: Algebraic Bethe Ansatz, Open Boundary}
\vspace{.15cm}
\centerline{November 2004}

\end{titlepage}


\pagestyle{empty}

\newpage

\pagestyle{plain}
\pagenumbering{arabic}

\renewcommand{\thefootnote}{\arabic{footnote}}

\section{Introduction}

The possibility of constructing $SU(2)$ invariant Heisenberg chain with arbitrary spin-$S$ solvable by
Bethe ansatz methods was a remarkable achievement of the representation theory underlying the associative
algebra describing the dynamical symmetry of quantum integrable systems \cite{RE}.
It turns out that the  Hamiltonian of such spin-$S$ XXX  Heisenberg
magnet \cite{BATA} commutes with the transfer matrix $T_S(\lambda)$ of a $2S+1$ state vertex on the
square $L \times L$ lattice \cite{RE,SOG}. This connection is based on well known relationships between one-dimensional
quantum spin chains and two-dimensional statistical mechanics
models whose Boltzmann weights satisfy the Yang-Baxter
equation \cite{BAX,QI}.

The row-to-row transfer matrix $T_S(\lambda)$ of such $2S+1$ state vertex model
can be conveniently written as the trace, over an auxiliary space
${\cal A} \equiv {\cal C}^{2S+1}$,
of an ordered product of Boltzmann weights. More specifically,
\EQ
T_{S}(\lambda)= \tr_{\cal A}[ {\cal{T}}_{\cal A}^{(S)}(\lambda)] ~~~~~
{\cal T}_{\cal A}^{(S)}(\lambda)={\cal L}_{{\cal A} L}^{(S)}(\lambda) {\cal L}_{{\cal A} L-1}^{(S)}(\lambda) \dots
{\cal L}_{{\cal A} 1}^{(S)}(\lambda),
\label{transfermatrix}
\EN
where $\lambda$ is the spectral parameter and ${\cal A}$ represents the horizontal
degrees of freedom of the vertex model.

The Boltzmann weight ${\cal L}_{ab}^{(S)}(\lambda)$ is solution of the Yang-Baxter equation
\EQ
{\cal L}_{ab}^{(S)}(\lambda-\mu)
{\cal T}_{a}^{(S)}(\lambda)
{\cal T}_{b}^{(S)}(\mu)=
{\cal T}_{b}^{(S)}(\mu)
{\cal T}_{a}^{(S)}(\lambda)
{\cal L}_{ab}^{(S)}(\lambda-\mu),
\label{YBA}
\EN
invariant
relative to the $SU(2)$ Lie algebra. It can be viewed
as $(2S+1)\times(2S+1)$ matrix on the auxiliary space whose elements are operators acting non-trivially only
on the $b$-th factor of the Hilbert space $\displaystyle \prod_{b=1}^{L}\otimes {\cal C}_{b}^{2S+1}$.
Its explicit expression in terms of the spin-$S$ $SU(2)$ generators
$\vec{S}_{\alpha}=(S_{a}^{x},S_{a}^{y},S_{a}^{z})$ is \cite{RE,BATA,SOG}
\EQ
{\cal L}_{ab}^{(S)}(\lambda)= (\lambda+2 \eta S) \sum_{l=0}^{2S} \prod_{k=l+1}^{2S} \frac{\lambda- \eta k}{\lambda+ \eta k} \prod_{\stackrel{n=0}{n
\neq l}}^{2S} \frac{\vec{S}_{a}\otimes \vec{S}_{b}-x_{n}}{x_{l}-x_{n}},
\label{Hsmodel}
\EN
where $x_{l}=\frac{1}{2}l(l+1)-S(S+1)$ and $\eta$ is the so-called quasi-classical parameter.

Besides the Yang-Baxter equation the operator
${\cal L}_{12}^{(S)}(\lambda)$ satisfies  other relevant properties such as
\begin{align}
\mbox{Unitarity: } & {\cal L}_{12}^{(S)}(\lambda) {\cal L}_{21}^{(S)}(-\lambda)= \zeta_{S}(\lambda)
\mbox{Id} \otimes \mbox{Id}; \label{uni} \\
\mbox{Parity invariance: } & P_{12}{\cal L}_{12}^{(S)}(\lambda)P_{12}={\cal L}_{12}^{(S)}(\lambda); \label{pari}\\
\mbox{Temporal invariance: } & {\cal L}_{12}^{(S)}(\lambda)^{t_{1}t_{2}}={\cal L}_{12}^{(S)}(\lambda);\label{tempo} \\
\mbox{Crossing symmetry: } & {\cal L}_{12}^{(S)}(\lambda) = (-1)^{2S} \frac{\varsigma_{S}(\lambda)}{\varsigma_{S}(-\lambda-\eta)}
\stackrel{1}{V}{\cal L}_{12}^{(S)}(-\lambda-\eta)^{t_{2}} \stackrel{1}{V^{-1}};
\label{cross}
\end{align}
where functions $\zeta_{S}(\lambda)=(2S\eta)^2-\lambda^2$ and
$\displaystyle \varsigma_{S}(\lambda)=\prod_{k=1}^{2S-1}(\lambda+k\eta)$. Here
$\mbox{Id}$ is the $(2S+1)\times(2S+1)$ identity matrix, $P_{12}$ is the permutation operator,
$t_{\alpha}$ denotes transposition on the $\alpha$-th space, $\stackrel{1}{V}=V \otimes \mbox{Id}$
and $\stackrel{2}{V}= \mbox{Id} \otimes V$. The matrix $V$ is
anti-diagonal whose non-null elements are $V_{i,j}=-(-1)^{i}\delta_{i,2S+2-j}$.

This notion of integrability has been extended to include integrable open boundary conditions \cite{CH,SK}.
In addition to the Yang-Baxter solution  ${\cal L}_{ab}^{(S)}(\lambda)$ determining the dynamics of the
bulk one has to introduce $(2S+1) \times (2S+1)$ $K$-matrices $K_{S}(\lambda)$ whose elements represent
the interactions at the left and right ends of the open spin chain.
Compatibility with bulk integrability demands that these matrices should satisfy the reflection equation given by \cite{SK}
\EQ
{\cal L}_{12}^{(S)}(\lambda-\mu) \stackrel{1}{K}_{S}(\lambda) {\cal L}_{21}^{(S)}(\lambda+\mu) \stackrel{2}{K}_{S}(\mu)
= \stackrel{2}{K}_{S}(\mu) {\cal L}_{12}^{(S)}(\lambda+\mu) \stackrel{1}{K}_{S}(\lambda) {\cal L}_{21}^{(S)}(\lambda-\mu),
\label{eqref}
\EN
where $\stackrel{1}{K}_{S}(\lambda)=K_{S}(\lambda) \otimes \mbox{Id}$ and $\stackrel{2}{K}_{S}(\lambda)=\mbox{Id}
\otimes K_{S}(\lambda)$.

In the case of open boundaries the analogue of the transfer matrix
is the following double-row operator \cite{SK}
\EQ
t_{S}(\lambda)=\tr_{\cal A}\left[ K_{S}^{(+)}(\lambda) {\cal T}_{\cal A}^{(S)}(\lambda) K_{S}^{(-)}(\lambda) \left[{{\cal
T}_{\cal A}^{(S)}}(-\lambda)\right]^{-1} \right],
\label{monodp}
\EN
where $K_{S}^{(-)}(\lambda)$ can be chosen as one of the solutions of the reflection equation (\ref{eqref}). The other
matrix $K_{S}^{(+)}(\lambda)$ can be directly obtained from
$K_{S}^{(-)}(\lambda)$  thanks to the extra relations (\ref{uni}-\ref{cross}) satisfied by the operator
${\cal L}_{ab}^{(S)}(\lambda)$. Following a scheme devised in ref.\cite{NEP}
this isomorphism becomes
\EQ
K_{S}^{(+)}(\lambda)= \left [ K_{S}^{(-)}(-\lambda-\eta) \right ]^{t}.
\label{isomor}
\EN

The understanding of the physical properties of the XXX-$S$ open
chain includes necessarily the exact diagonalization of the double-row operator (\ref{monodp}).
If the $K$-matrices are diagonal this problem can be tackled, for example,
by an extension of the quantum inverse scattering method \cite{SK}
and the use of fusion hierarchy procedures \cite{MA,DOI}.
The same does not occur when the $K$-matrices are non-diagonal due
to an apparent lack of simple reference states to start Bethe ansatz analysis. In spite of
this difficulty, progresses have recently been made for the anisotropic
version of the $S=\frac{1}{2}$ Heisenberg magnet usually denominated the XXZ spin chain. These achievements
have been made either by a functional Bethe ansatz analysis \cite{NEP1} or by means of
the algebraic Bethe ansatz method \cite{CA}. The latter approach has been based on earlier ideas developed
in the context of the eight vertex model \cite{SA}.
In particular, it was argued that the spectrum of the open XXZ chain can be parameterized
by Bethe ansatz equation provided certain constraint between the parameters of the Hamiltonian is satisfied.
Part of the conclusions were achieved with the help of a numerical
study of the spectrum for finite values of $L$ \cite{NEP2}. More recently, new results have been obtained in
ref.\cite{DEJ} by exploring the description of the open XXZ spin chain in terms of the Temperley-Lieb algebra.
The extension of all such analysis for integrable Heisenberg chains with arbitrary spin-$S$  appears to be
highly non-trivial and it is indeed an interesting open problem in the field of integrable models.

In this paper we would like to take some steps
towards the direction of solving the isotropic higher
spin Heisenberg model (\ref{Hsmodel}) with non-diagonal open boundaries.
We show that  the double-row transfer matrix operator
associated to the integrable XXX-$S$ Heisenberg chain can be diagonalized
by Bethe ansatz at least when the respective $K$-matrices
parameters satisfy one out of two possible types of constraints.
We find that the roots of the Bethe ansatz equations
are fixed by integers $n \leq 2SL$ that play the role of standard particle number sectors.
This feature shows that the Hilbert space has a multiparticle
structure which should be useful to determine the nature of the ground state and excitations unambiguously.

The outline of this paper is as follows. In section \ref{Kprop} we
argue that the non-diagonal $K$-matrices of the XXX-$S$ Heisenberg
model are diagonalizable by spectral independent similarity
transformations. In section \ref{Eigenprob} suitable quantum space transformations are used to
show that the diagonalization of $t_{S}(\lambda)$ is similar to an
eigenvalue problem with diagonal and triangular $K$-matrices
provided that certain constraints are satisfied. In section
\ref{BAalg} we discuss the quantum inverse scattering method
for the latter system, presenting the corresponding eigenvalues
and Bethe ansatz equations. Explicit
expressions for the eigenvectors in terms of similarity
transformation acting on creation fields can be written for spin
$\frac{1}{2}$ and $1$. In section \ref{Conclusion} our conclusions and
further perspectives are discussed. In Appendix A we
summarize certain properties of the $K$-matrices. In Appendices B and C we discuss
the one and two particle analysis of the eigenspectrum as well as
auxiliary expressions for $S=\frac{3}{2}$, respectively. Finally, in Appendix D we exhibit 
general relations concerning the one-particle unwanted terms 
and the two-particle state construction for arbitrary $S$.

\section{The $K$-matrices properties}\label{Kprop}

The most general reflection $K$-matrix associated to
the open XXX-$S$ Heisenberg chain possesses three free parameters.
For $S=\frac{1}{2}$ it is given by \cite{DEV}
\EQ
K_{\frac{1}{2}}^{(-)}(\lambda)=\left(\begin{array}{cc}
       \xi_{-}+\frac{\lambda}{\eta} & c_{-} \frac{\lambda}{\eta} \\
       d_{-} \frac{\lambda}{\eta} & \xi_{-}-\frac{\lambda}{\eta}\end{array} \right),
\label{Km}
\EN
while the isomorphism (\ref{isomor}) implies that
\EQ
K_{\frac{1}{2}}^{(+)}(\lambda)=\left(\begin{array}{cc}
       \xi_{+}-1-\frac{\lambda}{\eta} & -c_{+} \left(\frac{\lambda}{\eta}+1\right) \\
       -d_{+} \left(\frac{\lambda}{\eta}+1\right) & \xi_{+}+1+\frac{\lambda}{\eta}\end{array} \right),
\label{Kp}
\EN
where $\xi_{\pm}$, $c_{\pm}$ and $d_{\pm}$ are six free parameters.

A remarkable characteristic of these $K$-matrices is that they can
be diagonalized by similarity transformations which are
independent of the spectral parameter $\lambda$. More precisely,
it is possible to rewrite the equations (\ref{Km},\ref{Kp}) as
\EQ
K_{\frac{1}{2}}^{(-)}(\lambda)=\rho_{\frac{1}{2}}^{(-)} {\cal G}_{\frac{1}{2}}^{(-)} \left(\begin{array}{cc}
       \bar{\xi}_{-}+\frac{\lambda}{\eta} & 0 \\
       0 & \bar{\xi}_{-}-\frac{\lambda}{\eta}\end{array} \right) \left[{\cal G}_{\frac{1}{2}}^{(-)}\right]^{-1},
\label{KmD}
\EN
and
\EQ
K_{\frac{1}{2}}^{(+)}(\lambda)=\rho_{\frac{1}{2}}^{(+)}{\cal G}_{\frac{1}{2}}^{(+)} \left(\begin{array}{cc}
       \bar{\xi}_{+}-1-\frac{\lambda}{\eta} & 0 \\
       0 & \bar{\xi}_{+}+1+\frac{\lambda}{\eta}\end{array} \right) \left[{\cal G}_{\frac{1}{2}}^{(+)}\right]^{-1},
\label{KpD}
\EN
where ${\cal G}_{S}^{(\pm)}$ refer to appropriate $(2S+1) \times (2S+1)$ matrices. In what follows we
will represent them in terms 
of the standard Weyl basis $\hat{e}_{ij}$ by the expression
\EQ
{\cal G}_{S}^{(\pm)}= \sum_{i,j=1}^{2S+1} g_{i,j}^{(\pm)} \hat{e}_{ij}.
\label{gssmatrix}
\EN

In the specific case of $S=\frac{1}{2}$ the expressions relating the off-diagonal and the diagonal
elements of ${\cal G}_{\frac{1}{2}}^{(\pm)}$ are
\EQ
\frac{g_{2,1}^{(\pm)}}{g_{1,1}^{(\pm)}}=-\frac{1+\epsilon_{\pm}\sqrt{1+c_{\pm}d_{\pm}}}{c_{\pm}};
~~~~~\frac{g_{1,2}^{(\pm)}}{g_{2,2}^{(\pm)}}=\frac{1+\epsilon_{\pm}\sqrt{1+c_{\pm}d_{\pm}}}{d_{\pm}},
\EN
where $\epsilon_{+}=\epsilon_{-}= \pm 1$. The other variables $\bar{\xi}_{\pm}$ and
$\rho_{\frac{1}{2}}^{(\pm)}$ entering in
the formulae (\ref{KmD},\ref{KpD}) are given by
\EQ
\bar{\xi}_{\pm}=-\frac{\epsilon_{\pm}
\xi_{\pm}}{\sqrt{1+c_{\pm}d_{\pm}}};~~~~~~~\rho_{\frac{1}{2}}^{(\pm)}=-\epsilon_{\pm}
\sqrt{1+c_{\pm}d_{\pm}}.
\EN

The $K$-matrices for $S > \frac{1}{2}$ can be computed either by brute
force analysis of the reflection equation \cite{JA} or constructed
by the so-called fusion procedure \cite{NEP3}. Their matrix
elements expressions become very cumbersome as one increases the
value of the spin and this fact has been exemplified in Appendix
A for spin $1$ and $\frac{3}{2}$ cases. It turns out, however,
that we have found out that such $K$-matrices can be rewritten in
a rather compact and illuminating form with the help of
appropriate spectral independent similarity transformations,
namely
\EQ
K_{S}^{(\pm)}(\lambda)=\rho_{S}^{(\pm)} {\cal G}_{S}^{(\pm)} D_{S}^{(\pm)}(\lambda) \left[{\cal G}_{S}^{(\pm)}\right]^{-1}, \label{kdecompo}
\EN
where the overall normalizing constant is $\displaystyle \rho_{S}^{(\pm)} =
-(\epsilon_{\pm})^{2S} \left( \frac{\sqrt{1+c_{\pm} d_{\pm}}}{2S} \right)^{2S}$.

The diagonal matrix $D_{S}^{(\pm)}(\lambda)$ is defined by
\EQ
D_{S}^{(\pm)}(\lambda)=\sum_{j=1}^{2S+1}f_{j}^{(\pm)}(S;\lambda,\bar{\xi}_{\pm})
\hat{e}_{jj},
\label{diagonalD}
\EN
where the corresponding diagonal entries are
\begin{subequations}
\bear f_{\alpha}^{(-)}(S;\lambda,\bar{\xi}_{-})&=&\prod_{\beta=1}^{2S} \left[
\bar{\xi}_{-}+S+\frac{1}{2}-\beta-\mathrm{sign}\left(\alpha-\frac{1}{2}-\beta\right)\frac{\lambda}{\eta}
\right],\label{fsalphaminus}
\\
f_{\alpha}^{(+)}(S;\lambda,\bar{\xi}_{+}) &=& \prod_{\beta=1}^{2S} \left[
\bar{\xi}_{+}+S+\frac{1}{2}-\beta+\mathrm{sign}\left(\alpha-\frac{1}{2}-\beta\right)(\frac{\lambda}{\eta}+1)
\right].
\label{fsalphaplus} \ear
\end{subequations}

Interesting enough, the novel
parameters $\bar{\xi}_{\pm}$ encode both the dependence on the spin value and on the
variables describing the off-diagonal
$K$-matrices elements. Specifically, we have found that
\EQ
\bar{\xi}_{\pm}=-\frac{\epsilon_{\pm} 2S \xi_{\pm}}{\sqrt{1+c_{\pm} d_{\pm}}}.
\label{xirem}
\EN

Finally, the four elements $g_{1, 1}^{(\pm)}$, $g_{1, 2}^{(\pm)}$, $g_{2,
1}^{(\pm)}$ and $g_{2, 2}^{(\pm)}$ of ${\cal G}_{S}^{(\pm)}$ are related by the expressions
\bear
\frac{g_{2, 1}^{(\pm)}}{g_{1, 1}^{(\pm)}}&=&\frac{(-1)^{2S} \sqrt{2S}
\left( 1+\epsilon_{\pm} \sqrt{1+c_{\pm}d_{\pm}} \right)}{c_{\pm}}, \\
\frac{g_{2, 2}^{(\pm)}}{g_{1, 2}^{(\pm)}}&=&\frac{(-1)^{2S}\sqrt{2}}{\sqrt{S}c_{\pm}} \left[
\frac{(S-1) c_{\pm}d_{\pm}+(2S-1)(1+\epsilon_{\pm}
\sqrt{1+c_{\pm}d_{\pm}})}{1+\epsilon_{\pm}
\sqrt{1+c_{\pm}d_{\pm}}} \right],
\ear
and the remaining elements are obtained by the following recurrence relations
\EQ
g_{m,l}^{(\pm)}=\frac{\sqrt{2S(m-1)(2S+2-m)} g_{2, 1}^{(\pm)}
g_{m-1, l}^{(\pm)} - \sqrt{2S(l-1)(2S+2-l)} g_{1, 2}^{(\pm)} g_{m,
l-1}^{(\pm)}} {2S(m-l) g_{1, 1}^{(\pm)}},
\label{rec1}
\EN
for $l \neq m=1,\dots,2S+1$ while for $m=l$ we have
\EQ
g_{l,l}^{(\pm)}=\frac{g_{2, 2}^{(\pm)} g_{l-1, l-1}^{(\pm)}}{g_{1, 1}^{(\pm)}}-\frac{
\sqrt{2(2S-1)(l-2)(2S+3-l)} g_{3, 1}^{(\pm)}(S) g_{l-2, l}^{(\pm)} + 2(l-2) g_{2, 1}^{(\pm)} g_{l-1,l}^{(\pm)}}{\sqrt{2S(2S+2-l)(l-1)} g_{1, 1}^{(\pm)}}.
\label{rec2}
\EN

An important feature of our construction is that the matrices
${\cal G}_{S}^{(\pm)}$ are itself representations, without spectral
parameter, of the monodromy matrix associated to the Yang-Baxter
algebra (\ref{YBA}) generated by the operators ${\cal L}_{ab
}^{(S)}(\lambda)$. In fact, the matrix (\ref{gssmatrix},\ref{rec1},\ref{rec2})
with
four free parameters are the widest possible class of non-diagonal
twisted boundary conditions compatible with integrability for the
XXX-$S$ spin chain \cite{RM}. An immediate consequence of this symmetry is
the commutation relation
\EQ
\left[{\cal L}_{12}^{(S)}(\lambda),{\cal G}_{S}^{(\pm)} \otimes {\cal G}_{S}^{(\pm)} \right]=0,
\label{symmetry}
\EN
which will be of great use in next section.

\section{The eigenvalue problem}\label{Eigenprob}

The purpose of this section is to show that the eigenvalue problem
for the double-row transfer matrix operator $t_{S}(\lambda)$,
\EQ
t_{S}(\lambda)\ket{\psi}=\Lambda_{S}(\lambda)\ket{\psi},
\label{eigenvalueproblem}
\EN
associated to the XXX-$S$ chain with two general non-diagonal open
boundaries can be transformed into a similar problem with only one
genuine non-diagonal $K$-matrix.

In order to demonstrate that we use the decomposition property for the $K_{S}^{(+)}(\lambda)$ matrix described in section \ref{Kprop} and the operator $t_{S}(\lambda)$ becomes
\EQ
\frac{t_{S}(\lambda)}{\rho_{S}^{(+)}}=\tr_{\cal A}{\left[ {\cal G}_{S}^{(+)} D_{S}^{(+)}(\lambda) \left[{\cal G}_{S}^{(+)}\right]^{-1} {\cal T}_{\cal A}^{(S)}(\lambda) K_{S}^{(-)}(\lambda) \left[{{\cal T}_{\cal A}^{(S)}}(-\lambda) \right]^{-1} \right]}.
\label{isertident}
\EN

We now proceed by inserting identity terms of type  $\left[{\cal
G}_{S}^{(+)}\right]^{-1} {\cal G}_{S}^{(+)}$ in between all the
fundamental operators that appear in the trace (\ref{isertident}).
By using the invariance of the trace under cyclic permutation one
can rewrite Eq.(\ref{isertident}) as
\EQ
\frac{t_{S}(\lambda)}{\rho_{S}^{(+)}}=\tr_{\cal A}{\left[
D_{S}^{(+)}(\lambda)  \widetilde{{\cal T}}_{\cal A}^{(S)}(\lambda)
\widetilde{K}_{S}^{(-)}(\lambda) \left[ \widetilde{{\cal
T}}_{\cal A}^{(S)}(-\lambda) \right]^{-1} \right]},
\label{isertId}
\EN
where $\widetilde{{\cal T}}_{\cal A}^{(S)}(\lambda)=\widetilde{{\cal L}}_{{\cal A} L}^{(S)}(\lambda)
\widetilde{{\cal L}}_{{\cal A} L-1}^{(S)}(\lambda)\dots \widetilde{{\cal L}}_{{\cal A} 1}^{(S)}(\lambda)$. The new
operator $\widetilde{{\cal L}}_{{\cal A} j}^{(S)}(\lambda)$ and $K$-matrix $\widetilde{K}_{S}^{(-)}(\lambda)$ are given in
terms of unitary transformations acting on the auxiliary space by the expressions,
\EQ
\widetilde{{\cal L}}_{{\cal A} j}^{(S)}(\lambda)= \left[{\cal G}_{S}^{(+)}\right]^{-1} {\cal L}_{{\cal A} j}^{(S)}(\lambda)
{\cal G}_{S}^{(+)},
\label{gaugeT}
\EN
and
\EQ
\widetilde{K}_{S}^{(-)}(\lambda) = \left[{\cal G}_{S}^{(+)}\right]^{-1} K_{S}^{(-)}(\lambda) {\cal G}_{S}^{(+)}.
\label{gaugeL}
\EN

It turns out, however, that the gauge transformation (\ref{gaugeT}) on the
${\cal L}_{{\cal A} j}^{(S)}(\lambda)$
operators can be reversed with the help of a second transformation on the quantum space \cite{RM}. In fact, one can use property (\ref{symmetry})  to define quantum space matrices $V_{j}$ acting non-trivially only at the $j$-th site
\EQ
V_{j}=\mbox{Id} \otimes \dots \otimes \mbox{Id} \otimes \underbrace{{\cal G}_{S}^{(+)}}_{j\mbox{-th}}
\otimes \mbox{Id} \otimes \dots \otimes \mbox{Id},
\label{tapa}
\EN
such that they are able to undo the
transformation (\ref{gaugeT}), namely
\EQ
V_{j}^{-1} \widetilde{{\cal L}}_{{\cal A}j}^{(S)}(\lambda) V_{j} = {\cal L}_{{\cal A}j}^{(S)}(\lambda).
\EN

We now can use this property in order to define a new double-row transfer matrix operator $\bar{t}_{S}(\lambda)$
\EQ
\bar{t}_{S}(\lambda)= \prod_{j=1}^{L} V_{j}^{-1} t_{S}(\lambda) \prod_{j=1}^{L} V_{j},
\EN
having only one non-diagonal $K$-matrix
\EQ
\frac{\bar{t}_{S}(\lambda)}{\rho_{S}^{(+)}}=\tr_{\cal A}{\left[ D_{S}^{(+)}(\lambda)  {\cal T}_{\cal A}^{(S)}(\lambda) \widetilde{K}_{S}^{(-)}(\lambda) \left[ {\cal T}_{\cal A}^{(S)}(-\lambda) \right]^{-1} \right]}.
\label{tbar}
\EN

Clearly, the operators $t_{S}(\lambda)$ and $\bar{t}_{S}(\lambda)$ have the same eigenvalues while their eigenstates $\ket{\psi}$ and $\ket{\bar{\psi}}$ are related by a similarity transformation,
\EQ
\ket{\psi}=\prod_{j=1}^{L} V_{j} \ket{\bar{\psi}}.
\label{statesrel}
\EN

Though this framework clearly brings a considerable simplification
in the original eigenvalue problem, it is not enough to make 
the diagonalization of the double-row
operator $\bar{t}_{S}(\lambda)$ with six
free boundary parameters amenable to a standard Bethe ansatz analysis. This 
is because the $K$-matrix $\widetilde{K}_{S}^{(-)}$
is generally non-diagonal which still
imposes us the difficulty of finding suitable reference states needed
to begin the Bethe ansatz computations. However, a great advantage of 
this formulation is that one can easily identify
the existence
of at least three cases of physical interest in
which the standard $SU(2)$ highest weight states could be used as
pseudovacuums to build up the whole Hilbert space. The simplest
occurs when one of the boundaries is free, say $K_{S
}^{(-)}(\lambda)=\mbox{Id}$ while the other is still arbitrary with three
free parameters. The next one is when the $K$-matrices $K_{S
}^{(\pm)}(\lambda)$ are diagonalizable in the same basis, i.e.
${\cal G}_{S}^{(+)}={\cal G}_{S}^{(-)}$ which implies that
we have altogether four distinct couplings say $c_{-}$, $d_{-}$ and
$\xi_{\pm}$. This includes the important symmetric situation where
the left and right $K$-matrices are the same but arbitrarily
non-diagonal. As far as the Bethe ansatz technicalities are
concerned the most general case in which $SU(2)$ highest weight vectors
can be used as a reference state is when the effective $\widetilde{K}_{S
}^{(-)}(\lambda)$ $K$-matrix becomes either upper or lower
triangular. This leads us to an open integrable system with five free couplings since such
condition imposes certain constraint between the parameters
$c_{\pm}$ and $d_{\pm}$. Substituting the representation (\ref{kdecompo}) in Eq.(\ref{gaugeL})
and after some algebra we find that there are
two possible classes of restrictions satisfying the above mentioned triangularity property. It turns out that
these constraints depend only upon the variables $c_{\pm}$, $d_{\pm}$
and their expressions are,
\EQ
(\mbox{I}) ~~~ \frac{1+\epsilon_{-} \sqrt{1+c_{-} d_{-}}}{c_{-}} = \frac{1+\epsilon_{+} \sqrt{1+c_{+} d_{+}}}{c_{+}},
\label{constrain1}
\EN
\EQ
(\mbox{II}) ~~~ \frac{-d_{-} }{1+\epsilon_{-} \sqrt{1+c_{-} d_{-}}} = \frac{1+\epsilon_{+} \sqrt{1+c_{+} d_{+}}}{c_{+}}.
\label{constrain2}
\EN

Depending on the ratio $\displaystyle
\varepsilon=\frac{\epsilon_{+}}{\epsilon_{-}}$ the zeros entries
of $\widetilde{K}_{S}^{(-)}(\lambda)$ are either bellow or
above the principal diagonal. This feature has been summarized in
Table \ref{tab1} for each manifold. Note that the diagonal
elements of the triangular matrix $\widetilde{K}_{S
}^{(-)}(\lambda)$ will necessarily be the eigenvalues of $K_{S
}^{(-)}(\lambda)$. By considering  decomposition
(\ref{kdecompo}) we conclude that such eigenvalues are exactly the
entries of the diagonal matrix $\rho_{S}^{(-)}
D_{S}^{(-)}(\lambda)$.

Considering the above discussions, we find that the formulation of 
a Bethe ansatz solution
for the eigenspectrum of $\bar{t}_{S}(\lambda)$ on the
parameters manifolds $(\mbox{I})$ and $(\mbox{II})$ is certainly worthwhile to pursue.  It will leads us
to benefit from the knowledge of the exact spectrum with five out of six possible boundary couplings,
a considerable number of free parameters at our disposal.
A fundamental ingredient in the algebraic Bethe ansatz is the
quadratic relations satisfied by the matrix elements of the
double-row monodromy matrix defined by \cite{SK}
\EQ
\overline{{\cal T}}_{\cal A}^{(S)}(\lambda)={\cal T}_{\cal A}^{(S)}(\lambda) \widetilde{K}_{S}^{(-)}(\lambda)
\left[{\cal T}_{\cal A}^{(S)}(-\lambda)\right]^{-1},
\EN
and
consequently the double-row operator $\bar{t}_{S}(\lambda)$ can be
written in the form
\EQ
\frac{\bar{t}_{S}(\lambda)}{\rho_{S}^{(+)}}=\tr_{\cal A}{\left[ D_{S}^{(+)}(\lambda) \overline{{\cal T}}_{\cal A}^{(S)}(\lambda) \right]}.
\EN

Taking into account the property (\ref{symmetry}) we see that the
effective $\widetilde{K}_{S}^{(-)}(\lambda)$ matrix satisfies
the same reflection equation (\ref{eqref}) as the original
$K$-matrix $K_{S}^{(-)}(\lambda)$. As a consequence of that
and the fact the entries of $\widetilde{K}_{
S}^{(-)}(\lambda)$ are $c$-numbers it follows that
$\overline{{\cal T}}_{\cal A}^{(S)}(\lambda)$ is also a solution
of the reflection equation, namely
\EQ
{\cal L}_{12}^{(S)}(\lambda-\mu) \stackrel{1}{{\overline{\cal T}}_{\cal A}^{(S)}}(\lambda) {\cal L}_{21}^{(S)}(\lambda+\mu)
\stackrel{2}{{\overline{\cal T}}_{\cal A}^{(S)}}(\mu) =
\stackrel{2}{{\overline{{\cal T}}}_{\cal A}^{(S)}}(\mu) {\cal L}_{12}^{(S)}(\lambda+\mu)
\stackrel{1}{{\overline{{\cal
T}}}_{\cal A}^{(S)}}(\lambda) {\cal L}_{21}^{(S)}(\lambda-\mu).
\label{intertwREL}
\EN

In the next section we will explore such quadratic algebra
together with the existence of a pseudovacuum on which
$\overline{{\cal T}}_{\cal A}^{(S)}(\lambda)$ acts triangularly to
present the expressions for eigenvalues of $\bar{t}_{S}(\lambda)$
as well as the corresponding Bethe ansatz equations.

\section{Algebraic Bethe ansatz}\label{BAalg}

In the next subsections we will consider the diagonalization of
the operator $\bar{t}_{S}(\lambda)$ in the most general
restrictive condition (I) or (II) by an algebraic formulation of
the Bethe ansatz. The other two situations mentioned in section
\ref{Eigenprob} are special cases and the corresponding
eigenvalues and Bethe ansatz results can be derived from the
results, for example, obtained for manifold (I). This is obvious when ${\cal
G_{S}}^{(+)}={\cal G_{S}}^{(-)}$ and in the case $K_{
S}^{(-)}(\lambda)=\mbox{Id}$ one needs to consider the limit
$\bar{\xi}_{-} \rightarrow \infty$ with fixed $\rho_{S}^{(-)}=1$
in the results to be given bellow.

\subsection{The spin-$\frac{1}{2}$ solution} \label{submeio}

Here we shall consider the diagonalization of the double-row
transfer matrix $\bar{t}_{\frac{1}{2}}(\lambda)$ by means of the quantum
inverse scattering method \cite{QI,SK}. The corresponding bulk
Boltzmann weights (\ref{Hsmodel}) are those of the isotropic
six-vertex model,
\EQ
{\cal L}_{12}^{(\frac{1}{2})}(\lambda)=\left(\begin{array}{cccc}
\lambda+\eta & 0 & 0 & 0 \\
0 & \lambda & \eta & 0 \\
0 & \eta & \lambda & 0 \\
0 & 0& 0 & \lambda+\eta \end{array}\right).
\EN

Following the remarks of section \ref{Eigenprob} we are
assuming that the boundary couplings $c_{\pm}$ and $d_{\pm}$ satisfy one
of the two possible constraints described by Eqs.(\ref{constrain1},\ref{constrain2}). In this situation the
effective $\widetilde{K}_{S}^{(-)}(\lambda)$ $K$-matrix is
triangular and its diagonal entries are proportional to the 
eigenvalues $f_{j}^{(-)}(\frac{1}{2};\lambda,\bar{\xi}_{-})$. Without losing generality
one can clearly consider the case in which
$\widetilde{K}_{\frac{1}{2}}^{(-)}(\lambda)$ is upper triangular, and after some simplifications in Eq.(\ref{gaugeL}) we find that
\EQ
\widetilde{K}_{\frac{1}{2}}^{(-)}(\lambda)=\varepsilon \rho_{\frac{1}{2}}^{(-)}
\left(\begin{array}{cc}
f_{1}^{(-)}(\frac{1}{2};\lambda,\varepsilon \bar{\xi}_{-}) & \sigma_{12}\frac{g_{22}^{(+)}}{g_{11}^{(+)}}\frac{\lambda}{\eta} \\
0 & f_{2}^{(-)}(\frac{1}{2};\lambda,\varepsilon \bar{\xi}_{-})  \end{array}\right).
\label{km1p2tilde}
\EN

The off-diagonal term in Eq.(\ref{km1p2tilde}) is not expected to
affect the eigenvalues of $\bar{t}_{\frac{1}{2}}(\lambda)$ but it will
certainly be relevant in the structure of the eigenvectors. The explicit expression for
$\sigma_{12}$ has been 
presented in Appendix A.
As discussed in section \ref{Eigenprob} a direct consequence of the upper triangular property of $\widetilde{K}_{S}(\lambda)$ is that the following $SU(2)$ highest state vector
\EQ
\ket{\bar{0}_{S}}= \prod_{j=1}^{L}\otimes \ket{S,S}_j, ~~~~~
\ket{S,S}=\left(\begin{array}{c}
1 \\ 0 \\ \vdots \\ 0 \end{array}\right)_{2S+1},
\EN
is an exact eigenvector of the double transfer matrix $\bar{t}_{S}(\lambda)$.

This means that the state $\displaystyle \ket{\bar{0}_{\frac{1}{2}}}
$ can be used as pseudovacuum to build up the other
eigenvectors of $\bar{t}_{\frac{1}{2}}(\lambda)$ following the strategy of
the algebraic Bethe ansatz approach \cite{QI,SK}. A main step in this
method involves writing the double-row monodromy matrix
$\overline{\cal T}_{\cal A}^{(\frac{1}{2})}(\lambda)$ in the $2 \times 2$
form
\EQ
\overline{{\cal T}}_{\cal A}^{(1/2)}=\left(
\begin{array}{cc}
A(\lambda) & B(\lambda) \\
C(\lambda) & D(\lambda)
\end{array} \right).
\EN

By using the intertwining relation (\ref{intertwREL}) and
following the procedure devised first by Sklyanin \cite{SK} one
can derive the commutation rules
\bear
A(\lambda)B(\mu)&=&\frac{(\mu-\lambda+\eta)(\mu+\lambda)}{(\mu+\lambda+\eta)(\mu-\lambda)}B(\mu)A(\lambda)-\frac{2\mu}{(2\mu+\eta)}\frac{\eta}{(\mu-\lambda)}B(\lambda)A(\mu)
\nonumber\\
&-&\frac{\eta}{(\mu+\lambda+\eta)}B(\lambda)\widetilde{D}(\mu),
\label{commutationrel1}
\\
\widetilde{D}(\lambda)B(\mu)&=&\frac{(\lambda+\mu+2\eta)(\lambda-\mu+\eta)}{(\lambda-\mu)(\lambda+\mu+\eta)}B(\mu)\widetilde{D}(\lambda)-\frac{2\eta(\lambda+\eta)}{(2\lambda+\eta)(\lambda-\mu)}B(\lambda)\widetilde{D}(\mu)
\nonumber \\
&+&\frac{4\eta\mu(\lambda+\eta)}{(\lambda+\mu+\eta)(2\lambda+\eta)(2\mu+\eta)}B(\lambda)A(\mu),
\label{commutationrel2}
\ear
where the new field $\widetilde{D}(\lambda)$ is introduced in order to simplify the commutation relations. It
is given by the following combination between the operators $A(\lambda)$ and $D(\lambda)$,
\EQ
\widetilde{D}(\lambda)=D(\lambda)-\frac{\eta}{2\lambda+\eta}A(\lambda).
\label{shift}\EN

In terms of the operators $A(\lambda)$ and
$\widetilde{D}(\lambda)$ the double-row transfer matrix eigenvalue
problem can now be written as
\EQ
\left[f_{1}^{(+)}(\frac{1}{2};\lambda,\bar{\xi}_{+})+ f_{2}^{(+)}(\frac{1}{2};\lambda,\bar{\xi}_{+}) \frac{\eta}{2\lambda+\eta} \right]
A(\lambda)\ket{\bar{\psi}} + f_{2}^{(+)}(\frac{1}{2};\lambda,\bar{\xi}_{+}) \widetilde{D}(\lambda)\ket{\bar{\psi}}=
\frac{\Lambda_{\frac{1}{2}}(\lambda)}{\rho_{\frac{1}{2}}^{(+)}} \ket{\bar{\psi}},
\label{doublerowT}
\EN
while the action of the fields $A(\lambda)$, $\widetilde{D}(\lambda)$ and $C(\lambda)$ on
the reference state $\ket{\bar{0}_{\frac{1}{2}}}$ are given by
\begin{subequations}
\bear
A(\lambda)\ket{\bar{0}_{\frac{1}{2}}} &=& \varepsilon \rho_{\frac{1}{2}}^{(-)}
f_{1}^{(-)}(\frac{1}{2};\lambda,\varepsilon \bar{\xi}_{-}) \left[ \frac{(\lambda+\eta )^2}{\zeta_{\frac{1}{2}}(\lambda)} \right]^L
\ket{\bar{0}_{\frac{1}{2}}}, \label{Anovacuo}
\\
\widetilde{D}(\lambda) \ket{\bar{0}_{\frac{1}{2}}} &=& \varepsilon \rho_{\frac{1}{2}}^{(-)} \left[
f_{2}^{(-)}(\frac{1}{2};\lambda,\varepsilon \bar{\xi}_{-})
-f_{1}^{(-)}(\frac{1}{2};\lambda,\varepsilon \bar{\xi}_{-})\frac{\eta}{2\lambda+\eta} \right]
\left[ \frac{\lambda^2}{\zeta_{\frac{1}{2}}(\lambda)} \right]^{L}  \ket{\bar{0}_{\frac{1}{2}}},\label{Dtnovacuo}
\\
C(\lambda)\ket{\bar{0}_{\frac{1}{2}}} &=& 0.
\label{Cnovacuo}
\ear
\end{subequations}

The fields $B(\lambda)$ are interpreted as a kind of creation
operators over the pseudovacuum $\ket{\bar{0}_{\frac{1}{2}}}$ and the multiparticle Bethe
states
$\ket{\bar{\psi}_n(\lambda_1,\dots,\lambda_n)}$
are supposed to be given by
\EQ
\ket{\bar{\psi}_n(\lambda_1,\dots,\lambda_n)}=B(\lambda_{1})\dots B(\lambda_{n}) \ket{\bar{0}_{\frac{1}{2}}}.
\label{eigen}
\EN

The rapidities $\lambda_{j}$ will be determined by solving the
eigenvalue problem with the above ansatz for the eigenvectors.
This is done with the  help of the commutation relations
(\ref{commutationrel1}, \ref{commutationrel2}) to move
$A(\lambda)$ and $\widetilde{D}(\lambda)$ in Eq.(\ref{doublerowT}) over the creation fields until they reach the
reference state $\ket{\bar{0}_{\frac{1}{2}}}$. The terms proportional to the eigenvectors (\ref{eigen})
are easily collected by keeping only the first pieces of the commutation rules. After using
expressions
(\ref{Anovacuo}-\ref{Dtnovacuo}) and some simplifications we find that the final
result for the eigenvalues are
\bear
\frac{\Lambda_{\frac{1}{2}}(\lambda)}{\rho_{\frac{1}{2}}^{(+)}\rho_{\frac{1}{2}}^{(-)}}=
\left[ \frac{(\lambda+\eta)^2}{\zeta_{\frac{1}{2}}(\lambda)} \right]^L
\left[\frac{\varepsilon2(\lambda+\eta)(\lambda+\varepsilon \eta \bar{\xi}_{-})(-\lambda+\eta
\bar{\xi}_{+})}{(2\lambda+\eta)\eta^{2}} \right] \prod_{j=1}^{n}
\frac{(\lambda_{j}-\lambda+\frac{\eta}{2})}{(\lambda_{j}-\lambda-\frac{\eta}{2})}
\frac{(\lambda_{j}+\lambda-\frac{\eta}{2})}{(\lambda_{j}+\lambda+\frac{\eta}{2})}+
\nonumber \\
+\left[ \frac{\lambda^2}{\zeta_{\frac{1}{2}}(\lambda)} \right]^{L}
\left[\frac{\varepsilon2\lambda(-\lambda+\varepsilon \eta
\bar{\xi}_{-}-\eta)(\lambda+\eta\bar{\xi}_{+}+\eta)}{(2\lambda+\eta)\eta^{2}}
\right] \prod_{j=1}^{n}
\frac{(\lambda_{j}-\lambda-\frac{3\eta}{2})}{(\lambda_{j}-\lambda-\frac{\eta}{2})}
\frac{(\lambda_{j}+\lambda+\frac{3\eta}{2})}{(\lambda_{j}+\lambda+\frac{\eta}{2})}
\ear
where we have used the values of $f_{j}^{(\pm)}(\frac{1}{2};\lambda,\bar{\xi}_{\pm})$
taken from Eqs.(\ref{fsalphaminus}, \ref{fsalphaplus}) and performed the 
displacements $\displaystyle \lambda_{i}\rightarrow
\lambda_{i}-\frac{\eta}{2}$ on the rapidities.

The remaining terms that are not proportional
to $\ket{\bar{\psi}(\lambda_1,\dots,\lambda_n)}$ can be canceled out by imposing further restrictions on the
rapidities $\lambda_j$. These are known as the Bethe ansatz equations which in our case are
given by
\EQ
\left[\frac{\lambda_{j}+\frac{\eta}{2}}{\lambda_{j}-\frac{\eta}{2}}
\right]^{2L}= \left(\frac{\lambda_{j}-\varepsilon
\eta\bar{\xi}_{-}+\frac{\eta}{2}}{\lambda_{j}+\varepsilon
\eta\bar{\xi}_{-}-\frac{\eta}{2}} \right)
\left(\frac{\lambda_{j}+\eta\bar{\xi}_{+}+\frac{\eta}{2}}{\lambda_{j}-\eta\bar{\xi}_{+}-\frac{\eta}{2}}
\right)\prod_{\stackrel{i=1}{i\neq j}}^{n}
\frac{(\lambda_{j}-\lambda_{i}+\eta)}{(\lambda_{j}-\lambda_{i}-\eta)}
\frac{(\lambda_{j}+\lambda_{i}+\eta)}{(\lambda_{j}+\lambda_{i}-\eta)}
\label{BAspin1p2}
\EN

We now can derive similar results for the open spin-$\frac{1}{2}$ chain that commutes with
the double-row transfer matrix $t_{\frac{1}{2}}(\lambda)$. The corresponding Hamiltonian is proportional to the first-order expansion of $t_{\frac{1}{2}}(\lambda)$ in the spectral parameter \cite{SK,DEV}
\EQ
{\cal H}_{\frac{1}{2}}=\frac{1}{\eta}\sum_{i=1}^{L-1} \left[ \sigma_{i}^{x}\sigma_{i+1}^{x} + \sigma_{i}^{y}\sigma_{i+1}^{y} + \sigma_{i}^{z}\sigma_{i+1}^{z} \right]
+ \frac{1}{\eta \xi_{-}} \left[ \sigma_{1}^{z} + c_{-} \sigma_{1}^{+} + d_{-}\sigma_{1}^{-} \right] - \frac{1}{\eta \xi_{+}}\left[\sigma_{L}^{z} + c_{+} \sigma_{L}^{+} + d_{+}\sigma_{L}^{-} \right],
\label{Hspin1p2}
\EN
where $\sigma_{\alpha}^{x}$, $\sigma_{\alpha}^{y}$ and $\sigma_{\alpha}^{z}$ are the
Pauli matrices with $\sigma_{\alpha}^{\pm}=\frac{1}{2} \left(\sigma_{\alpha}^{x} \pm i \sigma_{\alpha}^{y}\right)$. Its
eigenvalues $E_n(\lambda)$ are obtained in terms of the rapidities
$\lambda_{j}$ that satisfy the Bethe ansatz equation (\ref{BAspin1p2}) by the following expression
\EQ
E_n(\lambda)=2\eta \sum_{k=1}^{n}\frac{1}{\lambda_{k}^{2}-\frac{\eta^{2}}{4}} + \frac{L}{\eta}-\frac{1}{\eta} \left[ 1 + \frac{\rho_{\frac{1}{2}}^{(+)}}{\xi_{+}}  - \varepsilon \frac{\rho_{\frac{1}{2}}^{(-)}}{\xi_{-}} \right].
\EN

We would like to close this section with the following remark. The ferromagnetic $\eta <0$ Hamiltonian  (\ref{Hspin1p2}) is known to describe the stochastic dynamics of symmetric hopping of particles in one dimension provided that certain relations are satisfied by the boundary parameters \cite{SHU}. More specifically, letting $\alpha(\gamma)$ be the rate of injection (ejection) of particles at the left boundary and $\delta(\beta)$ the corresponding rate at the right boundary we have \cite{SHU,MUR}
\EQ
\alpha-\gamma = \frac{1}{\eta \xi_{-}}, ~~~~ \alpha= \frac{d_{-}}{2\eta \xi_{-}}, ~~~~ \gamma=\frac{c_{-}}{2\eta \xi_{-}},
\EN
and
\EQ
\beta-\delta = \frac{1}{\eta \xi_{+}}, ~~~~ \beta= -\frac{c_{+}}{2\eta \xi_{+}}, ~~~~ \delta=-\frac{d_{+}}{2\eta \xi_{+}}.
\EN

The above particular parameterization of the boundary parameters
$c_{\pm}$ and $d_{\pm}$ satisfies the constraints (I) or (II)
for arbitrary values of the particle injection and ejection rates. 
Though
the spectrum at this special case have been determined before \cite{SHU,MUR}
not much is known about the behavior of the wave functions. This information
can now be in principle extracted by combining the unitary
transformation (\ref{tapa}, \ref{statesrel}) with the multiparticle
state structure (\ref{eigen}). This knowledge of the eigenvectors
can be used to calculate correlation functions, thanks to
recent developments made in the quantum inverse scattering
method \cite{MAI,WAN} which allows us to reconstruct local
spin operators in terms of the monodromy matrix fields. We hope to return to this problem
since this could provide us with new insights on the physics of stochastic dynamics of interacting particle systems.

\subsection{The spin-$1$ solution}

The statistical system associated to the integrable XXX-Heisenberg model with spin-$1$ is a three-state vertex model with nineteen non-null Boltzmann weights given by
\EQ
{\cal L}_{12}^{(1)}(\lambda)=\left( \begin{array}{ccccccccc}
\bar{a}(\lambda) & 0 & 0 & 0 & 0 & 0 & 0 & 0 & 0 \\
0 & \bar{b}(\lambda) & 0 & \bar{h}(\lambda) & 0 & 0 & 0 & 0 & 0 \\
0 & 0 & \bar{e}(\lambda) & 0 & \bar{d}(\lambda) & 0 & \bar{c}(\lambda) & 0 & 0 \\
0 & \bar{h}(\lambda) & 0 & \bar{b}(\lambda) & 0 & 0 & 0 & 0 & 0 \\
0 & 0 & \bar{d}(\lambda) & 0 & \bar{g}(\lambda) & 0 & \bar{d}(\lambda) & 0 & 0 \\
0 & 0 & 0 & 0 & 0 & \bar{b}(\lambda) & 0 & \bar{h}(\lambda) & 0 \\
0 & 0 & \bar{c}(\lambda) & 0 & \bar{d}(\lambda) & 0 & \bar{e}(\lambda) & 0 & 0 \\
0 & 0 & 0 & 0 & 0 & \bar{h}(\lambda) & 0 & \bar{b}(\lambda) & 0 \\
0 & 0 & 0 & 0 & 0 & 0 & 0 & 0 & \bar{a}(\lambda)
\end{array}\right),
\label{Loperatorspin1}
\EN
with
\bear
\bar{a}(\lambda)=\lambda+ 2 \eta, ~~~ \bar{b}(\lambda)=\lambda, ~~~ \bar{c}(\lambda)=\frac{2\eta^2}{\lambda+\eta}, ~~~ \bar{d}(\lambda)=\frac{2\eta \lambda}{\lambda +\eta},  \\
\bar{e}(\lambda)=\frac{\lambda(\lambda-\eta)}{\lambda+\eta}, ~~~ \bar{g}(\lambda)=\bar{b}(\lambda)+\bar{c}(\lambda), ~~~ \bar{h}(\lambda)=2\eta.
\ear

As before we can consider the situation when the effective $\widetilde{K}_{1}(\lambda)$
matrix is upper triangular. In this case, carrying on few algebraic simplifications in Eq.(\ref{gaugeL}) we find that
\EQ
\widetilde{K}_{1}(\lambda)=\rho_{1}^{(-)} \left( \begin{array}{ccc}
f_{1}(1;\lambda,\varepsilon \bar{\xi}_{-}) & 2\frac{g_{22}^{(+)}}{g_{11}^{(+)}}
\frac{\lambda}{\eta}\left[ \kappa_{12}(\frac{1}{2}-\frac{\lambda}{\eta})-\kappa_{23} \right]
& 2\kappa_{13} \left(\frac{g_{22}^{(+)}}{g_{11}^{(+)}}\right)^{2} \frac{\lambda}{\eta}
\left(\frac{1}{2}-\frac{\lambda}{\eta} \right) \\
0 & f_{2}(1;\lambda,\varepsilon \bar{\xi}_{-}) & -2\frac{g_{22}^{(+)}}{g_{11}^{(+)}}
\frac{\lambda}{\eta}
\left[\kappa_{12}(\frac{1}{2}-\frac{\lambda}{\eta})+\kappa_{23} \right]
\\
0 & 0 & f_{3}(1;\lambda,\varepsilon \bar{\xi}_{-})
\end{array} \right),
\label{kmtildes1}
\EN
where the off-diagonal coefficients $\kappa_{12}$, $\kappa_{13}$ and $\kappa_{23}$ have been collected in Appendix A.

At this point we need to start introducing suitable notation for the double monodromy
operator $\overline{\cal T}_{\cal A}^{(S)}(\lambda)$. Here we shall use a representation which
can be easily extended to accommodate arbitrary spin-$S$ case,
\EQ
\overline{\cal T}_{\cal A}^{(1)}(\lambda)=\left(\begin{array}{ccc}
A_{1}(\lambda) & B_{12}(\lambda) & B_{13}(\lambda) \\
C_{21}(\lambda) & A_{2}(\lambda) & B_{23}(\lambda) \\
C_{31}(\lambda) & C_{32}(\lambda) & A_{3}(\lambda)
\end{array} \right).
\label{barmonodromyS1}
\EN

The next step is to rewrite the eigenvalue problem in terms of the double monodromy matrix elements. To perform  this task is convenient to define new diagonal operators $\widetilde{A}_{i}(\lambda)$ in terms of appropriate linear combinations of the fields $A_{i}(\lambda)$ \cite{SK,FAN}. This is done in such way that the action of the new fields on the state $\ket{\bar{0}_{1}}$ will be proportional to a single bulk term. Keeping in mind possible extension to general values of the spin we define,
\bear
A_{1}(\lambda)&=&\widetilde{A}_{1}(\lambda), \\
A_{2}(\lambda)&=&\widetilde{A}_{2}(\lambda) + \frac{\bar{h}(2\lambda)}{\bar{a}(2\lambda)} \widetilde{A}_{1}(\lambda), \\
A_{3}(\lambda)&=&\widetilde{A}_{3}(\lambda) + \frac{\bar{c}(2\lambda)}{\bar{a}(2\lambda)} \widetilde{A}_{1}(\lambda) + \frac{\bar{h}_{1}(2\lambda)}{\bar{h}_{2}(2\lambda)} \widetilde{A}_{2}(\lambda),
\ear
where the functions $\bar{h}_{1}(\lambda)$ and $\bar{h}_{2}(\lambda)$ are the following determinants
\EQ
\bar{h}_{1}(\lambda)=\left| \begin{array}{cc}
\bar{a}(\lambda) & \bar{c}(\lambda) \\ \bar{h}(\lambda) & \bar{h}(\lambda)
\end{array} \right|, ~~~~
\bar{h}_{2}(\lambda)=\left| \begin{array}{cc}
\bar{a}(\lambda) & \bar{h}(\lambda) \\ \bar{h}(\lambda) & \bar{g}(\lambda)
\end{array} \right|.
\EN

Taking into account the representation (\ref{barmonodromyS1})  and the above redefinitions of the diagonal fields, the diagonalization of the doubled transfer matrix $\bar{t}_{1}(\lambda)$ becomes equivalent to the problem
\EQ
\sum_{i=1}^{3} \omega_{i}^{(+)}(\lambda) \widetilde{A}_{i}(\lambda) \ket{\bar{\psi}_n(\lambda_1,\dots,\lambda_n)} = \frac{\Lambda_{1}(\lambda)}{\rho_{1}^{(+)}} \ket{\bar{\psi}_n(\lambda_1,\dots,\lambda_n)},
\EN
with
\bear
\omega_{1}^{(+)}(\lambda)&=&f_{1}^{(+)}(1;\lambda,\bar{\xi}_{+})+\frac{\bar{h}(2\lambda)}{\bar{a}(2\lambda)}f_{2}^{(+)}(1;\lambda,\bar{\xi}_{+}) + \frac{\bar{c}(2\lambda)}{\bar{a}(2\lambda)} f_{3}^{(+)}(1;\lambda,\bar{\xi}_{+}), \\
\omega_{2}^{(+)}(\lambda)&=&f_{2}^{(+)}(1;\lambda,\bar{\xi}_{+}) + \frac{\bar{h}_{1}(2\lambda)}{\bar{h}_{2}(2\lambda)}
f_{3}^{(+)}(1;\lambda,\bar{\xi}_{+}), \\
\omega_{3}^{(+)}(\lambda)&=&f_{3}^{(+)}(1;\lambda,\bar{\xi}_{+}).
\ear

Another important ingredient is to determine the action of the double monodromy matrix elements on the
pseudovacuum $\ket{\bar{0}_{1}}$. This can be done with the help of the Yang-Baxter algebra \cite{SK,FAN} and the
triangularity properties of both ${\cal L}_{ab}^{(1)}(\lambda)$ and
$\widetilde{K}_{1}^{(-)}(\lambda)$ operators upon $\ket{\bar{0}_{1}}$. Following ref.\cite{FAN} and
taking into account Eq.(\ref{kmtildes1}) we have
\bear
\widetilde{A}_{1}(\lambda)\ket{\bar{0}_{1}} &=&\rho_{1}^{(-)} \omega_{1}^{(-)}(\lambda) \left[ \frac{\bar{a}(\lambda)^{2}}{\zeta_{1}(\lambda)} \right]^{L} \ket{\bar{0}_{1}}, \\
\widetilde{A}_{2}(\lambda)\ket{\bar{0}_{1}} &=&\rho_{1}^{(-)} \omega_{2}^{(-)}(\lambda) \left[ \frac{\bar{b}(\lambda)^{2}}{\zeta_{1}(\lambda)} \right]^{L} \ket{\bar{0}_{1}}, \\
\widetilde{A}_{3}(\lambda)\ket{\bar{0}_{1}} &=&\rho_{1}^{(-)} \omega_{3}^{(-)}(\lambda) \left[ \frac{\bar{e}(\lambda)^{2}}{\zeta_{1}(\lambda)} \right]^{L} \ket{\bar{0}_{1}}, \nonumber \\
C_{21}(\lambda)\ket{\bar{0}_{1}} &=& C_{31}(\lambda)\ket{\bar{0}_{1}}=C_{32}(\lambda)\ket{\bar{0}_{1}}=0,
\ear
with
\bear
\omega_{1}^{(-)}(\lambda)&=&f_{1}^{(-)}(1;\lambda,\varepsilon\bar{\xi}_{-}), \\
\omega_{2}^{(-)}(\lambda)&=&f_{2}^{(-)}(1;\lambda,\varepsilon\bar{\xi}_{-}) - \frac{\bar{h}(2\lambda)}{\bar{a}(2\lambda)}
f_{1}^{(-)}(1;\lambda,\varepsilon\bar{\xi}_{-}), \\
\omega_{3}^{(-)}(\lambda)&=&f_{3}^{(-)}(1;\lambda,\varepsilon\bar{\xi}_{-}) -\frac{\bar{h}_{1}(2\lambda)}{\bar{h}_{2}(2\lambda)} f_{2}^{(-)}(1;\lambda,\varepsilon\bar{\xi}_{-})
- \frac{\bar{h}_{3}(2\lambda)}{\bar{h}_{2}(2\lambda)} f_{1}^{(-)}(1;\lambda,\varepsilon\bar{\xi}_{-}),
\ear
where the new function
$\bar{h}_{3}(\lambda)=\left| \begin{array}{cc} \bar{c}(\lambda) & \bar{h}(\lambda) \\  \bar{h}(\lambda) & \bar{g}(\lambda) \end{array} \right| $.

Also one expects that the operators $B_{12}(\lambda)$,
$B_{13}(\lambda)$ and $B_{23}(\lambda)$ play the role of creation
operators over the reference state $\ket{\bar{0}_{1}}$. Therefore
it is natural to seek for other eigenvectors of
$\bar{t}_{1}(\lambda)$ as linear combinations of products of these
creation fields acting on $\ket{\bar{0}_{1}}$. This is done by
exploring the commutation rules between the diagonal
$\widetilde{A}_{i}(\lambda)$ and the creation fields which can be
derived from the boundary Yang-Baxter algebra (\ref{intertwREL}).
A careful analysis of these relations reveals us that the
construction of the eigenvectors can be based on either
$B_{12}(\lambda)$ and $B_{13}(\lambda)$ or $B_{23}(\lambda)$ and
$B_{13}(\lambda)$ pair of fields rather than on arbitrary
combination of the three possible creation operators. We remark
that this redundancy is not particular of this system, but it is a
general feature of the algebraic Bethe ansatz framework developed
in ref.\cite{TA,PM} for a large family of integrable vertex models
with periodic boundary. This formalism has been generalized by
Li et al
ref.\cite{LI} to include vertex models with open boundaries based on ideas
first envisaged by Fan
\cite{FAN} and later extended for systems solvable
by nested Bethe ansatz \cite{GUA}. We also recall that in the context
of three state
vertex models this approach was recently reviewed in
ref.\cite{AL}. Considering that such algebraic framework has
already been described in these references, we shall not repeat
the details here, and in what follows we will present only the
main results for the eigenvectors and the eigenvalues. Here we
consider that the eigenvectors will be constructed in terms of a
linear combination of products of the creation fields
$B_{12}(\lambda)$ and $B_{13}(\lambda)$ acting on the vector
$\ket{\bar{0}_{1}}$.
It turns out that the eigenstates
$\ket{\bar{\psi}_n(\lambda_{1},\dots,\lambda_{n})}$
form a multiparticle structure  and they can be constructed as
\EQ
\ket{\bar{\psi}_n(\lambda_{1},\dots,\lambda_{n})}=
{\varphi}_n(\lambda_{1},\dots,\lambda_{n}) \ket{\bar{0}_1}
\EN
such that the vector 
${\varphi}_n(\lambda_{1},\dots,\lambda_{n})$ satisfy
a second order
recursion relation of the form
\bear
{\varphi}_{n}(\lambda_{1},\dots,\lambda_{n}) &=&
B_{12}(\lambda_{1}) {\varphi}_{n-1}(\lambda_{2},\dots,\lambda_{n}) \nonumber \\
&+& B_{13}(\lambda_{1}) \sum_{i=2}^{n}
{\varphi}_{n-2}(\lambda_{2},\dots,\lambda_{i-1},\lambda_{i+1},\dots,\lambda_n) \nonumber \\
&\times& \left (
\Gamma_{1}^{(i)}(\lambda_{1},\dots,\lambda_{n})  \widetilde{A}_{1}(\lambda_{i})
+\Gamma_{2}^{(i)}(\lambda_{1},\dots,\lambda_{n})  \widetilde{A}_{2}(\lambda_{i}) \right ),
\ear

Here we are assuming the identification $\ket{\bar{\psi}_0} \equiv \ket{\bar{0}_1} $. The functions
$\Gamma_{1}^{(i)}(\lambda_{1},\dots,\lambda_{n})$ and
$\Gamma_{2}^{(i)}(\lambda_{1},\dots,\lambda_{n})$ are given by
\EQ
\Gamma_{1}^{(i)}(\lambda_{1},\dots,\lambda_{n}) = -\frac{\bar{d}(\lambda_{1}-\lambda_{i})}{\bar{b}(\lambda_{1}+\lambda_{i})} \bar{p}(\lambda_{1},\lambda_{i}) \prod_{j=2}^{i-1} \frac{\bar{h}_{4}(\lambda_{j}-\lambda_{i})}{\bar{a}(\lambda_{j}-\lambda_{i})\bar{e}(\lambda_{j}-\lambda_{i})} \prod_{\stackrel{k=2}{k \neq i}}^{n} \frac{\bar{a}(\lambda_{k}-\lambda_{i})\bar{b}(\lambda_{k}+\lambda_{i})}{\bar{b}(\lambda_{k}-\lambda_{i})\bar{a}(\lambda_{k}+\lambda_{i})},
\EN
and
\EQ
\Gamma_{2}^{(i)}(\lambda_{1},\dots,\lambda_{n}) = \frac{\bar{d}(\lambda_{1}+\lambda_{i})}{\bar{b}(\lambda_{1}+\lambda_{i})}  \prod_{j=2}^{i-1} \frac{\bar{h}_{4}(\lambda_{j}-\lambda_{i})}{\bar{a}(\lambda_{j}-\lambda_{i})\bar{e}(\lambda_{j}-\lambda_{i})} \prod_{\stackrel{k=2}{k \neq i}}^{n} \frac{\bar{h}_{4}(\lambda_{i}-\lambda_{k})}{\bar{b}(\lambda_{i}-\lambda_{k})\bar{e}(\lambda_{i}-\lambda_{k})}\frac{\bar{h}_{2}(\lambda_{k}+\lambda_{i})}{\bar{a}(\lambda_{k}+\lambda_{i})\bar{b}(\lambda_{k}+\lambda_{i})},
\EN
with $\displaystyle \bar{p}(x,y)=\frac{\bar{e}(x+y)}{\bar{e}(x-y)}-\frac{\bar{h}(2y)}{\bar{a}(2y)}\frac{\bar{d}(x+y)}{\bar{d}(x-y)}$ and $\bar{h}_{4}(\lambda)=\left|\begin{array}{cc} \bar{g}(\lambda) & \bar{d}(\lambda) \\ \bar{d}(\lambda) & \bar{e}(\lambda)
\end{array} \right|$.

The action of the doubled transfer matrix $\bar{t}_{1}(\lambda)$ on the state
$\ket{\bar{\psi}(\lambda_{1},\dots,\lambda_{n})}$ is performed relying on
similar data for the $(n-1)$ and $(n-2)$ particle states and with help of mathematical induction. Adapting
the discussion of refs.\cite{LI,AL} to our case we can infer that the eigenvalue expression is
\bear
\frac{\Lambda_{1}(\lambda)}{\rho_{1}^{(+)}\rho_{1}^{(-)}}&=&
\omega_{1}^{(+)}(\lambda) \omega_{1}^{(-)}(\lambda) \left[ \frac{\bar{a}(\lambda)^2}{\zeta_{1}(\lambda)} \right]^L \prod_{j=1}^{n}
\frac{\bar{a}(\lambda_{j}-\lambda) \bar{b}(\lambda_{j}+\lambda)}{\bar{b}(\lambda_{j}-\lambda)\bar{a}(\lambda_{j}+\lambda)}
\nonumber \\
&+& \omega_{2}^{(+)}(\lambda)\omega_{2}^{(-)}(\lambda) \left[ \frac{\bar{b}(\lambda)^2}{\zeta_{1}(\lambda)} \right]^{L} \prod_{j=1}^{n}
\frac{\bar{h}_{4}(\lambda-\lambda_{j}) \bar{h}_{2}(\lambda+\lambda_{j})}{\bar{e}(\lambda-\lambda_{j}) \bar{b}(\lambda-\lambda_{j}) \bar{a}(\lambda+\lambda_{j}) \bar{b}(\lambda+\lambda_{j})}
\nonumber \\
&+& \omega_{3}^{(+)}(\lambda)\omega_{3}^{(-)}(\lambda) \left[\frac{\bar{e}(\lambda)^{2}}{\zeta_{1}(\lambda)} \right]^{L} \prod_{j=1}^{n}
\frac{\bar{b}(\lambda-\lambda_{j}) \bar{h}_{5}(\lambda+\lambda_{j})}{\bar{e}(\lambda-\lambda_{j}) \bar{e}(\lambda+\lambda_{j}) \bar{b}(\lambda+\lambda_{j})},
\label{eingS1}
\ear
where $\bar{h}_{5}(\lambda)=\left|\begin{array}{cc}
\bar{b}(\lambda) & \bar{d}(\lambda) \\ \bar{d}(\lambda) & \bar{b}(\lambda)
\end{array} \right|$ and provided that the rapidities $\lambda_{j}$ satisfy the following Bethe ansatz equations
\EQ
\left[\frac{\bar{a}(\lambda_{j})}{\bar{b}(\lambda_{j})} \right]^{2L} \frac{\omega_{1}^{(+)}(\lambda_{j}) \omega_{1}^{(-)}(\lambda_{j}) }{\omega_{2}^{(+)}(\lambda_{j}) \omega_{2}^{(-)}(\lambda_{j}) } \frac{\left[\bar{b}(2\lambda_{j})\right]^{2}}{\bar{h}_{2}(2\lambda_{j})} = \prod_{\stackrel{i=1}{i\neq j}}^{n}
\frac{\bar{b}(\lambda_{j}-\lambda_{i}) \bar{h}_{2}(\lambda_{j}+\lambda_{i})}{\bar{e}(\lambda_{j}-\lambda_{i}) \left[\bar{b}(\lambda_{j}+\lambda_{i})\right]^{2}}.
\label{BAS1}
\EN

Now we are almost ready to get standard expressions for the
eigenvalues and Bethe ansatz equations. By introducing the new
set of variables $\lambda_{i}=\lambda_{i}-\eta$
and performing many simplifications
in the functions
entering Eqs.(\ref{eingS1}, \ref{BAS1})
we conclude that the eigenvalues are
\bear
\frac{\Lambda_{1}(\lambda)}{\rho_{1}^{(+)}\rho_{1}^{(-)}}&=&
\left[\frac{(2\lambda+3\eta)(\lambda+\varepsilon\eta\bar{\xi}_{-}-\frac{\eta}{2})(\lambda+\varepsilon \eta \bar{\xi}_{-}+\frac{\eta}{2})(\lambda-\eta \bar{\xi}_{+}-\frac{\eta}{2})(\lambda-\eta\bar{\xi}_{+}+\frac{\eta}{2})}{(2\lambda+\eta) \eta^{4} } \right] \nonumber \\
&\times& \left[ \frac{(\lambda+2\eta)^2}{\zeta_{1}(\lambda)} \right]^L \prod_{j=1}^{n}
\frac{(\lambda_{j}-\lambda+\eta)}{(\lambda_{j}-\lambda-\eta)}
\frac{(\lambda_{j}+\lambda-\eta)}{(\lambda_{j}+\lambda+\eta)}
\nonumber \\
&+& \left[\frac{(\lambda+\varepsilon\eta\bar{\xi}_{-}-\frac{\eta}{2})(\lambda-\varepsilon\eta\bar{\xi}_{-}+\frac{3\eta}{2}) (\lambda-\eta\bar{\xi}_{+}-\frac{\eta}{2})(\lambda+\eta\bar{\xi}_{+}+\frac{3\eta}{2})}{\eta^{4}} \right] \nonumber \\
&\times& \left[ \frac{\lambda^2}{\zeta_{1}(\lambda)} \right]^{L} \prod_{j=1}^{n}
\frac{(\lambda_{j}-\lambda+\eta)}{(\lambda_{j}-\lambda-\eta)}
\frac{(\lambda+\lambda_{j}-\eta)}{(\lambda+\lambda_{j}+\eta)}
\frac{(\lambda-\lambda_{j}+2\eta)}{(\lambda-\lambda_{j})}
\frac{(\lambda+\lambda_{j}+2\eta)}{(\lambda+\lambda_{j})} \nonumber \\
&+& \left[\frac{(2\lambda-\eta)(\lambda-\varepsilon\eta\bar{\xi}_{-}+\frac{\eta}{2})(\lambda-\varepsilon \eta \bar{\xi}_{-}+\frac{3\eta}{2}) (\lambda+\eta \bar{\xi}_{+}+\frac{\eta}{2})(\lambda+\eta\bar{\xi}_{+}+\frac{3\eta}{2})}{(2\lambda+\eta)\eta^{4}} \right] \nonumber \\
&\times& \left[\frac{\left(\frac{\lambda(\lambda-\eta)}{\lambda+\eta} \right)^{2}}{\zeta_{1}(\lambda)} \right]^{L} \prod_{j=1}^{n}
\frac{(\lambda-\lambda_{j}+2\eta)}{(\lambda-\lambda_{j})}
\frac{(\lambda+\lambda_{j}+2\eta)}{(\lambda+\lambda_{j})},
\ear
while the Bethe ansatz equations for the new rapidities $\lambda_{i}$ becomes
\EQ
\left[\frac{\lambda_{j}+\eta}{\lambda_{j}-\eta}
\right]^{2L}= \left(\frac{\lambda_{j}-\varepsilon
\eta\bar{\xi}_{-}+\frac{\eta}{2}}{\lambda_{j}+\varepsilon
\eta\bar{\xi}_{-}-\frac{\eta}{2}} \right)
\left(\frac{\lambda_{j}+\eta\bar{\xi}_{+}+\frac{\eta}{2}}{\lambda_{j}-\eta\bar{\xi}_{+}-\frac{\eta}{2}}
\right)
\prod_{\stackrel{i=1}{i\neq j}}^{n}
\frac{(\lambda_{j}-\lambda_{i}+\eta)}{(\lambda_{j}-\lambda_{i}-\eta)}
\frac{(\lambda_{j}+\lambda_{i}+\eta)}{(\lambda_{j}+\lambda_{i}-\eta)}.
\label{BAspin1}
\EN
where ${\bar{\xi}}_{\pm}$ are taken from Eq.(\ref{xirem}) with $S=1$.

We finally remark that the results of this subsection offers us in principle the basis to solve
the $O(3)$ non-linear sigma model with non-diagonal open boundaries.  Due to the isomorphism $O(3) \sim SU(2)_2$
the elements of the operator (\ref{Loperatorspin1}) can indeed be interpreted as the scattering amplitudes
of the $S$-matrix associated to the $O(3)$ field theory \cite{ZAM}. One expects that similar relation is
also valid for the boundary scattering matrices \cite{MO}. In this case, we need to adapt our results to
include the solution of the eigenspectrum of an open transfer matrix in the presence of inhomogeneities,
following for example the lines of ref.\cite{SAL}. It would be interesting to exploit this possibility 
to determine the effects of the boundaries in the physics of the $O(3)$ model.

\subsection{The spin-$S$ solution}\label{spinSsol}

The classical analogue of the solvable spin-$S$ XXX model is the
$2S+1$-state vertex model (\ref{Hsmodel}) having the total number of
$\frac{1}{3}(2S+1)\left[2(2S+1)^2+1 \right]$ non-null Boltzmann weights. The
transformed upper triangular $\widetilde{K}_{S}^{(-)}(\lambda)$
matrix corresponding to the left boundary in
$\bar{t}_{S}(\lambda)$ is \EQ
\widetilde{K}_{S}^{(-)}(\lambda)=(\varepsilon)^{2S} \rho_{S}^{(-)}
\left(\begin{array}{cccccc}
 f_{1}^{(-)}(S;\lambda,\varepsilon \bar{\xi}_{-}) & * & * &  \cdots & * \\
0 & f_{2}^{(-)}(S;\lambda,\varepsilon \bar{\xi}_{-})  & * &  \cdots & * \\
0 & 0 &  f_{3}^{(-)}(S;\lambda,\varepsilon \bar{\xi}_{-})  & &  \vdots \\
\vdots & \vdots & & \ddots &  * \\
0 & 0 &  \cdots & 0 & f_{2S+1}^{(-)}(S;\lambda,\varepsilon \bar{\xi}_{-})
 \end{array} \right),
\EN
where $*$ denotes non-vanishing values that can be directly determined from Eq.(\ref{gaugeL}).

To implement the quantum inverse scattering framework we will represent the
doubled monodromy matrix by the following structure
\EQ
\overline{\cal T}_{\cal A}^{(S)}(\lambda)=\left(\begin{array}{ccccc}
A_{1}(\lambda) & B_{12}(\lambda) &  & \cdots & B_{1 (2S+1)}(\lambda) \\
C_{21}(\lambda) & A_{2}(\lambda) &  & \cdots & B_{2 (2S+1)}(\lambda )\\
 &  &  &       & \vdots \\
\vdots & \vdots & & \ddots & B_{2S (2S+1)}(\lambda) \\
C_{(2S+1) 1}(\lambda) & C_{(2S+1) 2}(\lambda) & \cdots &  C_{(2S+1) 2S}(\lambda)     &  A_{2S+1}(\lambda) \\
\end{array} \right).
\label{barmonodromyS}
\EN

The next step in the algebraic formulation consists in
determining the action of the $\overline{\cal T}_{\cal A}^{(S)}(\lambda)$ elements 
on the reference state $\ket{\bar{0}_{S}}$ which helps us to distinguish 
creation and annihilation fields as well as to reformulate the eigenvalues 
problem in terms of appropriate linear combinations of diagonal fields.  To perform
that we need to know certain commutation relations 
between the operators ${\cal T}_{\cal A}^{(S)}(\lambda)$ and $\left[ {\cal T}_{\cal A}^{(S)}(\lambda) \right]^{-1}$. This can be obtained by using Eq.(\ref{YBA}) with $\mu=-\lambda$ \cite{FAN} to get the following general matrix relation
\EQ
\left[ {\cal T}_{2}^{(S)}(-\lambda) \right]^{-1} {\cal L}_{12}^{(S)}(2\lambda) {\cal T}_{1}^{(S)}(\lambda) =
{\cal T}_{1}^{(S)}(\lambda) {\cal L}_{12}^{(S)}(2\lambda) \left[ {\cal T}_{2}^{(S)}(-\lambda) \right]^{-1}.
\label{relastar}
\EN

By applying both sides of relation (\ref{relastar}) on the pseudovacuum $\ket{\bar{0}_{S}}$ and by taking into account the upper triangular property of both ${\cal L}_{12}^{(S)}(2\lambda)$ and $\widetilde{K}_{S}^{(-)}(\lambda)$ when acting on the state $\ket{\bar{0}_{S}}$ we conclude that all the fields $C_{\alpha \beta}(\lambda)$ are annihilators,
\EQ
C_{\alpha \beta}(\lambda)\ket{\bar{0}_{S}}=0
\EN
while $B_{\alpha \beta}(\lambda)$ acts as creation fields upon $\ket{\bar{0}_{S}}$.

We also see that $A_{i}(\lambda)\ket{\bar{0}_{S}}$ for $i=2,\dots,2S+1$
turns out to be proportional to many distinct bulk terms of
the form $\left[ t_{i}(\lambda) \right]^{2L}$
since it involves the action of upper elements of
the operator $\left[ \widetilde{{\cal T}}_{\cal A}^{(S)}(-\lambda) \right]^{-1}$ on $\ket{\bar{0}_{S}}$.
In the specific case of the $XXX-S$ model the functions $t_{i}(\lambda)$ are
\EQ
t_{i}(\lambda)=(\lambda+2\eta S)\prod_{k=S-i+2}^{S} \frac{\lambda+\eta k-\eta S}{\lambda+\eta k+\eta S},
\EN

As remarked in previous sections this difficulty can be
circumvented by writing the fields $A_{i}(\lambda)$ as linear
combinations of new operators $\widetilde{A}_{i}(\lambda)$ such
that their action on $\ket{\bar{0}_{S}}$ is proportional only to
$\left[ t_{i}(\lambda) \right]^{2L}$ term. The solution of this
problem involves a considerable amount of algebraic work but the
final answer can fortunately be given in terms of the determinants of
certain $j \times j$ matrices that shall be denoted by $M_{j,i}^{(+)}(\lambda)$. Its elements are
determined in terms of the entries of the ${\cal L}_{12}^{(S)}(\lambda)$
operator. More precisely, by writing $\displaystyle {\cal
L}_{12}^{(S)}(\lambda)=\sum_{abcd=1}^{2S+1} R_{a,b}^{c,d}(\lambda)
\hat{e}_{cb} \otimes \hat{e}_{ad}$ we find that such linear
combination is
\EQ
A_{i}(\lambda)=\sum_{j=1}^{i} \frac{\left|M_{j,i}^{(+)}(2\lambda)\right|}{\left|M_{j,j}^{(+)}(2\lambda)\right|}
\widetilde{A}_{j}(\lambda)
\label{Acombination}
\EN
where the $j \times j$ matrix $M_{j,i}^{(+)}(\lambda)$ is given by
\EQ
M_{j,i}^{(+)}(\lambda)=\left(\begin{array}{ccccc}
R_{1,1}^{1,1}(\lambda) & R_{1,1}^{2,2}(\lambda) & \cdots & R_{1,1}^{j-1,j-1}(\lambda) & R_{1,1}^{i,i}(\lambda) \\
R_{2,2}^{1,1}(\lambda) & R_{2,2}^{2,2}(\lambda) & \cdots & R_{2,2}^{j-1,j-1}(\lambda) & R_{2,2}^{i,i}(\lambda) \\
\vdots & \vdots &  & \vdots & \vdots \\
R_{j,j}^{1,1}(\lambda) & R_{j,j}^{2,2}(\lambda) & \cdots & R_{j,j}^{j-1,j-1}(\lambda) & R_{j,j}^{i,i}(\lambda) \\
\end{array} \right)_{j \times j}.
\EN

By using the relation (\ref{Acombination}) and the action of all $A_{i}(\lambda)$ on the reference state we find that
\EQ
\widetilde{A}_{i}(\lambda)\ket{\bar{0}_{S}}= \rho_{S}^{(-)} \omega_{i}^{(-)}(\lambda) \left[ \frac{t_{i}^{2}(\lambda)}{\zeta_{S}(\lambda)} \right]^{L} \ket{\bar{0}_{S}},
\label{actionome}
\EN
where
\EQ
\omega_{i}^{(-)}(\lambda)=(\varepsilon)^{2S} \left[ f_{i}^{(-)}(S;\lambda,\varepsilon \bar{\xi}_{-}) -\sum_{k=1}^{i-1} \frac{\left| M_{i-1,k}^{(-)}(2\lambda) \right| }{\left| M_{i-1,i-1}^{(+)}(2\lambda) \right|}
f_{k}^{(-)}(S; \lambda,\varepsilon \bar{\xi}_{-}) \right],
\label{omegam}
\EN
while the entries of a second $j \times j$ auxiliary matrix $M_{j,i}^{(-)}(\lambda)$ are given by
\EQ
M_{j,i}^{(-)}(\lambda)=\left(\begin{array}{cccccccc}
R_{1,1}^{1,1}(\lambda) & R_{1,1}^{2,2}(\lambda) & \cdots & R_{1,1}^{i-1,i-1}(\lambda) & R_{1,1}^{j+1,j+1}(\lambda) & R_{1,1}^{i+1,i+1}(\lambda) & \cdots & R_{1,1}^{j,j}(\lambda) \\
R_{2,2}^{1,1}(\lambda) & R_{2,2}^{2,2}(\lambda) & \cdots & R_{2,2}^{i-1,i-1}(\lambda) & R_{2,2}^{j+1,j+1}(\lambda) & R_{2,2}^{i+1,i+1}(\lambda) & \cdots & R_{2,2}^{j,j}(\lambda) \\
\vdots & \vdots &   & \vdots & \vdots & \vdots &  & \vdots \\
R_{j,j}^{1,1}(\lambda) & R_{j,j}^{2,2}(\lambda) & \cdots & R_{j,j}^{i-1,i-1}(\lambda) & R_{j,j}^{j+1,j+1}(\lambda) & R_{j,j}^{i+1,i+1}(\lambda) & \cdots & R_{j,j}^{j,j}(\lambda) \\
\end{array} \right)_{j \times j},
\EN

Equipped with Eq.(\ref{Acombination}) one can now write the eigenvalue
problem (\ref{eigenvalueproblem}) in terms of the new diagonal
fields $\widetilde{A}_{i}(\lambda)$, namely
\EQ
\sum_{k=1}^{2S+1} \omega_{k}^{(+)}(\lambda) \widetilde{A}_{k}(\lambda) \ket{\bar{\psi}} = \frac{\Lambda_{S}(\lambda)}{\rho_{S}^{(+)}} \ket{\bar{\psi}},
\EN
where
\EQ
\omega_{k}^{(+)}(\lambda)=\sum_{i=k}^{2S+1}  \frac{\left| M_{k,i}^{(+)}(2\lambda)
\right|}{\left| M_{k,k}^{(+)}(2\lambda) \right|} f_{i}(S;\lambda,\bar{\xi}_{+}).
\label{omegap}
\EN

At this stage we would like to
emphasize that the role construction presented above
is  applicable to any multistate vertex model whose
Boltzmann weights  are invariant by one
$U(1)$ charge conservation symmetry. In order to get manageable expressions for the eigenvalues, however,
one still needs to carry on cumbersome simplifications on the general formulae given in Eqs.(\ref{omegam},\ref{omegap}).
In the case of the $XXX-S$ model
we are able to show that all the contributions to
$\omega_{i}^{(\pm)}(\lambda)$ miraculously factorized in the following product forms
\EQ
\omega_{i}^{(\pm)}(\lambda)=\tau_{i}^{(\pm)}(\lambda) \chi_{i}^{(\pm)}(\lambda,\bar{\xi}_{\pm})
\EN
where
\bear
\tau_{i}^{(+)}(\lambda)=\prod_{k=1}^{i} \frac{2\lambda+\left[2S+3-i-k \right]\eta}{2\lambda+\left[2-k \right]\eta}, \\
\tau_{i}^{(-)}(\lambda)=(\varepsilon)^{2S}\prod_{k=i}^{2S+1} \frac{2\lambda+\left[2-2i+k \right]\eta}{2\lambda+\left[1+k-i \right]\eta},
\ear
and
\bear
\chi_{i}^{(+)}(\lambda,\bar{\xi}_{+})&=&
\prod_{j=1}^{2S+1-i} \left[ \bar{\xi}_{+} +S+\frac{1}{2}-j-
\frac{\lambda}{\eta} \right] 
\prod_{j=1}^{i-1} \left[ \bar{\xi}_{+} +S+\frac{3}{2}-j+\frac{\lambda}{\eta} \right], \\
\chi_{i}^{(-)}(\lambda,\bar{\xi}_{-})&=&\prod_{j=i}^{2S} \left[ \varepsilon \bar{\xi}_{-} + S +\frac{1}{2}-j+
\frac{\lambda}{\eta} \right]
\prod_{j=2S+2-i}^{2S} \left[ \varepsilon \bar{\xi}_{-} +S-\frac{1}{2}-j-\frac{\lambda}{\eta} \right].
\ear

Before proceeding with further results we stress that the above explicit expressions for
$\omega_{i}^{(\pm)}(\lambda)$
with arbitrary $S$ are  novel
in the literature since they were unknown even in the case of diagonal boundaries \cite{DOI}.
Now we reached a point in which we have gathered the basic
ingredients to start an algebraic Bethe ansatz analysis of the eigenspectrum of $\bar{t}_{S}(\lambda)$.  In
particular the vector $\ket{\bar{0}_{S}}$ is itself an eigenstate of $\bar{t}_{S}(\lambda)$ with the eigenvalue
\EQ
\frac{\Lambda_{S}^{(0)}(\lambda)}{\rho_{S}^{(+)} \rho_{S}^{(-)}} = \sum_{i=1}^{2S+1} \omega_{i}^{(+)}(\lambda) \omega_{i}^{(-)}(\lambda)  \left[ \frac{t_{i}^{2}(\lambda)}{\zeta_{S}(\lambda)} \right]^{L}.
\EN

The other eigenvectors of $\bar{t}_{S}(\lambda)$ are looked as states created by
the action of the fields $B_{\alpha \beta}(\lambda)$ on the reference state
$\ket{\bar{0}_{S}}$.  A single particle excitation is made by lowering
the value of  the azimuthal spin component by an unity on
the ferromagnetic pseudovacuum $\ket{\bar{0}_{S}}$. From the point of view
of the algebraic Bethe ansatz framework
this excitation can be represented by
$B_{jj+1}(\lambda_1)
\ket{\bar{0}_{S}}$ for any $j=1,\dots,2S$. As far as commutation relations are
concerned we find that it is simpler
to choose the one-particle state as
\EQ
\ket{\bar{\psi}_{1}(\lambda_1)}= B_{12}(\lambda_{1}) \ket{\bar{0}_{S}}.
\EN

The action of the double transfer matrix $\bar{t}_{S}(\lambda)$ on
this state can be computed with the aid of the
commutation relations between the fields $\widetilde{A}_{i}(\lambda)$
and $B_{12}(\lambda)$ that can be obtained from the boundary
Yang-Baxter algebra (\ref{intertwREL}).  In Appendix B we present details  of our
analysis of the one-particle eigenvalue problem for $S=\frac{3}{2}$. This study together with
the previous results for $S=1$ \cite{LI,AL} and the help of mathematical
induction lead us to the  following general expression
\bear
\frac{\bar{t}_{S}(\lambda)}{\rho_{S}^{(+)} \rho_{S}^{(-)}}
\ket{\bar{\psi}_{1}(\lambda_1)} &=& \sum_{i=1}^{2S+1} \left[ \frac{t_{i}^{2}(\lambda)}{\zeta_{S}(\lambda)} \right]^{L}
\omega_{i}^{(+)}(\lambda) \omega_{i}^{(-)}(\lambda)
Q_{i}(\lambda, \lambda_{1})\ket{\bar{\psi}_{1}(\lambda_1)} \nonumber \\
&+& \sum_{i=1}^{2S} B_{i i+1}(\lambda)
\left[q_i^{(1)}(\lambda,\lambda_{1})\widetilde{A}_{1}(\lambda_{1})
+ q_i^{(2)}(\lambda,\lambda_{1}) \widetilde{A}_{2}(\lambda_{1}) \right]
\ket{\bar{0}_{S}},
\label{EigVecSpS}
\ear
where function $Q_{i}(\lambda,\lambda_{j})$ is given by
\EQ
Q_{i}(\lambda,\lambda_{j})=\begin{cases}
\displaystyle \frac{R_{1,1}^{1,1}(\lambda_{j}-\lambda)R_{1,2}^{2,1}
(\lambda_{j}+\lambda)}{R_{1,2}^{2,1}(\lambda_{j}-\lambda)
R_{1,1}^{1,1}(\lambda_{j}+\lambda)}, ~~~~ i=1 \cr
\displaystyle \frac{\left| \begin{array}{cc}
R_{1,i+1}^{i+1,1}(\lambda-\lambda_{j}) & R_{1,i}^{i+1,2}(\lambda-\lambda_{j}) \\
R_{2, i+1}^{i, 1}(\lambda-\lambda_{j}) & R_{2, i}^{i, 2}(\lambda-\lambda_{j})
\end{array} \right|}{R_{1, i}^{i, 1}(\lambda-\lambda_{j})
R_{1,i+1}^{i+1,1}(\lambda-\lambda_{j})} \frac{\left| \begin{array}{cc}
R_{1,i-1}^{i-1,1}(\lambda+\lambda_{j}) & R_{1,i-1}^{i,2}(\lambda+\lambda_{j}) \\
R_{2, i}^{i-1, 1}(\lambda+\lambda_{j}) & R_{2, i}^{i,2}(\lambda+\lambda_{j}) \end{array} \right|}{R_{1,
i-1}^{i-1,1}(\lambda+\lambda_{j}) R_{1, i}^{i, 1}(\lambda+\lambda_{j})}, ~~~~ i=2,\dots,2S \cr
\displaystyle \frac{R_{2,2S+1}^{2S+1, 2}(\lambda-\lambda_{j})}{R_{1, 2S+1}^{2S+1, 1}(\lambda-\lambda_{j})} \frac{\left|
\begin{array}{cc}
R_{1, 2S}^{2S, 1}(\lambda+\lambda_{j}) & R_{1, 2S}^{2S+1, 2}(\lambda+\lambda_{j}) \\
R_{2, 2S+1}^{2S, 1}(\lambda+\lambda_{j}) & R_{2, 2S+1}^{2S+1, 2}(\lambda+\lambda_{j})
\end{array} \right|}{R_{1, 2S}^{2S, 1}(\lambda+\lambda_{j}) R_{1, 2S+1}^{2S+1, 1}(\lambda+\lambda_{j})}, ~~~~ i=2S+1
\end{cases}
\EN

From (\ref{EigVecSpS}) we see that the unwanted terms proportional to $B_{ii+1}(\lambda)$ can be eliminated by imposing that the rapidity $\lambda_{1}$ satisfies the following one-particle Bethe ansatz equation,
\EQ
\left[\frac{t_{1}(\lambda_{1})}{t_{2}(\lambda_{1})} \right]^{2L} \frac{\omega_{1}^{(-)}(\lambda_{1}) }{\omega_{2}^{(-)}(\lambda_{1}) } = -\frac{q_{i}^{(2)}(\lambda,\lambda_{1})}{q_{i}^{(1)}(\lambda,\lambda_{1})}, ~~~~i=1,\dots,2S.
\label{BASS}
\EN

We note that though the expressions  for $q_{i}^{(1)}(\lambda,\lambda_{1})$ 
and $q_{i}^{(2)}(\lambda,\lambda_{1})$ have in general a very involved dependence on the i-th index, 
see for instance Appendix B, we have found out that the ratio 
$\frac{q_{i}^{(2)}(\lambda,\lambda_{1})}{q_{i}^{(1)}(\lambda,\lambda_{1})}$ does not depend 
of such index. Its expression for arbitrary S, coming directly from the commutation 
rules, involves many complicated terms and 
it has been collected in Appendix D. It turns out, however, 
that it is possible to carry out further simplifications in equations (\ref{q1},\ref{q2}) 
thanks to several identities between the Boltzmann weights $R_{a,b}^{c,d}(\lambda)$. This also leads us to 
conclude that the ratio $\frac{q_{i}^{(2)}(\lambda,\lambda_{1})}{q_{i}^{(1)}(\lambda,\lambda_{1})}$ does not 
depend on the spectral parameter $\lambda$. This is consistent to what one would expect from a 
standard Bethe ansatz analysis and the
simplified expression for such ratio reads
\EQ
\frac{q_{i}^{(2)}(\lambda,\lambda_{1})}{q_{i}^{(1)}(\lambda,\lambda_{1})}=- 
\frac{\omega_{2}^{(+)}(\lambda_{1}) }{\omega_{1}^{(+)}(\lambda_{1}) } \Theta(\lambda_{1})
\label{RATIOT}
\EN
where for later convenience we define function $\Theta(\lambda)$ separately, namely 
\EQ
\Theta(\lambda)=\frac{R_{1,1}^{1,1}(\lambda)R_{2,2}^{2,2}(\lambda)-R_{1,1}^{2,2}(\lambda)R_{2,2}^{1,1}(\lambda)}{\left[R_{1,2}^{2,1}(\lambda)\right]^{2}}.
\EN

Putting all these results together we find that $\ket{\bar{\psi}_{1}(\lambda_1)}$ is an 
eigenvector of $\bar{t}_{S}(\lambda)$ with eigenvalue $\Lambda_{1}(\lambda,\lambda_1)$ given by
\EQ
\frac{\Lambda_{S}^{(1)}(\lambda,\lambda_{1})}{\rho_S^{(+)} \rho_S^{(-)}}=
\sum_{i=1}^{2S+1} \left[ \frac{t_{i}^{2}(\lambda)}{\zeta_{S}(\lambda)} \right]^{L}
\omega_{i}^{(+)}(\lambda) \omega_{i}^{(-)}(\lambda)
Q_{i}(\lambda, \lambda_{1})
\EN
provided that the variable $\lambda_{1}$ satisfies the restriction
\EQ
\left[\frac{t_{1}(\lambda_{1})}{t_{2}(\lambda_{1})} \right]^{2L} \frac{\omega_{1}^{(+)}(\lambda_{1})\omega_{1}^{(-)}(\lambda_{1})}{\omega_{2}^{(+)}(\lambda_{1})\omega_{2}^{(-)}(\lambda_{1})} = \Theta(\lambda_{1}).
\label{BASSsimp}
\EN

Here we remark that the equation (\ref{BASSsimp}) is equivalent to the condition of analyticity of $\Lambda_{1}(\lambda,\lambda_{1})$ as a function of the rapidity $\lambda_{1}$. This fact is indeed an extra verification of the validity of our Bethe ansatz analysis.

We now turn to the analysis of the two-particle state. In this case one expects that this 
state should be given in terms of two linearly independent vectors 
$B_{12}(\lambda_{1})B_{12}(\lambda_{2})\ket{\bar{0}_{S}}$ and $B_{13}(\lambda_{1})\ket{\bar{0}_{S}}$. Previous 
experience in determining two-particle states \cite{FAN,TA,PM} suggests us to look first 
for the commutation rule between the fields $B_{12}(\lambda_{1})$ and $B_{12}(\lambda_{2})$. To avoid overcrowding 
this section with more heavier formulae we have exhibited this relation for $S\ge1$ in Appendix D. From
equations (\ref{Acombination}) and the observations made in Appendix D we clearly see that the state
\bear
\ket{\bar{\psi}_{2}(\lambda_{1},\lambda_{2})} ~=~
B_{12}(\lambda_{1}) B_{12}(\lambda_{2})&+&B_{13}(\lambda_{1}) \Biggl[\frac{R_{1,2}^{3,2}(\lambda_{1}+\lambda_{2})}{R_{1,2}^{2,1}(\lambda_{1}+\lambda_{2})}\widetilde{A}_{2}(\lambda_{2})~+
\nonumber\\
+~\Biggl(\frac{R_{1,2}^{3,2}(\lambda_{1}+\lambda_{2})}{R_{1,2}^{2,1}(\lambda_{1}+\lambda_{2})}\frac{\left| M_{1,2}^{(+)}(2\lambda_{2})
\right|}{\left| M_{1,1}^{(+)}(2\lambda_{2}) \right|}&-&\frac{R_{1,2}^{3,2}(\lambda_{1}-\lambda_{2})}{R_{1,3}^{3,1}(\lambda_{1}-\lambda_{2})} \frac{R_{1,3}^{3,1}(\lambda_{1}+\lambda_{2})}{R_{1,2}^{2,1}(\lambda_{1}+\lambda_{2})}
\Biggr)\widetilde{A}_{1}(\lambda_{2})\Biggr]\ket{\bar{0}_{S}},
\label{TWOS}
\ear
is symmetric under the exchange of the rapidities $\lambda_{1}$ and $\lambda_{2}$. In other words we have that
\EQ
\ket{\bar{\psi}_{2}(\lambda_{1},\lambda_{2})} = Z_{S}(\lambda_{1},\lambda_{2}) \ket{\bar{\psi}_{2}(\lambda_{2},\lambda_{1})}
\EN
where $Z_{S}(\lambda_{1},\lambda_{2})$ is the following function,
\EQ
Z_{S}(\lambda_1,\lambda_2)=- \frac{R_{2,1}^{1,2}(\lambda+\lambda_1)}{R_{1,2}^{2,1}(\lambda+\lambda_1)}
\frac{\left|
\begin{array}{cc}
R_{1, 2}^{3, 2}(\lambda-\lambda_{1}) & R_{1, 3}^{3, 1}(\lambda-\lambda_{1}) \\
R_{2, 2}^{2, 2}(\lambda-\lambda_{1}) & R_{2, 3}^{2, 1}(\lambda-\lambda_{1})
\end{array} \right|}{R_{1, 1}^{1, 1}(\lambda-\lambda_{1}) R_{1, 3}^{3, 1}(\lambda-\lambda_{1})}
\label{ZZU}
\EN

This state is therefore an educated ansatz for the two-particle vector for general $S\ge1$. Note that it 
reproduces the previous state for $S=1$ \cite{FAN,LI,AL} and in Appendix B we have presented all the  
needed evidences that it is indeed a suitable eigenvector for $S=\frac{3}{2}$. The corresponding 
eigenvalue can be calculated by keeping only the terms proportional to the vector 
$B_{12}(\lambda_{1}) B_{12}(\lambda_{2})$ coming  from the first part of the
commutation relations between the fields $\tilde{A}_i(\lambda)$ and $B_{12}(\lambda_i)$. Taking into 
account our previous experience with the one-particle state and the structure of the commutation rules  discussed in
Appendices B and D we find that
\EQ
\frac{\Lambda_{S}^{(2)}(\lambda,\lbrace\lambda_{1},\lambda_{2}\rbrace)}{\rho_S^{(+)}\rho_S^{(-)}}=\sum_{i=1}^{2S+1} \left[ \frac{t_{i}^{2}(\lambda)}{\zeta_{S}(\lambda)} \right]^{L}
\omega_{i}^{(+)}(\lambda) \omega_{i}^{(-)}(\lambda) \prod_{j=1}^{n=2} Q_{i}(\lambda, \lambda_{j})
\label{EigVa2SpS}
\EN

The associated Bethe ansatz equations are expected to be the condition on the rapidities such that the residues at the simple poles $\lambda=\lambda_{1},\lambda_{2}$ present in functions $Q_{i}(\lambda, \lambda_{j})$ vanish. This condition is equivalent to the following system of equations
\EQ
\left[\frac{t_{j}(\lambda_{j})}{t_{2}(\lambda_{j})} \right]^{2L} \frac{\omega_{1}^{(+)}(\lambda_{j})\omega_{j}^{(-)}(\lambda_{j})}{\omega_{2}^{(+)}(\lambda_{j})\omega_{2}^{(-)}(\lambda_{j})} = \Theta(\lambda_{j}) \prod_{\stackrel{i=1}{i\neq j}}^{n=2}\frac{Q_{2}(\lambda_{j}, \lambda_{i})}{Q_{1}(\lambda_{j}, \lambda_{i})} ~~~~ j=1,\dots,n=2.
\label{BA2SSsimp}
\EN

By the some token, one expects that general multiparticle states can 
in principle be constructed in terms of a recurrence relation of order 
$2S$ that involves the creation fields $B_{1j}(\lambda)~~j=2,\dots,2S+1$. The precise 
structure of such relation for arbitrary $S$ has however eluded us so far. This by no means 
prevents us to propose general expressions for the corresponding eigenvalues and Bethe ansatz equations. In 
any factorizable theory, it is believed that the two-particle results already contain 
the main flavour about the content of the spectrum. This means that the expressions (\ref{EigVa2SpS}) and (\ref{BA2SSsimp}) are expected to be valid for general values of $n \le2LS$. Considering these observations 
and after working out the explicit expressions 
for functions $Q_{i}(\lambda, \lambda_{j})$ we find that the 
n-particle eigenvalue $\Lambda_{S}^{(n)}(\lambda,\lbrace\lambda_{i}\rbrace)$ is given by
\bear
\frac{\Lambda_{S}^{(n)}(\lambda,\lbrace\lambda_{i}\rbrace)}{\rho_{S}^{(+)}\rho_{S}^{(-)}}=\sum_{i=1}^{2S+1} \left[\frac{t_{i}^{2}(\lambda)}{\zeta_{S}(\lambda)} \right]^{L} \omega_{i}^{(+)}(\lambda) \omega_{i}^{(-)}(\lambda) ~~~~~~~~~~~~~~
\nonumber \\
\times \prod_{j=1}^{n} \frac{[\lambda-\lambda_{j}+\eta(S+1)][\lambda-\lambda_{j}-\eta S]}{[\lambda-\lambda_{j}+\eta (S+2-i)][\lambda-\lambda_{j}+\eta(S+1-i)]} \frac{[\lambda+\lambda_{j}+\eta(S+1)][\lambda+\lambda_{j}-\eta S]}{[\lambda+\lambda_{j}+\eta (S+2-i)][\lambda+\lambda_{j}+\eta(S+1-i)]}
\nonumber \\
\label{EigVASSpS}
\ear
while the Bethe ansatz equations are given by
\EQ
\left[\frac{\lambda_{j}+\eta S}{\lambda_{j}-\eta S}
\right]^{2L}= \left(\frac{\lambda_{j}-\varepsilon
\eta\bar{\xi}_{-}+\frac{\eta}{2}}{\lambda_{j}+\varepsilon
\eta\bar{\xi}_{-}-\frac{\eta}{2}} \right)
\left(\frac{\lambda_{j}+\eta\bar{\xi}_{+}+\frac{\eta}{2}}{\lambda_{j}-\eta\bar{\xi}_{+}-\frac{\eta}{2}}
\right)
\prod_{\stackrel{i=1}{i\neq j}}^{n}
\frac{(\lambda_{j}-\lambda_{i}+\eta)}{(\lambda_{j}-\lambda_{i}-\eta)}
\frac{(\lambda_{j}+\lambda_{i}+\eta)}{(\lambda_{j}+\lambda_{i}-\eta)},
\label{BAspinS}
\EN
where we have performed the displacement $\lambda_{i} \rightarrow \lambda_{i} - \eta S$ in order to bring 
the Bethe ansatz equations in a more symmetrical form.

At this point it should be emphasized that 
the right hand side of the Bethe ansatz equations (\ref{BAspinS}) depend on
both the spin $S$ and the off-diagonal elements $c_{\pm},d_{\pm}$ through 
the renormalized variable ${\bar{\xi}}_{\pm}$ defined in Eq.(\ref{xirem}).
We also mention that we have verified numerically for several values of $L$ and $S$ that 
the equations (\ref{EigVASSpS},\ref{BAspinS}) indeed reproduces 
the ground state and few low-lying excitations of the 
double-row transfer matrix $\bar{t}_{S}(\lambda)$. In particular, we have 
been able to check the completeness of the Bethe ansatz solution for $L=2$ up to $S=\frac{3}{2}$.
Finally,  we remark
that the final results for the eigenvalues (\ref{EigVa2SpS}) and 
Bethe ansatz equations (\ref{BA2SSsimp}) are expected to be valid for any integrable vertex model whose
underlying $R$-matrix possesses an unique $U(1)$ charge symmetry and the invariance (\ref{pari},\ref{tempo}).

\section{Conclusions}\label{Conclusion}

The purpose of this paper was to solve the integrable XXX-$S$ Heisenberg model with 
open boundary conditions by means of the quantum inverse 
scattering approach. We first argued that the 
corresponding $K$-matrices are diagonalizable by special 
similarity transformations without a dependence on the spectral parameter.  This fact together 
with the property of reversing gauge transformed Boltzmann weights leads us 
to an eigenvalue problem with only one non-diagonal effective $K$-matrix. In the cases 
when such $K$-matrix are either upper or lower triangular we 
managed to present explicit expressions for the eigenvalues of 
the doubled transfer matrix operator $t_{S}(\lambda)$ as well 
as the associated Bethe ansatz equations for arbitrary values of 
the spin-$S$.  This condition was shown 
to be equivalent to two possible constraints between 
the four off-diagonal boundary parameters, leading us
with five free parameters out of six possible ones.

We hope that the ideas developed in this paper will be also suitable 
to solve a broad class of isotropic integrable systems 
with non-diagonal open boundaries. In fact, the method devised 
here has been first applied to the fundamental $SU(N)$ isotropic 
vertex model under more restrictive open boundary conditions \cite{GA}. We expect 
that the nested Bethe ansatz approach could be further generalized 
to tackle effective triangular $K$-matrices 
which will provide us the solution of the associated 
doubled transfer matrix 
operator with fewer constrained boundary 
parameters as compared to that presented in ref.\cite{GA}. We also 
hope that other vertex models based on higher rank 
symmetries such as $O(N)$ and $sp(2N)$ Lie algebras could 
be dealt by the framework discussed in this work. This  assumes that certain classes of
non-diagonal $K$-matrices of these vertex models can be classified in 
terms of similarity transformations that are 
itself symmetries of the corresponding 
$\cal L$-operator, acting on spectral 
dependent diagonal solutions 
for the reflection equation. This would means that our observation of section (\ref{Kprop}) 
for $SU(2)$ could be generalized to other Lie algebras as well. We plan to investigate 
such rather interesting possibility in a future work.

\section*{Acknowledgements}
The authors C.S. de Melo and G.A.P. Ribeiro thank  FAPESP (Funda\c c\~ao de Amparo \`a Pesquisa do Estado de S\~ao Paulo)
for financial support. The work of M.J. Martins has been supported by the Brazilian Research Council-CNPq and FAPESP.

\newpage

\addcontentsline{toc}{section}{Appendix A}
\section*{\bf Appendix A: The $K$-matrix properties}
\setcounter{equation}{0}
\renewcommand{\theequation}{A.\arabic{equation}}

In this Appendix we briefly summarize the explicit expressions
of the $K$-matrix elements satisfying the reflection
equation (\ref{eqref}) for $S=1$ and $\frac{3}{2}$. For $S=1$ \cite{JA} the corresponding matrix is given by
\bear
K_{1}(\lambda)=\left( \begin{array}{ccc}
        k_{11}(\lambda) & k_{12}(\lambda) & k_{13}(\lambda) \\
        k_{21}(\lambda) & k_{22}(\lambda) & k_{23}(\lambda) \\
        k_{31}(\lambda) & k_{32}(\lambda) & k_{33}(\lambda)
                \end{array} \right),
\ear
where the elements $k_{ij}(\lambda)$ are given by
\bear
k_{11}(\lambda)&=&-\frac{1}{4}\left(2\xi - \frac{1}{2} +\frac{\lambda}{\eta} \right) \left(2\xi+\frac{1}{2} + \frac{\lambda}{\eta} \right) +\frac{cd}{8}\left(\frac{1}{2}-\frac{\lambda}{\eta} \right),
\\
k_{12}(\lambda)&=&\frac{c}{2\sqrt{2}}\left(2\xi-\frac{1}{2}+\frac{\lambda}{\eta}\right)\frac{\lambda}{\eta},
\\
k_{13}(\lambda)&=&\frac{c^{2}}{4}\left(\frac{1}{2}-\frac{\lambda}{\eta}\right)\frac{\lambda}{\eta},
\\
k_{21}(\lambda)&=&\frac{d}{2\sqrt{2}}\left(2\xi-\frac{1}{2}+\frac{\lambda}{\eta}\right)\frac{\lambda}{\eta},
\\
k_{22}(\lambda)&=&-\frac{1}{4}\left(2\xi + \frac{1}{2}
-\frac{\lambda}{\eta} \right) \left(2\xi-\frac{1}{2} +
\frac{\lambda}{\eta} \right)
+\frac{cd}{4}\left(\frac{1}{2}-\frac{\lambda}{\eta}
\right)\left(\frac{1}{2}+\frac{\lambda}{\eta} \right),
\\
k_{23}(\lambda)&=&\frac{c}{2\sqrt{2}}\left(2\xi+\frac{1}{2}-\frac{\lambda}{\eta}\right)\frac{\lambda}{\eta},
\\
k_{31}(\lambda)&=&\frac{d^{2}}{4}\left(\frac{1}{2}-\frac{\lambda}{\eta}\right)\frac{\lambda}{\eta},
\\
k_{32}(\lambda)&=&\frac{d}{2\sqrt{2}}\left(2\xi+\frac{1}{2}-\frac{\lambda}{\eta}\right)\frac{\lambda}{\eta},
\\
k_{33}(\lambda)&=&-\frac{1}{4}\left(2\xi - \frac{1}{2}
-\frac{\lambda}{\eta} \right) \left(2\xi+\frac{1}{2} -
\frac{\lambda}{\eta} \right)
+\frac{cd}{8}\left(\frac{1}{2}-\frac{\lambda}{\eta}
\right). \ear

On the other hand for $S=\frac{3}{2}$ we have
\bear
K_{\frac{3}{2}}(\lambda)=\left( \begin{array}{cccc}
        \bar{k}_{11}(\lambda) & \bar{k}_{12}(\lambda) & \bar{k}_{13}(\lambda) & \bar{k}_{14}(\lambda) \\
        \bar{k}_{21}(\lambda) & \bar{k}_{22}(\lambda) & \bar{k}_{23}(\lambda) & \bar{k}_{24}(\lambda) \\
        \bar{k}_{31}(\lambda) & \bar{k}_{32}(\lambda) & \bar{k}_{33}(\lambda) & \bar{k}_{34}(\lambda) \\
    \bar{k}_{41}(\lambda) & \bar{k}_{42}(\lambda) & \bar{k}_{43}(\lambda) & \bar{k}_{44}(\lambda)
                \end{array} \right),
\ear where the elements $\bar{k}_{ij}(\lambda)$ are given by
\begin{align}
\bar{k}_{11}(\lambda)&=\frac{cd}{18}\left[\xi- \left(3\xi+\frac{\lambda}{\eta}\right)\left( 1-\frac{\lambda}{\eta}\right)\right]+\frac{1}{27}\left(3\xi+\frac{\lambda}{\eta}\right)\left(3\xi-1+\frac{\lambda}{\eta}\right)\left(3\xi+1+\frac{\lambda}{\eta}\right),
\\
\bar{k}_{12}(\lambda)&=-\frac{c}{18\sqrt{3}}\left[cd\left(1-\frac{\lambda}{\eta}\right)-2\left(3\xi-1+\frac{\lambda}{\eta}\right)\left(3\xi+\frac{\lambda}{\eta}\right)\right]\frac{\lambda}{\eta},
\\
\bar{k}_{13}(\lambda)&=-\frac{c^{2}}{9\sqrt{3}}\left(\frac{1}{2}-\frac{\lambda}{\eta}\right)\left(3\xi-1+\frac{\lambda}{\eta}\right)\frac{\lambda}{\eta},
\\
\bar{k}_{14}(\lambda)&=\frac{c^{3}}{27}\left(\frac{1}{2}-\frac{\lambda}{\eta}\right)\left(1-\frac{\lambda}{\eta}\right)\frac{\lambda}{\eta},
\\
\bar{k}_{21}(\lambda)&=-\frac{d}{18\sqrt{3}}\left[cd\left(1-\frac{\lambda}{\eta}\right)-2\left(3\xi-1+\frac{\lambda}{\eta}\right)\left(3\xi+\frac{\lambda}{\eta}\right)\right]\frac{\lambda}{\eta},
\\
\bar{k}_{22}(\lambda)&=\frac{2cd}{27}\left[\left(\frac{\lambda}{\eta}\right)^{3}+\left(3\xi-\frac{5}{4}\right)\left(\frac{\lambda}{\eta}\right)^{2}+\left(\frac{1}{4}+\frac{3\xi}{4}\right)\frac{\lambda}{\eta}-\frac{3\xi}{2}\right]
\\
&+\frac{1}{27}\left(3\xi+1-\frac{\lambda}{\eta}\right)\left(3\xi+\frac{\lambda}{\eta}\right)\left(3\xi-1+\frac{\lambda}{\eta}\right),
\\
\bar{k}_{23}(\lambda)&=-\frac{c}{54}\left[cd\left(1-\frac{\lambda}{\eta}\right)\left(1+\frac{2\lambda}{\eta}\right)+4\left(\left(1-\frac{\lambda}{\eta}\right)^{2}-(3\xi)^{2}\right)\right]\frac{\lambda}{\eta},
\\
\bar{k}_{24}(\lambda)&=-\frac{c^{2}}{9\sqrt{3}}\left(\frac{1}{2}-\frac{\lambda}{\eta}\right)\left(3\xi+1-\frac{\lambda}{\eta}\right)\frac{\lambda}{\eta},
\\
\bar{k}_{31}(\lambda)&=-\frac{d^{2}}{9\sqrt{3}}\left(\frac{1}{2}-\frac{\lambda}{\eta}\right)\left(3\xi-1+\frac{\lambda}{\eta}\right)\frac{\lambda}{\eta},
\\
\bar{k}_{32}(\lambda)&=-\frac{d}{54}\left[cd\left(1-\frac{\lambda}{\eta}\right)\left(1+\frac{2\lambda}{\eta}\right)+4\left(\left(1-\frac{\lambda}{\eta}\right)^{2}-(3\xi)^{2}\right)\right]\frac{\lambda}{\eta},
\\
\bar{k}_{33}(\lambda)&=-\frac{2cd}{27}\left[\left(\frac{\lambda}{\eta}\right)^{3}-\left(3\xi+\frac{5}{4}\right)\left(\frac{\lambda}{\eta}\right)^{2}+\left(\frac{1}{4}-\frac{3\xi}{4}\right)\frac{\lambda}{\eta}+\frac{3\xi}{2}\right]
\\
&+\frac{1}{27}\left(3\xi-\frac{\lambda}{\eta}\right)\left(3\xi+1-\frac{\lambda}{\eta}\right)\left(3\xi-1+\frac{\lambda}{\eta}\right),
\\
\bar{k}_{34}(\lambda)&=-\frac{c}{18\sqrt{3}}\left[cd\left(1-\frac{\lambda}{\eta}\right)-2\left(3\xi-\frac{\lambda}{\eta}\right)\left(3\xi+1-\frac{\lambda}{\eta}\right)\right]\frac{\lambda}{\eta},
\\
\bar{k}_{41}(\lambda)&=\frac{d^{3}}{27}\left(\frac{1}{2}-\frac{\lambda}{\eta}\right)\left(1-\frac{\lambda}{\eta}\right)\frac{\lambda}{\eta},
\\
\bar{k}_{42}(\lambda)&=-\frac{d^{2}}{9\sqrt{3}}\left(\frac{1}{2}-\frac{\lambda}{\eta}\right)\left(3\xi+1-\frac{\lambda}{\eta}\right)\frac{\lambda}{\eta},
\end{align}
\begin{align}
\bar{k}_{43}(\lambda)&=-\frac{d}{18\sqrt{3}}\left[cd\left(1-\frac{\lambda}{\eta}\right)-2\left(3\xi-\frac{\lambda}{\eta}\right)\left(3\xi+1-\frac{\lambda}{\eta}\right)\right]\frac{\lambda}{\eta},
\\
\bar{k}_{44}(\lambda)&=\frac{cd}{18}\left[\xi-\left(3\xi-\frac{\lambda}{\eta}\right)\left( 1-\frac{\lambda}{\eta}\right)\right] +\frac{1}{27}\left(3\xi-\frac{\lambda}{\eta}\right)\left(3\xi-1-\frac{\lambda}{\eta}\right)\left(3\xi+1-\frac{\lambda}{\eta}\right),
\end{align}

Next we list the dependence of the off-diagonal coefficients of the transformed $K$-matrix
$\widetilde{K}_{S}^{(-)}(\lambda)$ on the parameters $c_{\pm}$ and $d_{\pm}$. For $S=\frac{1}{2}$ we find
\EQ
\sigma_{12}=\frac{-\epsilon_{+}\left[c_{+}d_{-}+c_{-}d_{+} -2c_{+}d_{+} \right]
+\left(c_{-}d_{+}-d_{-}c_{+} \right)\sqrt{1+c_{+}d_{+}}}{2 d_{+}\sqrt{1+c_{+}d_{+}}},
\EN
while for $S=1$ we have
\bear
\kappa_{12}&=&\frac{(2+c_{-}d_{+}+c_{+}d_{-})}{32\sqrt{2} (1+c_{+}d_{+})}
\left(-\epsilon_{+}2 (c_{+}-c_{-})\sqrt{1+c_{+}d_{+}} - \Delta \right),  \\
\kappa_{13}& =& \frac{-\epsilon_{+} \Theta^{0}+ 2 c_{+}^{4}\left( d_{+} - d_{-} \right)^{2} + \frac{c_{+}^{4}}{4}\left[ d_{+}^{2}\Theta^{1} - 2d_{+}d_{-}\Theta^{2}
+ d_{-}^{2}\Theta^{3} \right]}{32(1+c_{+}d_{+})(2+c_{+}d_{+}+2\epsilon_{+}\sqrt{1+c_{+}d_{+}})^{2}},  \\
\kappa_{23}& = &
\frac{\xi_{-}}{8\sqrt{2}\sqrt{1+c_{+}d_{+}}} \left(2 (c_{+}-c_{-})\sqrt{1+c_{+}d_{+}} + \epsilon_{+} \Delta \right),
\ear
where $\Delta$ and $\Theta^{i}$ are given by
\bear
\Delta     &=& 2 c_{-} - c_{+} (2-c_{-}d_{+}+c_{+}d_{-}), \\
\Theta^{0} &=& c_{+}^{4}(d_{+}-d_{-}) \sqrt{1+c_{+}d_{+}} \left[d_{-}(2+c_{+}d_{+})-d_{+}(2+c_{-}d_{+})\right], \\
\Theta^{1} &=& 4c_{+}d_{+}+c_{-}d_{+}(4+c_{-}d_{+}), \\
\Theta^{2} &=& 2 c_{-}d_{+}+c_{+}d_{+}(6+c_{-}d_{+}), \\
\Theta^{3} &=& c_{+}d_{+}(8+c_{+}d_{+}).
\ear

\addcontentsline{toc}{section}{Appendix B}
\section*{\bf Appendix B: One and Two particle states for $S=\frac{3}{2}$ }
\setcounter{equation}{0}
\renewcommand{\theequation}{B.\arabic{equation}}

The purpose of this Appendix is to present some of the technical details entering the 
analysis of the one and two particle states for $S=\frac{3}{2}$. In order to do that it is convenient to work with a
new matrix $\check{R}_{ab}^{(S)}(\lambda)=P_{ab} {\cal{L}}_{ab}^{(S)}(\lambda)$ where $P_{ab}$ is the permutator.
This matrix plays a direct role in the quantum inverse scattering method and 
Eq.(\ref{intertwREL}) is rewritten in terms of 
$\check{R}_{ab}^{(S)}(\lambda)$ as
\EQ
\check{R}_{12}^{(S)}(u-v) \stackrel{1}{{\overline{\cal T}}_{\cal A}^{(S)}}(u) \check{R}^{(S)}_{12}(u+v)
\stackrel{1}{{\overline{\cal T}}_{\cal A}^{(S)}}(v) =
\stackrel{1}{{\overline{{\cal T}}}_{\cal A}^{(S)}}(v) \check{R}^{(S)}_{12}(u+v)
\stackrel{1}{{\overline{{\cal
T}}}_{\cal A}^{(S}}(u) \check{R}^{(S)}_{12}(u-v),
\label{intertwRELcodificada}
\EN

In the specific case of a  
44 vertex model, the $\check{R}_{ab}^{(\frac{3}{2})}(\lambda)$ operator can be expressed in terms of
the following matrix: 
{\scriptsize
\bear
\left(\begin{array}{cccccccccccccccc}
a(\lambda) & 0 & 0 & 0 & 0 & 0 & 0 & 0 & 0 & 0 & 0 & 0 & 0 & 0 & 0 & 0 \\
0 & b(\lambda) & 0 & 0 & e(\lambda) & 0 & 0 & 0 & 0 & 0 & 0 & 0 & 0 & 0 & 0 & 0 \\
0 & 0 & c(\lambda) & 0 & 0 & f(\lambda) & 0 & 0 & h(\lambda) & 0 & 0 & 0 & 0 & 0 & 0 & 0 \\
0 & 0 & 0 & d(\lambda) & 0 & 0 & g(\lambda) & 0 & 0 & i(\lambda) & 0 & 0 & j(\lambda) & 0 & 0 & 0 \\
0 & e(\lambda) & 0 & 0 & b(\lambda) & 0 & 0 & 0 & 0 & 0 & 0 & 0 & 0 & 0 & 0 & 0 \\
0 & 0 & r(\lambda) & 0 & 0 & l(\lambda) & 0 & 0 & r(\lambda) & 0 & 0 & 0 & 0 & 0 & 0 & 0 \\
0 & 0 & 0 & g_{1}(\lambda) & 0 & 0 & m(\lambda) & 0 & 0 & q(\lambda) & 0 & 0 & i_{1}(\lambda) & 0 & 0 & 0 \\
0 & 0 & 0 & 0 & 0 & 0 & 0 & n(\lambda) & 0 & 0 & r_{1}(\lambda) & 0 & 0 & h_{1}(\lambda) & 0 & 0 \\
0 & 0 & h(\lambda) & 0 & 0 & f(\lambda) & 0 & 0 & c(\lambda) & 0 & 0 & 0 & 0 & 0 & 0 & 0 \\
0 & 0 & 0 & i_{1}(\lambda) & 0 & 0 & q(\lambda) & 0 & 0 & m(\lambda) & 0 & 0 & g_{1}(\lambda) & 0 & 0 & 0 \\
0 & 0 & 0 & 0 & 0 & 0 & 0 & f_{1}(\lambda) & 0 & 0 & l_{1}(\lambda) & 0 & 0 & f_{1}(\lambda) & 0 & 0 \\
0 & 0 & 0 & 0 & 0 & 0 & 0 & 0 & 0 & 0 & 0 & b_{1}(\lambda) & 0 & 0 & e_{1}(\lambda) & 0 \\
0 & 0 & 0 & j(\lambda) & 0 & 0 & i(\lambda) & 0 & 0 & g(\lambda) & 0 & 0 & d(\lambda) & 0 & 0 & 0 \\
0 & 0 & 0 & 0 & 0 & 0 & 0 & h_{1}(\lambda) & 0 & 0 & r_{1}(\lambda) & 0 & 0 & n(\lambda) & 0 & 0 \\
0 & 0 & 0 & 0 & 0 & 0 & 0 & 0 & 0 & 0 & 0 & e_{1}(\lambda) & 0 & 0 & b_{1}(\lambda) & 0 \\
0 & 0 & 0 & 0 & 0 & 0 & 0 & 0 & 0 & 0 & 0 & 0 & 0 & 0 & 0 & a_{1}(\lambda) \\
\end{array} \right),
\nonumber 
\ear  \EQ \EN }

In order to solve the one-particle problem one first needs to obtain the 
appropriate commutation rules between the fields $A_{i}(u)$ and $B_{12}(v)$ coming 
from the boundary Yang-Baxter equation (\ref{intertwRELcodificada}). Using the symbol [i,j] to represent  
the i-th row and the j-th column of 
Eq.(\ref{intertwRELcodificada}) we conclude that such suitable commutation rules are derivate  from the
entries
[1,2],[2,3], [3,4], [2,6], [3,7] and [4,8]. 
Further progress are made replacing the fields 
$A_{i}(u)$ by $\widetilde{A}_{i}(u)$ in these equations with the help of the relations (\ref{Acombination}).
After several algebraic manipulations we obtain the following structure
\bear
\widetilde{A}_{1}(u)B_{12}(v) &=& a_{1}^{1}(u,v)B_{12}(v)\widetilde{A}_{1}(u)+a_{2}^{1}(u,v)B_{12}(u)\widetilde{A}_{1}(v)+a_{3}^{1}(u,v)B_{12}(u)\widetilde{A}_{2}(v)
\nonumber \\
&+& a_{4}^{1}(u,v)B_{13}(v)C_{21}(u)+a_{5}^{1}(u,v)B_{13}(u)C_{21}(v)+a_{6}^{1}(u,v)B_{13}(u)C_{32}(v)
\nonumber \\
&+& a_{7}^{1}(u,v)B_{14}(v)C_{31}(u)+a_{8}^{1}(u,v)B_{14}(u)C_{31}(v)+a_{9}^{1}(u,v)B_{14}(u)C_{42}(v)
\nonumber \\
\label{B3}
\ear
\bear
\widetilde{A}_{2}(u)B_{12}(v) &=& a_{1}^{2}(u,v)B_{12}(v)\widetilde{A}_{2}(u)+a_{2}^{2}(u,v)B_{12}(u)\widetilde{A}_{1}(v)+a_{3}^{2}(u,v)B_{12}(u)\widetilde{A}_{2}(v)
\nonumber \\
&+&a_{4}^{2}(u,v)B_{23}(u)\widetilde{A}_{1}(v)+a_{5}^{2}(u,v)B_{23}(u)\widetilde{A}_{2}(v)+a_{6}^{2}(u,v)B_{13}(v)C_{21}(u)
\nonumber \\
&+& a_{7}^{2}(u,v)B_{13}(v)C_{32}(u)+a_{8}^{2}(u,v)B_{13}(u)C_{21}(v)+a_{9}^{2}(u,v)B_{13}(u)C_{32}(v)
\nonumber\\
&+&a_{10}^{2}(u,v)B_{24}(u)C_{21}(v)+a_{11}^{2}(u,v)B_{24}(u)C_{32}(v)+a_{12}^{2}(u,v)B_{14}(v)C_{31}(u)
\nonumber\\
&+&a_{13}^{2}(u,v)B_{14}(v)C_{42}(u)+a_{14}^{2}(u,v)B_{14}(u)C_{31}(v)+a_{15}^{2}(u,v)B_{14}(u)C_{42}(v)
\nonumber \\
\label{B4}
\ear
\bear
\widetilde{A}_{3}(u)B_{12}(v) &=& a_{1}^{3}(u,v)B_{12}(v)\widetilde{A}_{3}(u)+a_{2}^{3}(u,v)B_{12}(u)\widetilde{A}_{1}(v)+a_{3}^{3}(u,v)B_{12}(u)\widetilde{A}_{2}(v)
\nonumber \\
&+&a_{4}^{3}(u,v)B_{23}(u)\widetilde{A}_{1}(v)+a_{5}^{3}(u,v)B_{23}(u)\widetilde{A}_{2}(v)+a_{6}^{3}(u,v)B_{34}(u)\widetilde{A}_{1}(v)
\nonumber \\
&+&a_{7}^{3}(u,v)B_{34}(u)\widetilde{A}_{2}(v)+a_{8}^{3}(u,v)B_{13}(v)C_{21}(u)+a_{9}^{3}(u,v)B_{13}(v)C_{32}(u)
\nonumber \\
&+&a_{10}^{3}(u,v)B_{13}(v)C_{43}(u)+a_{11}^{3}(u,v)B_{13}(u)C_{21}(v)+a_{12}^{3}(u,v)B_{13}(u)C_{32}(v)
\nonumber\\
&+&a_{13}^{3}(u,v)B_{24}(u)C_{21}(v)+a_{14}^{3}(u,v)B_{24}(u)C_{32}(v)+a_{15}^{3}(u,v)B_{14}(v)C_{31}(u)
\nonumber\\
&+&a_{16}^{3}(u,v)B_{14}(v)C_{42}(u)+a_{17}^{3}(u,v)B_{14}(u)C_{31}(v)+a_{18}^{3}(u,v)B_{14}(u)C_{42}(v)
\nonumber \\
\label{B5}
\ear
\bear
\widetilde{A}_{4}(u)B_{12}(v) &=& a_{1}^{4}(u,v)B_{12}(v)\widetilde{A}_{4}(u)+a_{2}^{4}(u,v)B_{12}(u)\widetilde{A}_{1}(v)+a_{3}^{4}(u,v)B_{12}(u)\widetilde{A}_{2}(v)
\nonumber \\
&+&a_{4}^{4}(u,v)B_{23}(u)\widetilde{A}_{1}(v)+a_{5}^{4}(u,v)B_{23}(u)\widetilde{A}_{2}(v)+a_{6}^{4}(u,v)B_{34}(u)\widetilde{A}_{1}(v)
\nonumber \\
&+&a_{7}^{4}(u,v)B_{34}(u)\widetilde{A}_{2}(v)+a_{8}^{4}(u,v)B_{13}(v)C_{21}(u)+a_{9}^{4}(u,v)B_{13}(v)C_{32}(u)
\nonumber \\
&+&a_{10}^{4}(u,v)B_{13}(v)C_{43}(u)+a_{11}^{4}(u,v)B_{13}(u)C_{21}(v)+a_{12}^{4}(u,v)B_{13}(u)C_{32}(v)
\nonumber\\
&+&a_{13}^{4}(u,v)B_{24}(u)C_{21}(v)+a_{14}^{4}(u,v)B_{24}(u)C_{32}(v)+a_{15}^{4}(u,v)B_{14}(v)C_{31}(u)
\nonumber\\
&+&a_{16}^{4}(u,v)B_{14}(v)C_{42}(u)+a_{17}^{4}(u,v)B_{14}(u)C_{31}(v)+a_{18}^{4}(u,v)B_{14}(u)C_{42}(v)
\nonumber \\
\label{B6}
\ear

Before proceeding we would like to remark that several identities between the 
Boltzmann weights have been used in order to obtain relations (\ref{B3}-\ref{B6}). We also note that many of the
coefficients $a_i^{j}(u,v)$ are proportional to annihilation operators and not all of them 
are relevant 
in the calculations. In Appendix C we have listed only those that indeed play an important role in our analysis
since in general they are
sufficiently cumbersome. 
By applying Eqs.(\ref{B3}-\ref{B6}) on the pseudovacuum $\ket{\bar{0}_{\frac{3}{2}}}$ we see that
one can rearrange the action of the double transfer matrix $\bar{t}_{\frac{3}{2}}(\lambda)$ on the one-particle
state $B_{12}(\lambda_1) 
\ket{\bar{0}_{\frac{3}{2}}}$ as 
in Eq.(\ref{EigVecSpS}). Furthermore, it turns out that the functions 
$Q_{i}(\lambda,\lambda_{1})$, $q_{i}^{(1)}(\lambda,\lambda_{1})$ and $q_{i}^{(2)}(\lambda,\lambda_{1})$ 
can therefore be explicitly read off, namely
\bear
Q_{i}(\lambda,\lambda_{1}) &=& a_{1}^{i}(\lambda,\lambda_{1})
\\
q_{i}^{(1)}(\lambda,\lambda_{1}) &=& \sum_{j=i}^{4} \omega_{j}^{(+)}(\lambda) a_{2 i}^{j}(\lambda,\lambda_{1})
\\
q_{i}^{(2)}(\lambda,\lambda_{1}) &=& \sum_{j=i}^{4}\omega_{j}^{(+)}(\lambda) a_{2 i+1}^{j}(\lambda,\lambda_{1}) 
\ear
where $i=1,\dots,4$ and function $\omega_{i}^{(+)}(\lambda)$ has been defined in Eq.(\ref{omegap}).

As mentioned in the main text the ratio
$\frac{q_{i}^{(2)}(\lambda,\lambda_{1})}{q_{i}^{(1)}(\lambda,\lambda_{1})}$ is independent of the i-th index
and of the spectral  parameter $\lambda$. 
In our case this ratio is given by
\EQ
\frac{q_{i}^{(2)}(\lambda,\lambda_{1})}{q_{i}^{(1)}(\lambda,\lambda_{1})}=- \frac{\omega_{2}^{(+)}(\lambda_{1}) }{\omega_{1}^{(+)}(\lambda_{1}) } \left(\frac{a(\lambda_{1}) l(\lambda_{1})- b(\lambda_{1}) b(\lambda_{1})}{\left[e(\lambda_{1})\right]^{2}}\right)
\EN

Next we turn to the two-particle state. In order to obtain an ansatz to this vector we have considered the
commutation rules [1,3] and [1,6] coming from Eq. (\ref{intertwRELcodificada}). Acting these relations on 
$\ket{\bar{0}_{\frac{3}{2}}}$ leads us to the following expression
\bear
\left[ B_{12}(u)B_{12}(v)+B_{13}(u)\left(\alpha_{2}(u,v)\widetilde{A}_{2}(v)+
\alpha_{1}(u,v)\widetilde{A}_{1}(v)\right) \right] \ket{\bar{0}_{\frac{3}{2}}}=~~~~
\nonumber \\
Z_{\frac{3}{2}}(u,v)\left[ B_{12}(v)B_{12}(u)+
B_{13}(v)\left(\alpha_{3}(u,v)\widetilde{A}_{2}(u)
+\alpha_{4}(u,v)\widetilde{A}_{1}(u)\right) \right] \ket{\bar{0}_{\frac{3}{2}}},
\label{ANZV}
\ear
where functions $\alpha_{i}(u,v)$ for $i=1,\dots,4$ and $Z_{\frac{3}{2}}(u,v)$ are given by
\EQ
\alpha_{1}(u,v)=\frac{f(u+v)}{e(u+v)}\frac{b(2v)}{a(2v)}-\frac{f(u-v)h(u+v)}{h(u-v)e(u+v)},
\EN
\EQ
\alpha_{2}(u,v)=\frac{f(u+v)}{e(u+v)},~~~~
\alpha_{3}(u,v)=\frac{f(u+v)}{e(u+v)},
\EN
\EQ
\alpha_{4}(u,v)=\frac{f(u+v)}{e(u+v)}\frac{b(2u)}{a(2u)}+\left(\frac{h(u-v)f(u-v)-f(u-v) c(u-v)}{h(u-v)l(u-v)-f(u-v) r(u-v)}\right)\frac{h(u+v)}{e(u+v)}
\EN
\EQ
Z_{\frac{3}{2}}(u,v)=\left(\frac{h(u-v) l(u-v)-f(u-v) r(u-v)}{a(u-v)h(u-v)}\right)\frac{e(u+v)}{e(u+v)}.
\EN

From Eq.(\ref{ANZV}) it follows that an appropriate two-particle state should be
\bear
\ket{\bar{\psi}_{2}(\lambda_{1},\lambda_{2})} = \left[
B_{12}(\lambda_{1}) B_{12}(\lambda_{2})+B_{13}(\lambda_{1}) \Bigl(\alpha_{2}(\lambda_{1},\lambda_{2})
\widetilde{A}_{2}(\lambda_{2})+\alpha_{1}(\lambda_{1},\lambda_{2})
\widetilde{A}_{1}(\lambda_{2})\Bigr) \right] \ket{\bar{0}_{\frac{3}{2}}},
\nonumber\\
\label{TWO}
\ear
since it is symmetric 
$\ket{\bar{\psi}_{2}(\lambda_{1},\lambda_{2})} = Z_{\frac{3}{2}}(\lambda_1,\lambda_2)
\ket{\bar{\psi}_{1}(\lambda_{2},\lambda_{1})}$ under the exchange of rapidities.

The next step is to solve the eigenvalue problem for the two-particle state (\ref{TWO}).  In order to do that
we need extra commutations  rules
between the fields $\widetilde{A}_{i}(u)$ and $B_{13}(v)$, $B_{12}(u)$ and $B_{jj+1}(v)$, $C_{j+1j}(u)$ and $B_{12}(v)$.
In the case of the fields $C_{j+1j}(u)$ and $B_{12}(v)$ the rules comes from the entries [2,5], [3,6] and [4,7] 
of Eq. (\ref{intertwRELcodificada}) and the ones for the other operators are obtained 
from [1,3],[2,4];[1,6],[2,7],[3,8];[2,10],[3,11] and [4,12] entries. 
After long algebraic
manipulations we are able to obtain the following expressions
\EQ
\widetilde{A}_{i}(u)B_{13}(v)\ket{\bar{0}_{\frac{3}{2}}}=b_{1}^{i}(u,v)B_{13}(v)\widetilde{A}_{i}(u)
\ket{\bar{0}_{\frac{3}{2}}}+\mathrm{unwanted~terms}~~~~~i=1,\dots,=4
\label{B16}
\EN
\bear
C_{21}(u)B_{12}(v)\ket{\bar{0}_{\frac{3}{2}}}&=&\left[ c_{22}^{1}(u,v)\widetilde{A}_{2}(v)\widetilde{A}_{2}(u)+c_{21}^{1}(u,v)\widetilde{A}_{2}(v)\widetilde{A}_{1}(u)+c_{12}^{1}(u,v)\widetilde{A}_{1}(v)\widetilde{A}_{2}(u) \right.
\nonumber\\
&+& \left. c_{11}^{1}(u,v)\widetilde{A}_{1}(v)\widetilde{A}_{1}(u)\right] \ket{\bar{0}_{\frac{3}{2}}} 
\label{B17}
\ear
\bear
C_{32}(u)B_{12}(v)\ket{\bar{0}_{\frac{3}{2}}}&=&\left[ c_{23}^{2}(u,v)\widetilde{A}_{2}(v)\widetilde{A}_{3}(u)+c_{22}^{2}(u,v)\widetilde{A}_{2}(v)\widetilde{A}_{2}(u)+c_{21}^{2}(u,v)\widetilde{A}_{2}(v)\widetilde{A}_{1}(u) \right.
\nonumber\\
&+& \left. c_{13}^{2}(u,v)\widetilde{A}_{1}(v)\widetilde{A}_{3}(u)+c_{12}^{2}(u,v)\widetilde{A}_{1}(v)\widetilde{A}_{2}(u)+c_{11}^{2}(u,v)\widetilde{A}_{1}(v)\widetilde{A}_{1}(u)\right] \ket{\bar{0}_{\frac{3}{2}}}
\nonumber \\
\label{B18}
\ear
\bear
C_{43}(u)B_{12}(v)\ket{\bar{0}_{\frac{3}{2}}}&=&\left[ c_{24}^{3}(u,v)\widetilde{A}_{2}(v)\widetilde{A}_{4}(u)+c_{23}^{3}(u,v)\widetilde{A}_{2}(v)\widetilde{A}_{3}(u)+c_{22}^{3}(u,v)\widetilde{A}_{2}(v)\widetilde{A}_{2}(u) \right.
\nonumber\\
&+& \left. c_{21}^{3}(u,v)\widetilde{A}_{2}(v)\widetilde{A}_{1}(u)+c_{14}^{3}(u,v)\widetilde{A}_{1}(v)\widetilde{A}_{4}(u)+c_{13}^{3}(u,v)\widetilde{A}_{1}(v)\widetilde{A}_{3}(u) \right.
\nonumber\\
&+& \left. c_{12}^{3}(u,v)\widetilde{A}_{1}(v)\widetilde{A}_{2}(u)+
c_{11}^{3}(u,v)\widetilde{A}_{1}(v)\widetilde{A}_{1}(u)\right] \ket{\bar{0}_{\frac{3}{2}}}
\label{B19}
\ear
\EQ
B_{12}(v)B_{12}(u)\ket{\bar{0}_{\frac{3}{2}}}=\left[ d_{2}^{1}(u,v)B_{13}(v)\widetilde{A}_{2}(u)
+d_{1}^{1}(u,v)B_{13}(v)\widetilde{A}_{1}(u)+\mathrm{unwanted~terms}\right ] \ket{\bar{0}_{\frac{3}{2}}}
\label{B20}
\EN
\bear
B_{12}(v)B_{23}(u)\ket{\bar{0}_{\frac{3}{2}}}&=&\left[ d_{3}^{2}(u,v)B_{13}(v)\widetilde{A}_{3}(u)+d_{2}^{2}(u,v)B_{13}(v)\widetilde{A}_{2}(u)+d_{1}^{2}(u,v)B_{13}(v)\widetilde{A}_{1}(u) \right.
\nonumber\\
&+& \left. \mathrm{unwanted~terms} \right] \ket{\bar{0}_{\frac{3}{2}}}
\label{B21}
\ear
\bear
B_{12}(v)B_{34}(u)\ket{\bar{0}_{\frac{3}{2}}}&=&\left[ (d_{4}^{3}(u,v)B_{13}(v)\widetilde{A}_{4}(u)+d_{3}^{3}(u,v)B_{13}(v)\widetilde{A}_{3}(u)+d_{2}^{3}(u,v)B_{13}(v)\widetilde{A}_{2}(u) \right.
\nonumber\\
&+&\left. d_{1}^{3}(u,v)B_{13}(v)\widetilde{A}_{1}(u)+\mathrm{unwanted~terms}\right] \ket{\bar{0}_{\frac{3}{2}}}
\label{B22}
\ear
where  by ``unwanted terms'' we mean those that do not give contributions proportional to
$\ket{\bar{\psi}_{2}(\lambda_{1},\lambda_{2})}$. The functions $b_1^i(u,v)$, $c^{k}_{ij}(u,v)$ and $d_i^j(u,v)$ are
once again very involved and have been presented in Appendix C.

We have now the main ingredients to study the action of the 
operators $\widetilde{A}_{i}(\lambda)$ on the two-particle state $\ket{\bar{\psi}_{2}(\lambda_{1},\lambda_{2})}$.
Taking into account the commutation rules Eqs.(\ref{B3})-(\ref{B6}) and (\ref{B16})-(\ref{B22}) and after
some algebra we conclude that the two-particle wanted terms have the following  structure
\bear
\frac{\bar{t}_{\frac{3}{2}}(\lambda)}{\rho_{\frac{3}{2}}^{(+)} }\ket{\bar{\psi}_{2}(\lambda_{1},\lambda_{2})}
&=& B_{12}(\lambda_{1}) B_{12}(\lambda_{2}) \sum_{i=1}^{4} 
\omega_{i}^{(+)}(\lambda)\left(\prod_{j=1}^{n=2}a_{1}^{i}(\lambda,\lambda_{j})\right) 
\widetilde{A}_{i}(\lambda) \ket{\bar{0}_{\frac{3}{2}}}
\nonumber\\
&+&B_{13}(\lambda_{1}) \widetilde{A}_{4}(\lambda)\left(\Lambda_{2}^{42}(\lambda,\lbrace \lambda_{i} \rbrace)\widetilde{A}_{2}(\lambda_{2})+\Lambda_{2}^{41}(\lambda,\lbrace \lambda_{i} \rbrace)\widetilde{A}_{1}(\lambda_{2})\right) \ket{\bar{0}_{\frac{3}{2}}}
\nonumber\\
&+&B_{13}(\lambda_{1}) \widetilde{A}_{3}(\lambda)\left(\Lambda_{2}^{32}(\lambda,\lbrace \lambda_{i} \rbrace)\widetilde{A}_{2}(\lambda_{2})+\Lambda_{2}^{31}(\lambda,\lbrace \lambda_{i} \rbrace)\widetilde{A}_{1}(\lambda_{2})\right) \ket{\bar{0}_{\frac{3}{2}}}
\nonumber\\
&+&B_{13}(\lambda_{1}) \widetilde{A}_{2}(\lambda)\left(\Lambda_{2}^{22}(\lambda,\lbrace \lambda_{i} \rbrace)\widetilde{A}_{2}(\lambda_{2})+\Lambda_{2}^{21}(\lambda,\lbrace \lambda_{i} \rbrace)\widetilde{A}_{1}(\lambda_{2})\right) \ket{\bar{0}_{\frac{3}{2}}}
\nonumber\\
&+&B_{13}(\lambda_{1}) \widetilde{A}_{1}(\lambda)\left(\Lambda_{2}^{12}(\lambda,\lbrace \lambda_{i} \rbrace)\widetilde{A}_{2}(\lambda_{2})+\Lambda_{2}^{11}(\lambda,\lbrace \lambda_{i} \rbrace)\widetilde{A}_{1}(\lambda_{2})\right) \ket{\bar{0}_{\frac{3}{2}}}
\nonumber\\
+ \mathrm{unwanted~terms}
\label{SEST}
\ear
where functions
$\Lambda_{2}^{lk}(\lambda,\lbrace \lambda_{i} \rbrace)$ are given by
\bear
\Lambda_{2}^{42}(\lambda,\lbrace \lambda_{i} \rbrace)&=&\omega_{4}^{(+)}(\lambda)\biggl(b_{1}^{4}(\lambda,\lambda_{1}) \alpha_{2}(\lambda_{1},\lambda_{2}) + a_{10}^{4}(\lambda,\lambda_{1}) c_{24}^{3}(\lambda,\lambda_{2}) + a_{1}^{4}(\lambda,\lambda_{1}) a_{7}^{4}(\lambda,\lambda_{2}) d_{4}^{3}(\lambda,\lambda_{1}) \biggr)
\nonumber\\
&+&\omega_{3}^{(+)}(\lambda)\biggl(a_{10}^{3}(\lambda,\lambda_{1}) c_{24}^{3}(\lambda,\lambda_{2}) + a_{1}^{3}(\lambda,\lambda_{1}) a_{7}^{3}(\lambda,\lambda_{2}) d_{4}^{3}(\lambda,\lambda_{1}) \biggr)
\ear
\bear
\Lambda_{2}^{41}(\lambda,\lbrace \lambda_{i} \rbrace)&=&\omega_{4}^{(+)}(\lambda)\biggl(b_{1}^{4}(\lambda,\lambda_{1}) \alpha_{1}(\lambda_{1},\lambda_{2}) + a_{10}^{4}(\lambda,\lambda_{1}) c_{14}^{3}(\lambda,\lambda_{2}) + a_{1}^{4}(\lambda,\lambda_{1}) a_{6}^{4}(\lambda,\lambda_{2}) d_{4}^{3}(\lambda,\lambda_{1}) \biggr)
\nonumber\\
&+&\omega_{3}^{(+)}(\lambda)\biggl(a_{10}^{3}(\lambda,\lambda_{1}) c_{14}^{3}(\lambda,\lambda_{2}) + a_{1}^{3}(\lambda,\lambda_{1}) a_{6}^{3}(\lambda,\lambda_{2}) d_{4}^{3}(\lambda,\lambda_{1}) \biggr)
\ear
\bear
\Lambda_{2}^{32}(\lambda,\lbrace \lambda_{i} \rbrace) &=& \omega_{4}^{(+)}(\lambda) \biggl(a_{10}^{4}(\lambda,\lambda_{1}) c_{23}^{3}(\lambda,\lambda_{2}) + a_{9}^{4}(\lambda,\lambda_{1}) c_{23}^{2}(\lambda,\lambda_{2}) + a_{1}^{4}(\lambda,\lambda_{1}) a_{7}^{4}(\lambda,\lambda_{2}) d_{3}^{3}(\lambda,\lambda_{1})
\nonumber \\
&+& a_{1}^{4}(\lambda,\lambda_{1}) a_{5}^{4}(\lambda,\lambda_{2}) d_{3}^{2}(\lambda,\lambda_{1}) \biggr) + \omega_{3}^{(+)}(\lambda)\biggl(b_{1}^{3}(\lambda,\lambda_{1}) \alpha_{2}(\lambda_{1},\lambda_{2}) + a_{10}^{3}(\lambda,\lambda_{1}) c_{23}^{3}(\lambda,\lambda_{2})
\nonumber\\
&+& a_{9}^{3}(\lambda,\lambda_{1}) c_{23}^{2}(\lambda,\lambda_{2}) + a_{1}^{3}(\lambda,\lambda_{1}) a_{7}^{3}(\lambda,\lambda_{2}) d_{3}^{3}(\lambda,\lambda_{1}) + a_{1}^{3}(\lambda,\lambda_{1}) a_{5}^{3}(\lambda,\lambda_{2}) d_{3}^{2}(\lambda,\lambda_{1}) \biggr)
\nonumber\\
&+&\omega_{2}^{(+)}(\lambda)\biggl(a_{7}^{2}(\lambda,\lambda_{1}) c_{23}^{2}(\lambda,\lambda_{2}) + a_{1}^{2}(\lambda,\lambda_{1}) a_{5}^{2}(\lambda,\lambda_{2}) d_{3}^{2}(\lambda,\lambda_{1}) \biggr)
\ear
\bear
\Lambda_{2}^{31}(\lambda,\lbrace \lambda_{i} \rbrace) &=& \omega_{4}^{(+)}(\lambda) \biggl(a_{10}^{4}(\lambda,\lambda_{1}) c_{13}^{3}(\lambda,\lambda_{2}) + a_{9}^{4}(\lambda,\lambda_{1}) c_{13}^{2}(\lambda,\lambda_{2}) + a_{1}^{4}(\lambda,\lambda_{1}) a_{6}^{4}(\lambda,\lambda_{2}) d_{3}^{3}(\lambda,\lambda_{1})
\nonumber \\
&+& a_{1}^{4}(\lambda,\lambda_{1}) a_{4}^{4}(\lambda,\lambda_{2}) d_{3}^{2}(\lambda,\lambda_{1}) \biggr) + \omega_{3}^{(+)}(\lambda)\biggl(b_{1}^{3}(\lambda,\lambda_{1}) \alpha_{1}(\lambda_{1},\lambda_{2}) + a_{10}^{3}(\lambda,\lambda_{1}) c_{13}^{3}(\lambda,\lambda_{2})
\nonumber\\
&+& a_{9}^{3}(\lambda,\lambda_{1}) c_{13}^{2}(\lambda,\lambda_{2}) + a_{1}^{3}(\lambda,\lambda_{1}) a_{6}^{3}(\lambda,\lambda_{2}) d_{3}^{3}(\lambda,\lambda_{1}) + a_{1}^{3}(\lambda,\lambda_{1}) a_{4}^{3}(\lambda,\lambda_{2}) d_{3}^{2}(\lambda,\lambda_{1}) \biggr)
\nonumber\\
&+&\omega_{2}^{(+)}(\lambda)\biggl(a_{7}^{2}(\lambda,\lambda_{1}) c_{13}^{2}(\lambda,\lambda_{2}) + a_{1}^{2}(\lambda,\lambda_{1}) a_{4}^{2}(\lambda,\lambda_{2}) d_{3}^{2}(\lambda,\lambda_{1}) \biggr)
\ear
\bear
&&\Lambda_{2}^{22}(\lambda,\lbrace \lambda_{i} \rbrace) = \omega_{4}^{(+)}(\lambda) \biggl(a_{10}^{4}(\lambda,\lambda_{1}) c_{22}^{3}(\lambda,\lambda_{2}) + a_{9}^{4}(\lambda,\lambda_{1}) c_{22}^{2}(\lambda,\lambda_{2}) + a_{8}^{4}(\lambda,\lambda_{1}) c_{22}^{1}(\lambda,\lambda_{2})
\nonumber\\
&+& a_{1}^{4}(\lambda,\lambda_{1}) a_{7}^{4}(\lambda,\lambda_{2}) d_{2}^{3}(\lambda,\lambda_{1}) + a_{1}^{4}(\lambda,\lambda_{1}) a_{5}^{4}(\lambda,\lambda_{2}) d_{2}^{2}(\lambda,\lambda_{1}) + a_{1}^{4}(\lambda,\lambda_{1}) a_{3}^{4}(\lambda,\lambda_{2}) d_{2}^{1}(\lambda,\lambda_{1}) \biggr)
\nonumber\\
&+& \omega_{3}^{(+)}(\lambda)\biggl( a_{10}^{3}(\lambda,\lambda_{1}) c_{22}^{3}(\lambda,\lambda_{2}) + a_{9}^{3}(\lambda,\lambda_{1}) c_{22}^{2}(\lambda,\lambda_{2}) + a_{8}^{3}(\lambda,\lambda_{1}) c_{22}^{1}(\lambda,\lambda_{2}) + a_{1}^{3}(\lambda,\lambda_{1})
\nonumber\\
&\times&  a_{7}^{3}(\lambda,\lambda_{2}) d_{2}^{3}(\lambda,\lambda_{1}) + a_{1}^{3}(\lambda,\lambda_{1}) a_{5}^{3}(\lambda,\lambda_{2}) d_{2}^{2}(\lambda,\lambda_{1}) + a_{1}^{3}(\lambda,\lambda_{1}) a_{3}^{3}(\lambda,\lambda_{2}) d_{2}^{1}(\lambda,\lambda_{1})\biggr)
\nonumber\\
&+& \omega_{2}^{(+)}(\lambda)\biggl(b_{1}^{2}(\lambda,\lambda_{1}) \alpha_{2}(\lambda_{1},\lambda_{2}) +a_{7}^{2}(\lambda,\lambda_{1}) c_{22}^{2}(\lambda,\lambda_{2}) + a_{6}^{2}(\lambda,\lambda_{1}) c_{22}^{1}(\lambda,\lambda_{2})
\nonumber\\
&+& a_{1}^{2}(\lambda,\lambda_{1}) a_{5}^{2}(\lambda,\lambda_{2}) d_{2}^{2}(\lambda,\lambda_{1}) + a_{1}^{2}(\lambda,\lambda_{1}) a_{3}^{2}(\lambda,\lambda_{2}) d_{2}^{1}(\lambda,\lambda_{1})\biggr)
\nonumber\\
&+&\omega_{1}^{(+)}(\lambda)\biggl(a_{4}^{1}(\lambda,\lambda_{1}) c_{22}^{1}(\lambda,\lambda_{2}) + a_{1}^{1}(\lambda,\lambda_{1}) a_{3}^{1}(\lambda,\lambda_{2}) d_{2}^{1}(\lambda,\lambda_{1})\biggr)
\ear
\bear
&&\Lambda_{2}^{21}(\lambda,\lbrace \lambda_{i} \rbrace) = \omega_{4}^{(+)}(\lambda) \biggl(a_{10}^{4}(\lambda,\lambda_{1}) c_{12}^{3}(\lambda,\lambda_{2}) + a_{9}^{4}(\lambda,\lambda_{1}) c_{12}^{2}(\lambda,\lambda_{2}) + a_{8}^{4}(\lambda,\lambda_{1}) c_{12}^{1}(\lambda,\lambda_{2})
\nonumber\\
&+& a_{1}^{4}(\lambda,\lambda_{1}) a_{6}^{4}(\lambda,\lambda_{2}) d_{2}^{3}(\lambda,\lambda_{1}) + a_{1}^{4}(\lambda,\lambda_{1}) a_{4}^{4}(\lambda,\lambda_{2}) d_{2}^{2}(\lambda,\lambda_{1}) + a_{1}^{4}(\lambda,\lambda_{1}) a_{2}^{4}(\lambda,\lambda_{2}) d_{2}^{1}(\lambda,\lambda_{1}) \biggr)
\nonumber\\
&+& \omega_{3}^{(+)}(\lambda)\biggl( a_{10}^{3}(\lambda,\lambda_{1}) c_{12}^{3}(\lambda,\lambda_{2}) + a_{9}^{3}(\lambda,\lambda_{1}) c_{12}^{2}(\lambda,\lambda_{2}) + a_{8}^{3}(\lambda,\lambda_{1}) c_{12}^{1}(\lambda,\lambda_{2}) + a_{1}^{3}(\lambda,\lambda_{1})
\nonumber\\
&\times&  a_{6}^{3}(\lambda,\lambda_{2}) d_{2}^{3}(\lambda,\lambda_{1}) + a_{1}^{3}(\lambda,\lambda_{1}) a_{4}^{3}(\lambda,\lambda_{2}) d_{2}^{2}(\lambda,\lambda_{1}) + a_{1}^{3}(\lambda,\lambda_{1}) a_{2}^{3}(\lambda,\lambda_{2}) d_{2}^{1}(\lambda,\lambda_{1})\biggr)
\nonumber\\
&+& \omega_{2}^{(+)}(\lambda)\biggl(b_{1}^{2}(\lambda,\lambda_{1}) \alpha_{1}(\lambda_{1},\lambda_{2}) +a_{7}^{2}(\lambda,\lambda_{1}) c_{12}^{2}(\lambda,\lambda_{2}) + a_{6}^{2}(\lambda,\lambda_{1}) c_{12}^{1}(\lambda,\lambda_{2})
\nonumber\\
&+& a_{1}^{2}(\lambda,\lambda_{1}) a_{4}^{2}(\lambda,\lambda_{2}) d_{2}^{2}(\lambda,\lambda_{1}) + a_{1}^{2}(\lambda,\lambda_{1}) a_{2}^{2}(\lambda,\lambda_{2}) d_{2}^{1}(\lambda,\lambda_{1})\biggr)
\nonumber\\
&+&\omega_{1}^{(+)}(\lambda)\biggl(a_{4}^{1}(\lambda,\lambda_{1}) c_{12}^{1}(\lambda,\lambda_{2}) + a_{1}^{1}(\lambda,\lambda_{1}) a_{2}^{1}(\lambda,\lambda_{2}) d_{2}^{1}(\lambda,\lambda_{1})\biggr)
\ear
\bear
&&\Lambda_{2}^{12}(\lambda,\lbrace \lambda_{i} \rbrace) = \omega_{4}^{(+)}(\lambda) \biggl(a_{10}^{4}(\lambda,\lambda_{1}) c_{21}^{3}(\lambda,\lambda_{2}) + a_{9}^{4}(\lambda,\lambda_{1}) c_{21}^{2}(\lambda,\lambda_{2}) + a_{8}^{4}(\lambda,\lambda_{1}) c_{21}^{1}(\lambda,\lambda_{2})
\nonumber\\
&+& a_{1}^{4}(\lambda,\lambda_{1}) a_{7}^{4}(\lambda,\lambda_{2}) d_{1}^{3}(\lambda,\lambda_{1}) + a_{1}^{4}(\lambda,\lambda_{1}) a_{5}^{4}(\lambda,\lambda_{2}) d_{1}^{2}(\lambda,\lambda_{1}) + a_{1}^{4}(\lambda,\lambda_{1}) a_{3}^{4}(\lambda,\lambda_{2}) d_{1}^{1}(\lambda,\lambda_{1}) \biggr)
\nonumber\\
&+& \omega_{3}^{(+)}(\lambda)\biggl( a_{10}^{3}(\lambda,\lambda_{1}) c_{21}^{3}(\lambda,\lambda_{2}) + a_{9}^{3}(\lambda,\lambda_{1}) c_{21}^{2}(\lambda,\lambda_{2}) + a_{8}^{3}(\lambda,\lambda_{1}) c_{21}^{1}(\lambda,\lambda_{2}) + a_{1}^{3}(\lambda,\lambda_{1})
\nonumber\\
&\times&  a_{7}^{3}(\lambda,\lambda_{2}) d_{1}^{3}(\lambda,\lambda_{1}) + a_{1}^{3}(\lambda,\lambda_{1}) a_{5}^{3}(\lambda,\lambda_{2}) d_{1}^{2}(\lambda,\lambda_{1}) + a_{1}^{3}(\lambda,\lambda_{1}) a_{3}^{3}(\lambda,\lambda_{2}) d_{1}^{1}(\lambda,\lambda_{1})\biggr)
\nonumber\\
&+& \omega_{2}^{(+)}(\lambda)\biggl(a_{7}^{2}(\lambda,\lambda_{1}) c_{21}^{2}(\lambda,\lambda_{2}) + a_{6}^{2}(\lambda,\lambda_{1}) c_{21}^{1}(\lambda,\lambda_{2})
\nonumber\\
&+& a_{1}^{2}(\lambda,\lambda_{1}) a_{5}^{2}(\lambda,\lambda_{2}) d_{1}^{2}(\lambda,\lambda_{1}) + a_{1}^{2}(\lambda,\lambda_{1}) a_{3}^{2}(\lambda,\lambda_{2}) d_{1}^{1}(\lambda,\lambda_{1})\biggr)
\nonumber\\
&+&\omega_{1}^{(+)}(\lambda)\biggl(b_{1}^{1}(\lambda,\lambda_{1}) \alpha_{2}(\lambda_{1},\lambda_{2}) + a_{4}^{1}(\lambda,\lambda_{1}) c_{21}^{1}(\lambda,\lambda_{2}) + a_{1}^{1}(\lambda,\lambda_{1}) a_{3}^{1}(\lambda,\lambda_{2}) d_{1}^{1}(\lambda,\lambda_{1})\biggr)
\ear
\bear
&&\Lambda_{2}^{11}(\lambda,\lbrace \lambda_{i} \rbrace) = \omega_{4}^{(+)}(\lambda) \biggl(a_{10}^{4}(\lambda,\lambda_{1}) c_{11}^{3}(\lambda,\lambda_{2}) + a_{9}^{4}(\lambda,\lambda_{1}) c_{11}^{2}(\lambda,\lambda_{2}) + a_{8}^{4}(\lambda,\lambda_{1}) c_{11}^{1}(\lambda,\lambda_{2})
\nonumber\\
&+& a_{1}^{4}(\lambda,\lambda_{1}) a_{6}^{4}(\lambda,\lambda_{2}) d_{1}^{3}(\lambda,\lambda_{1}) + a_{1}^{4}(\lambda,\lambda_{1}) a_{4}^{4}(\lambda,\lambda_{2}) d_{1}^{2}(\lambda,\lambda_{1}) + a_{1}^{4}(\lambda,\lambda_{1}) a_{2}^{4}(\lambda,\lambda_{2}) d_{1}^{1}(\lambda,\lambda_{1}) \biggr)
\nonumber\\
&+& \omega_{3}^{(+)}(\lambda)\biggl( a_{10}^{3}(\lambda,\lambda_{1}) c_{11}^{3}(\lambda,\lambda_{2}) + a_{9}^{3}(\lambda,\lambda_{1}) c_{11}^{2}(\lambda,\lambda_{2}) + a_{8}^{3}(\lambda,\lambda_{1}) c_{11}^{1}(\lambda,\lambda_{2}) + a_{1}^{3}(\lambda,\lambda_{1})
\nonumber\\
&\times&  a_{6}^{3}(\lambda,\lambda_{2}) d_{1}^{3}(\lambda,\lambda_{1}) + a_{1}^{3}(\lambda,\lambda_{1}) a_{4}^{3}(\lambda,\lambda_{2}) d_{1}^{2}(\lambda,\lambda_{1}) + a_{1}^{3}(\lambda,\lambda_{1}) a_{2}^{3}(\lambda,\lambda_{2}) d_{1}^{1}(\lambda,\lambda_{1})\biggr)
\nonumber\\
&+& \omega_{2}^{(+)}(\lambda)\biggl(a_{7}^{2}(\lambda,\lambda_{1}) c_{11}^{2}(\lambda,\lambda_{2}) + a_{6}^{2}(\lambda,\lambda_{1}) c_{11}^{1}(\lambda,\lambda_{2})
\nonumber\\
&+& a_{1}^{2}(\lambda,\lambda_{1}) a_{4}^{2}(\lambda,\lambda_{2}) d_{1}^{2}(\lambda,\lambda_{1}) + a_{1}^{2}(\lambda,\lambda_{1}) a_{2}^{2}(\lambda,\lambda_{2}) d_{1}^{1}(\lambda,\lambda_{1})\biggr)
\nonumber\\
&+&\omega_{1}^{(+)}(\lambda)\biggl(b_{1}^{1}(\lambda,\lambda_{1}) \alpha_{1}(\lambda_{1},\lambda_{2}) + a_{4}^{1}(\lambda,\lambda_{1}) c_{11}^{1}(\lambda,\lambda_{2}) + a_{1}^{1}(\lambda,\lambda_{1}) a_{2}^{1}(\lambda,\lambda_{2}) d_{1}^{1}(\lambda,\lambda_{1})\biggr)
\ear

It turns out that many
identities between the Boltzmann weights can be used in order to show the following remarkable property
\EQ
\Lambda_{2}^{lk}(\lambda,\lbrace \lambda_{i} \rbrace)= \alpha_k(\lambda_1,\lambda_2) \omega_l^{(+)}(\lambda)
\prod_{i=1}^{n=2} a_1^{l}(\lambda,\lambda_i)
\label{IDD}
\EN

Considering Eq.(\ref{SEST}), Eq.(\ref{IDD}) and Eq.(\ref{actionome}) together it is not difficult to derive the expression
\bear
\frac{\bar{t}_{\frac{3}{2}}(\lambda)}{\rho_{\frac{3}{2}}^{(+)} \rho_{\frac{3}{2}}^{(-)}} 
\ket{\bar{\psi}_{2}(\lambda_{1},\lambda_{2})} &=& \sum_{i=1}^{4} \left[ \frac{t_{i}^{2}(\lambda)}{\zeta_{\frac{3}{2}}(\lambda)} \right]^{L} \omega_{i}^{(+)}(\lambda) \omega_{i}^{(-)}(\lambda) \left(\prod_{j=1}^{n=2}a_{1}^{i}(\lambda,\lambda_{j})\right) \ket{\bar{\psi}_{2}(\lambda_{1},\lambda_{2})}
\nonumber\\
&+& \mathrm{ unwanted~terms}
\ear

As a final comment we would like to stress that we have also performed extensive checks  verifying that in fact
the unwanted terms are canceled out provided the rapidities $\lambda_i$ satisfy the restriction (\ref{BA2SSsimp}).

\addcontentsline{toc}{section}{Appendix C}
\section*{\bf Appendix C: Auxiliary functions for $S=\frac{3}{2}$}
\setcounter{equation}{0}
\renewcommand{\theequation}{C.\arabic{equation}}

The purpose of this Appendix is to list the expressions of the functions $a_i^{j}(u,v)$, $b_1^{j}(u,v)$, $c_i^{lk}(u,v)$
and $d_i^{j}(u,v)$ used in the previous Appendix. 
To sort notation we shall used the symbol $u_{\pm}= u\pm v$ and we emphasize that 
the most complicated functions $a_2^4(u,v)$ and $a_8^4(u,v)$ have been collected at the end of this Appendix:
\bear
a_{1}^{1}(u,v)=\frac{a(-u_{-})\,e(u_{+})}{a(u_{+})\,e(-u_{-})}
\ear
\bear
a_{2}^{1}(u,v)=-\left( \frac{e(u_{+})\,b(-u_{-}) +
       \frac{b(2\,v)\,e(-u_{-})\,b(u_{+})}{a(2\,v)}}{a(u_{+})\,e(-u_{-})}
     \right)
\ear
\bear
a_{3}^{1}(u,v)=-\left( \frac{b(u_{+})}{a(u_{+})} \right)
\ear
\bear
a_{4}^{1}(u,v)=\frac{a(-u_{-})\,f(u_{+})}{a(u_{+})\,e(-u_{-})}
\ear
\bear
a_{1}^{2}(u,v)=\frac{\left( -\left( b(u_{+})\,b(u_{+}) \right)  + a(u_{+})\,l(u_{+})
       \right) \,\left( h(u_{-})\,l(u_{-}) - f(u_{-})\,r(u_{-}) \right) }
     {a(u_{+})\,e(u_{-})\,e(u_{+})\,h(u_{-})}
\ear
\bear
a_{2}^{2}(u,v)&=&\frac{e(u_{+})\,b(-u_{-})\,
      \left( \frac{b(2\,u)}{a(2\,u)} +
        \frac{b(u_{-})\,b(u_{+})}{e(u_{-})\,e(u_{+})} \right) }{a(u_{+})\,
      e(-u_{-})} + \frac{b(2\,v)\,
      \left( \frac{b(2\,u)\,b(u_{+})}{a(2\,u)\,a(u_{+})} +
        \frac{b(u_{-})\,\left( {b(u_{+})}^2 - a(u_{+})\,l(u_{+}) \right) }
         {a(u_{+})\,e(u_{-})\,e(u_{+})} \right) }{a(2\,v)}
 \nonumber \\
	& + & \frac{a(u_{-})\,b(u_{+})\,\left( h(u_{-})\,l(u_{-}) -
        f(u_{-})\,r(u_{-}) \right) }{a(u_{+})\,{e(u_{-})}^2\,h(u_{-})} +
   \frac{b(u_{-})\,f(u_{-})\,r(u_{+})}{e(u_{-})\,e(u_{+})\,h(u_{-})}
\ear
\bear
a_{3}^{2}(u,v)=\frac{b(2\,u)\,b(u_{+})}{a(2\,u)\,a(u_{+})} +
   \frac{b(u_{-})\,\left( {b(u_{+})}^2 - a(u_{+})\,l(u_{+}) \right) }
    {a(u_{+})\,e(u_{-})\,e(u_{+})}
\ear
\bear
a_{4}^{2}(u,v)=-\left( \frac{b(2\,v)\,f(u_{+})}{a(2\,v)\,e(u_{+})} \right)  +
   \frac{f(u_{-})\,h(u_{+})}{e(u_{+})\,h(u_{-})}
\ear
\bear
a_{5}^{2}(u,v)=-\left( \frac{f(u_{+})}{e(u_{+})} \right)
\ear
\bear
a_{6}^{2}(u,v)=- \frac{b(u_{+})\,b(u_{-})\,
      \left( h(u_{-})\,l(u_{-}) - f(u_{-})\,r(u_{-}) \right) \,
      f(u_{+})}{a(u_{+})\,{e(u_{-})}^2\,e(u_{+})\,h(u_{-})}
\nonumber \\
      -\left( \frac{a(-u_{-})\,f(u_{+})\,
        \left( \frac{b(2\,u)}{a(2\,u)} +
          \frac{b(u_{-})\,b(u_{+})}{e(u_{-})\,e(u_{+})} \right) }{a(u_{+})\,
        e(-u_{-})} \right)
	+ \frac{q(u_{+})\,
      \left( -\left( f(u_{-})\,c(u_{-}) \right)  +
        h(u_{-})\,f(u_{-}) \right) }{e(u_{-})\,e(u_{+})\,
      h(u_{-})}
 \ear
\bear
a_{7}^{2}(u,v)=\frac{\left( a(u_{+})\,m(u_{+}) -
       b(u_{+})\,c(u_{+}) \right) \,
     \left( h(u_{-})\,l(u_{-}) - f(u_{-})\,r(u_{-}) \right) }{a(u_{+})\,
     e(u_{-})\,e(u_{+})\,h(u_{-})}
\ear
\bear
a_{1}^{3}(u,v)=\frac{\left( -\left( i(u_{-})\,i_{1}(u_{-}) \right)  +
       j(u_{-})\,q(u_{-}) \right) \,
     \left( e(u_{+})\,q(u_{+}) - f(u_{+})\,r(u_{+}) \right) }{e(u_{+})\,
     h(u_{-})\,h(u_{+})\,j(u_{-})}
\ear
\bear
&&a_{2}^{3}(u,v)=\frac{c(u_{-})\,g_{1}(u_{+})\,i(u_{-})}
    {h(u_{-})\,h(u_{+})\,j(u_{-})} +
   \frac{a(u_{-})\,c(u_{+})\,e(u_{+})\,
      \left( -\left( g_{1}(u_{-})\,i(u_{-}) \right)  +
        j(u_{-})\,m(u_{-}) \right) }{a(u_{+})\,e(u_{-})\,
      h(u_{-})\,h(u_{+})\,j(u_{-})}
\nonumber \\
&&+ \frac{f(u_{+})\,\left( j(u_{-})\,q(u_{-}) - i(u_{-})\,i_{1}(u_{-}) \right) \,
      \left( a(u_{+})\,b(u_{-})\,e(u_{-})\,r(u_{+}) - a(u_{-})\,b(u_{+})\,e(u_{+})\,r(u_{-}) \right) }{a(u_{+})\,e(u_{-})\,
      e(u_{+})\,{h(u_{-})}^2\,h(u_{+})\,j(u_{-})}
\nonumber \\
&&- \left( \frac{e(u_{+})\,b(-u_{-})\,
         \left( \frac{b(2\,u)}{a(2\,u)} +
           \frac{b(u_{-})\,b(u_{+})}{e(u_{-})\,e(u_{+})} \right) }{a(u_{+})\,
         e(-u_{-})} + \frac{b(2\,v)\,
         \left( \frac{b(2\,u)\,b(u_{+})}{a(2\,u)\,a(u_{+})} +
           \frac{b(u_{-})\,\left( {b(u_{+})}^2 - a(u_{+})\,l(u_{+}) \right) }
            {a(u_{+})\,e(u_{-})\,e(u_{+})} \right) }{a(2\,v)} \right.
\nonumber \\
&&  + \left. \frac{a(u_{-})\,b(u_{+})\,
         \left( h(u_{-})\,l(u_{-}) - f(u_{-})\,r(u_{-}) \right) }{a(u_{+})\,
         {e(u_{-})}^2\,h(u_{-})} +
      \frac{b(u_{-})\,f(u_{-})\,r(u_{+})}{e(u_{-})\,e(u_{+})\,h(u_{-})}
      \right)
\nonumber \\
&& \times  \left( \frac{-\left( c(2\,u)\,b(2\,u) \right)  +
         a(2\,u)\,m(2\,u)}{-\left( b(2\,u)\,b(2\,u) \right)  +
         a(2\,u)\,l(2\,u)} + \frac{f(u_{-})\,r(u_{+})}{h(u_{-})\,h(u_{+})}
      \right) - \frac{b(2\,v)\,c(u_{-})\,m(u_{+})}
    {a(2\,v)\,h(u_{-})\,h(u_{+})}
\nonumber \\
&&  + \frac{\left( e(u_{+})\,b(-u_{-}) +
        \frac{b(2\,v)\,e(-u_{-})\,b(u_{+})}{a(2\,v)} \right) \,
      \left( \frac{c(2\,u)}{a(2\,u)} +
        \frac{c(u_{-})\,c(u_{+})}{h(u_{-})\,h(u_{+})} +
        \frac{b(2\,u)\,f(u_{-})\,r(u_{+})}{a(2\,u)\,h(u_{-})\,h(u_{+})}
        \right) }{a(u_{+})\,e(-u_{-})}
\ear
\bear
&& a_{3}^{3}(u,v)=-\left( \frac{c(u_{-})\,m(u_{+})}{h(u_{-})\,h(u_{+})}
      \right)  - \left( \frac{b(2\,u)\,b(u_{+})}{a(2\,u)\,a(u_{+})} +
      \frac{b(u_{-})\,\left( {b(u_{+})}^2 - a(u_{+})\,l(u_{+}) \right) }
       {a(u_{+})\,e(u_{-})\,e(u_{+})} \right) \,
\nonumber\\
&& \times  \left( \frac{-\left( c(2\,u)\,b(2\,u) \right)  + a(2\,u)\,m(2\,u)}
       {-\left( b(2\,u)\,b(2\,u) \right)  + a(2\,u)\,l(2\,u)} +
      \frac{f(u_{-})\,r(u_{+})}{h(u_{-})\,h(u_{+})} \right)
\nonumber \\
&& + \frac{b(u_{+})\,\left( \frac{c(2\,u)}{a(2\,u)} +
        \frac{c(u_{-})\,c(u_{+})}{h(u_{-})\,h(u_{+})} +
        \frac{b(2\,u)\,f(u_{-})\,r(u_{+})}{a(2\,u)\,h(u_{-})\,h(u_{+})}
        \right) }{a(u_{+})}
\ear
\bear
&& a_{4}^{3}(u,v)=\frac{f(u_{-})\,i(u_{-})\,i_{1}(u_{+})}
    {h(u_{-})\,h(u_{+})\,j(u_{-})} +
   \frac{e(u_{-})\,f(u_{+})\,\left( -\left( i(u_{-})\,
           i_{1}(u_{-}) \right)  + j(u_{-})\,q(u_{-}) \right) }
      {e(u_{+})\,{h(u_{-})}^2\,j(u_{-})}
\nonumber \\
&& -   \left( -\left( \frac{b(2\,v)\,f(u_{+})}{a(2\,v)\,e(u_{+})} \right)  +
      \frac{f(u_{-})\,h(u_{+})}{e(u_{+})\,h(u_{-})} \right) \,
    \left( \frac{-\left( c(2\,u)\,b(2\,u) \right)  + a(2\,u)\,m(2\,u)}
       {-\left( b(2\,u)\,b(2\,u) \right)  + a(2\,u)\,l(2\,u)} +
      \frac{f(u_{-})\,r(u_{+})}{h(u_{-})\,h(u_{+})} \right)
\nonumber \\
&& - \frac{b(2\,v)\,f(u_{-})\,q(u_{+})}{a(2\,v)\,h(u_{-})\,h(u_{+})}
\ear

\bear
a_{5}^{3}(u,v)=-\left( \frac{f(u_{-})\,q(u_{+})}{h(u_{-})\,h(u_{+})} \right)  +
   \frac{f(u_{+})\,\left( \frac{-\left( c(2\,u)\,b(2\,u) \right)  +
           a(2\,u)\,m(2\,u)}{-\left( b(2\,u)\,b(2\,u) \right)  +
           a(2\,u)\,l(2\,u)} + \frac{f(u_{-})\,r(u_{+})}{h(u_{-})\,h(u_{+})}
        \right) }{e(u_{+})}
\ear
\bear
a_{6}^{3}(u,v)=-\left( \frac{b(2\,v)\,i(u_{+})}{a(2\,v)\,h(u_{+})} \right)  +
   \frac{i(u_{-})\,j(u_{+})}{h(u_{+})\,j(u_{-})}
\ear
\bear
a_{7}^{3}(u,v)=-\left( \frac{i(u_{+})}{h(u_{+})} \right)
\ear

\bear
&&a_{8}^{3}(u,v)=-\left( \frac{a(-u_{-})\,f(u_{+})\,
        \left( \frac{c(2\,u)}{a(2\,u)} +
          \frac{c(u_{-})\,c(u_{+})}{h(u_{-})\,h(u_{+})} +
          \frac{b(2\,u)\,f(u_{-})\,r(u_{+})}{a(2\,u)\,h(u_{-})\,h(u_{+})}
          \right) }{a(u_{+})\,e(-u_{-})} \right)
\nonumber \\
&& + \frac{\left( g(u_{-})\,j(u_{-}) - d(u_{-})\,i(u_{-})  \right) \,
      f_{1}(u_{+})}{h(u_{-})\,h(u_{+})\,j(u_{-})} -  \frac{c(u_{+})\,b(u_{-})\,\left( j(u_{-})\,m(u_{-}) - g_{1}(u_{-})\,i(u_{-}) \right) \,
      f(u_{+})}{a(u_{+})\,e(u_{-})\,h(u_{-})\,h(u_{+})\,
      j(u_{-})} 
\nonumber \\
&& + \frac{f(u_{+})\, \left(  j(u_{-})\,q(u_{-}) - i(u_{-})\,i_{1}(u_{-}) \right) \,
      \left(  b(u_{+})\,b(u_{-})\,r(u_{-})\,f(u_{+}) - a(u_{+})\,e(u_{-})\,c(u_{-})\, q(u_{+}) \right) }
      {a(u_{+})\,e(u_{-})\,e(u_{+})\,{h(u_{-})}^2\,h(u_{+})\,j(u_{-})} 
\nonumber \\
&& - \left( \frac{-\left( c(2\,u)\,b(2\,u) \right)  + a(2\,u)\,m(2\,u)}
       {-\left( b(2\,u)\,b(2\,u) \right)  + a(2\,u)\,l(2\,u)} +
      \frac{f(u_{-})\,r(u_{+})}{h(u_{-})\,h(u_{+})} \right) \,
\nonumber \\
&& \times   \left( -\left( \frac{a(-u_{-})\,f(u_{+})\,
           \left( \frac{b(2\,u)}{a(2\,u)} +
             \frac{b(u_{-})\,b(u_{+})}{e(u_{-})\,e(u_{+})} \right) }{
           a(u_{+})\,e(-u_{-})} \right)  +
      \frac{q(u_{+})\,
         \left( -\left( f(u_{-})\,c(u_{-}) \right)  +
           h(u_{-})\,f(u_{-}) \right) }{e(u_{-})\,e(u_{+})\,
         h(u_{-})} \right.
\nonumber \\
&& \left. - \frac{b(u_{+})\,b(u_{-})\,
         \left( h(u_{-})\,l(u_{-}) - f(u_{-})\,r(u_{-}) \right) \,
         f(u_{+})}{a(u_{+})\,{e(u_{-})}^2\,e(u_{+})\,
         h(u_{-})} \right)
\ear
\bear
&& a_{9}^{3}(u,v)=\frac{\left( -\left( g_{1}(u_{-})\,i(u_{-}) \right)  +
        j(u_{-})\,m(u_{-}) \right) \,
      \left( a(u_{+})\,l_{1}(u_{+}) -
        c(u_{+})\,c(u_{+}) \right) }{a(u_{+})\,h(u_{-})\,
      h(u_{+})\,j(u_{-})} 
\nonumber \\
&& + \frac{f(u_{+})\, \left( j(u_{-})\,q(u_{-}) - i(u_{-})\,i_{1}(u_{-}) \right) \,
      \left( b(u_{+})\,e(u_{-})\,c(u_{+})\, r(u_{-}) - a(u_{+})\,e(u_{-})\,m(u_{+})\, r(u_{-}) \right) }{ a(u_{+})\,e(u_{-})\,e(u_{+})\,{h(u_{-})}^2\,
      h(u_{+})\,j(u_{-})} 
\nonumber \\
&&  - \frac{\left( a(u_{+})\,
         m(u_{+}) - b(u_{+})\,c(u_{+})
        \right) \,\left( h(u_{-})\,l(u_{-}) - f(u_{-})\,r(u_{-}) \right) }{a(u_{+})\,
      e(u_{-})\,e(u_{+})\,h(u_{-})}
\nonumber \\
&& \times \left( \frac{-\left( c(2\,u)\,b(2\,u) \right)  + a(2\,u)\,m(2\,u)}
         {-\left( b(2\,u)\,b(2\,u) \right)  + a(2\,u)\,l(2\,u)} +
        \frac{f(u_{-})\,r(u_{+})}{h(u_{-})\,h(u_{+})} \right)
\ear
\bear
a_{10}^{3}(u,v)=\frac{\left( e(u_{+})\,f_{1}(u_{+}) -
       f(u_{+})\,g_{1}(u_{+}) \right) \,
     \left( -\left( i(u_{-})\,i_{1}(u_{-}) \right)  +
       j(u_{-})\,q(u_{-}) \right) }{e(u_{+})\,h(u_{-})\,h(u_{+})\,j(u_{-})}
\ear
\bear
a_{1}^{4}(u,v)=\frac{h_{1}(u_{-})\,
     \left( h(u_{+})\,h_{1}(u_{+}) -
       i(u_{+})\,i_{1}(u_{+}) \right) }{h(u_{+})\,j(u_{-})\,
     j(u_{+})}
\ear
\bear
&& a_{3}^{4}(u,v)= \frac{\left( \frac{d(2\,u)}{a(2\,u)} +
        \frac{d(u_{-})\,d(u_{+})}{j(u_{-})\,j(u_{+})} +
        \frac{b(2\,u)\,g(u_{-})\,g_{1}(u_{+})}
         {a(2\,u)\,j(u_{-})\,j(u_{+})} +
        \frac{c(2\,u)\,i(u_{-})\,i_{1}(u_{+})}
         {a(2\,u)\,j(u_{-})\,j(u_{+})} \right) \,b(u_{+})}{a(u_{+})} - \left( \frac{n(u_{+})\,d(u_{-})}{j(u_{-})\,j(u_{+})}
      \right)
\nonumber \\
&&   - \left( \frac{b(2\,u)\,b(u_{+})}{a(2\,u)\,a(u_{+})} +
      \frac{b(u_{-})\,\left( {b(u_{+})}^2 - a(u_{+})\,l(u_{+}) \right) }
       {a(u_{+})\,e(u_{-})\,e(u_{+})} \right)
\nonumber \\
&& \times  \left( \frac{g(u_{-})\,g_{1}(u_{+})}{j(u_{-})\,j(u_{+})} +
      \frac{i(u_{-})\,i_{1}(u_{+})\,\left|M_{2,3}^{(+)}(2\,u)\right| }
         {j(u_{-})\,j(u_{+})\,\left|M_{2,2}^{(+)}(2\,u)\right| } +
      \frac{ \left|M_{2,4}^{(+)}(2\,u)\right| }{ \left|M_{2,2}^{(+)}(2\,u)\right| } \right)
\nonumber \\
&&	- \left( \frac{i(u_{-})\,i_{1}(u_{+})}{j(u_{-})\,j(u_{+})} +
      \frac{ \left|M_{3,4}^{(+)}(2\,u)\right| }{ \left|M_{3,3}^{(+)}(2\,u)\right| } \right) \,
    \left( -\left( \frac{c(u_{-})\,m(u_{+})}
         {h(u_{-})\,h(u_{+})} \right)  \right.
\nonumber \\
&&    -  \left( \frac{b(2\,u)\,b(u_{+})}{a(2\,u)\,a(u_{+})} +
         \frac{b(u_{-})\,\left( {b(u_{+})}^2 - a(u_{+})\,l(u_{+}) \right) }
          {a(u_{+})\,e(u_{-})\,e(u_{+})} \right) \,
       \left( \frac{\left|M_{2,3}^{(+)}(2\,u)\right|}{\left|M_{2,2}^{(+)}(2\,u)\right|} +
         \frac{f(u_{-})\,r(u_{+})}{h(u_{-})\,h(u_{+})} \right)
\nonumber \\
&& \left.   +   \frac{b(u_{+})\,\left( \frac{c(2\,u)}{a(2\,u)} +
           \frac{c(u_{-})\,c(u_{+})}{h(u_{-})\,h(u_{+})} +
           \frac{b(2\,u)\,f(u_{-})\,r(u_{+})}{a(2\,u)\,h(u_{-})\,h(u_{+})}
           \right) }{a(u_{+})} \right)
\ear
\bear
&& a_{4}^{4}(u,v)=\frac{h_{1}(u_{-})\,i(u_{+})\,
      \left( -\left( \frac{e(u_{-})\,f(u_{+})\,h(u_{+})\,
             i_{1}(u_{-})}{e(u_{+})\,h(u_{-})} \right)  +
        f(u_{-})\,i_{1}(u_{+}) \right) }{h(u_{+})\,{j(u_{-})}^2\,
      j(u_{+})} - \frac{b(2\,v)\,f_{1}(u_{+})\,g(u_{-})}
    {a(2\,v)\,j(u_{-})\,j(u_{+})}
\nonumber \\
&&    + \frac{e(u_{-})\,f_{1}(u_{-})\,g(u_{+})\,h(u_{+})}
    {e(u_{+})\,h(u_{-})\,j(u_{-})\,j(u_{+})} -
   \left(  \frac{f(u_{-})\,h(u_{+})}{e(u_{+})\,h(u_{-})} - \frac{b(2\,v)\,f(u_{+})}{a(2\,v)\,e(u_{+})} \right) 
\nonumber \\
&& \times  \left( \frac{g(u_{-})\,g_{1}(u_{+})}{j(u_{-})\,j(u_{+})} +
      \frac{i(u_{-})\,i_{1}(u_{+})\, \left|M_{2,3}^{(+)}(2\,u)\right| }{j(u_{-})\,j(u_{+})\,\left|M_{2,2}^{(+)}(2\,u)\right| } +
      \frac{\left|M_{2,4}^{(+)}(2\,u)\right|}{\left|M_{2,2}^{(+)}(2\,u)\right|} \right)  
\nonumber \\
&&    -  \left( \frac{i(u_{-})\,i_{1}(u_{+})}{j(u_{-})\,j(u_{+})} +
      \frac{ \left|M_{3,4}^{(+)}(2\,u)\right| }{ \left|M_{3,3}^{(+)}(2\,u)\right| } \right) \,
    \left( \frac{f(u_{-})\,i(u_{-})\,i_{1}(u_{+})}
       {h(u_{-})\,h(u_{+})\,j(u_{-})} \right.
\nonumber \\
&&    + \frac{e(u_{-})\,f(u_{+})\,
         \left( -\left( i(u_{-})\,i_{1}(u_{-}) \right)  +
           j(u_{-})\,q(u_{-}) \right) }{e(u_{+})\,{h(u_{-})}^2\,j(u_{-})} -
      \frac{b(2\,v)\,f(u_{-})\,q(u_{+})}{a(2\,v)\,h(u_{-})\,h(u_{+})}
      - \left( -\left( \frac{b(2\,v)\,f(u_{+})}{a(2\,v)\,e(u_{+})} \right) \right.
\nonumber \\
&& \left. \left.  +  \frac{f(u_{-})\,h(u_{+})}{e(u_{+})\,h(u_{-})} \right) \,
       \left( \frac{-\left( c(2\,u)\,b(2\,u) \right)  + a(2\,u)\,m(2\,u)}
          {-\left( b(2\,u)\,b(2\,u) \right)  + a(2\,u)\,l(2\,u)} +
         \frac{f(u_{-})\,r(u_{+})}{h(u_{-})\,h(u_{+})} \right)  \right)
\ear
\bear
&& a_{5}^{4}(u,v)=-\left( \frac{f_{1}(u_{+})\,g(u_{-})}{j(u_{-})\,j(u_{+})}
      \right)  + \frac{f(u_{+})\,
      \left( \frac{g(u_{-})\,g_{1}(u_{+})}{j(u_{-})\,j(u_{+})} +
        \frac{i(u_{-})\,i_{1}(u_{+})\,\left|M_{2,3}^{(+)}(2\,u)\right| }{j(u_{-})\,j(u_{+})\, \left|M_{2,2}^{(+)}(2\,u)\right|}
	+ \frac{\left|M_{2,4}^{(+)}(2\,u)\right|}{ \left|M_{2,2}^{(+)}(2\,u)\right| } \right) }{e(u_{+})}
\nonumber \\
&&   - \left( \frac{i(u_{-})\,i_{1}(u_{+})}{j(u_{-})\,j(u_{+})}
	 + \frac{ \left|M_{3,4}^{(+)}(2\,u)\right| }{ \left|M_{3,3}^{(+)}(2\,u)\right| } \right) \,
    \left( -\left( \frac{f(u_{-})\,q(u_{+})}{h(u_{-})\,h(u_{+})} \right)  +
      \frac{f(u_{+})\,\left( \frac{\left|M_{2,3}^{(+)}(2\,u)\right|}{\left|M_{2,2}^{(+)}(2\,u)\right|} +
           \frac{f(u_{-})\,r(u_{+})}{h(u_{-})\,h(u_{+})} \right) }{e(u_{+})}
      \right)
\nonumber \\
\ear
\bear
&& a_{6}^{4}(u,v)=\frac{h(u_{-})\,h_{1}(u_{-})\,i(u_{+})}
    {h(u_{+})\,{j(u_{-})}^2} - \frac{b(2\,v)\,h_{1}(u_{+})\,
      i(u_{-})}{a(2\,v)\,j(u_{-})\,j(u_{+})} -
   \left( -\left( \frac{b(2\,v)\,i(u_{+})}{a(2\,v)\,h(u_{+})} \right)  +
      \frac{i(u_{-})\,j(u_{+})}{h(u_{+})\,j(u_{-})} \right)
\nonumber \\
&& \times  \left( \frac{i(u_{-})\,i_{1}(u_{+})}{j(u_{-})\,j(u_{+})} +
      \frac{ \left|M_{3,4}^{(+)}(2\,u)\right| }{ \left|M_{3,3}^{(+)}(2\,u)\right| } \right)
\ear
\bear
&& a_{7}^{4}(u,v)=-\left( \frac{h_{1}(u_{+})\,i(u_{-})}{j(u_{-})\,j(u_{+})}
      \right)  + \frac{i(u_{+})\,
      \left( \frac{i(u_{-})\,i_{1}(u_{+})}{j(u_{-})\,j(u_{+})} +
        \frac{ \left|M_{3,4}^{(+)}(2\,u)\right| }{ \left|M_{3,3}^{(+)}(2\,u)\right| } \right) }{h(u_{+})}
\ear
\bear
&& a_{9}^{4}(u,v)=\frac{n(u_{-})\,
      \left( a(u_{+})\,b_{1}(u_{+}) -
        d(u_{+})\,c(u_{+}) \right) }{a(u_{+})\,j(u_{-})\,
      j(u_{+})}
\nonumber \\
&&  - \left( \frac{g(u_{-})\,g_{1}(u_{+})}
         {j(u_{-})\,j(u_{+})} +
        \frac{i(u_{-})\,i_{1}(u_{+})\,\left|M_{2,3}^{(+)}(2\,u)\right|}{j(u_{-})\,j(u_{+})\, \left|M_{2,2}^{(+)}(2\,u)\right|}
	+ \frac{\left|M_{2,4}^{(+)}(2\,u)\right|}{\left|M_{2,2}^{(+)}(2\,u)\right|} \right)
\nonumber \\
&& \times   \frac{ \left( a(u_{+})\,m(u_{+}) -
        b(u_{+})\,c(u_{+}) \right) \,
      \left( h(u_{-})\,l(u_{-}) - f(u_{-})\,r(u_{-}) \right) }{a(u_{+})\,
      e(u_{-})\,e(u_{+})\,h(u_{-})}
\nonumber \\
&&    +  \frac{f_{1}(u_{-})\,g(u_{+})\,
      \left( -\left( a(u_{+})\,e(u_{-})\,m(u_{+})\,
           r(u_{-}) \right)  + b(u_{+})\,e(u_{-})\,c(u_{+})\,
         r(u_{-}) \right) }{a(u_{+})\,e(u_{-})\,e(u_{+})\,h(u_{-})\,j(u_{-})\,
      j(u_{+})} - h_{1}(u_{-})\,i(u_{+})
\nonumber \\
&&  \times  \frac{\left( \frac{g_{1}(u_{-})\,
           \left( a(u_{+})\,l_{1}(u_{+}) -
             c(u_{+})\,c(u_{+}) \right) }{a(u_{+})\,
           h(u_{+})\,j(u_{-})} +
        \frac{f(u_{+})\,i_{1}(u_{-})\,
\left( b(u_{+})\,e(u_{-})\,c(u_{+})\,r(u_{-}) - a(u_{+})\,e(u_{-})\,m(u_{+})\, r(u_{-}) \right) }
           {a(u_{+})\,e(u_{-})\,e(u_{+})\,h(u_{-})\,h(u_{+})\,j(u_{-})}
        \right) }{j(u_{-})\,j(u_{+})}
\nonumber \\
&&  - \left( \frac{i(u_{-})\,i_{1}(u_{+})}{j(u_{-})\,j(u_{+})} +
      \frac{ \left|M_{3,4}^{(+)}(2\,u)\right| }{ \left|M_{3,3}^{(+)}(2\,u)\right| } \right)
\nonumber \\
&& \times  \left( \frac{\left( -\left( g_{1}(u_{-})\,i(u_{-}) \right)
               + j(u_{-})\,m(u_{-}) \right) \,
         \left( a(u_{+})\,l_{1}(u_{+}) -
           c(u_{+})\,c(u_{+}) \right) }{a(u_{+})\,h(u_{-})\,
         h(u_{+})\,j(u_{-})} + \right.
\nonumber \\
&&  \frac{f(u_{+})\,
         \left( j(u_{-})\,q(u_{-}) - i(u_{-})\,i_{1}(u_{-}) \right) \,
\left( b(u_{+})\,e(u_{-})\,c(u_{+})\,r(u_{-}) - a(u_{+})\,e(u_{-})\,m(u_{+})\, r(u_{-}) \right) }
         {a(u_{+})\,e(u_{-})\,e(u_{+})\,{h(u_{-})}^2\,h(u_{+})\,j(u_{-})}
\nonumber \\
&& \left. -  \frac{\left( a(u_{+})\,m(u_{+}) -
           b(u_{+})\,c(u_{+}) \right) \,
         \left( h(u_{-})\,l(u_{-}) - f(u_{-})\,r(u_{-}) \right) \,
         \left( \frac{ \left|M_{2,3}^{(+)}(2\,u)\right| }{ \left|M_{2,2}^{(+)}(2\,u)\right| }
     +  \frac{f(u_{-})\,r(u_{+})}{h(u_{-})\,h(u_{+})} \right) }{a(u_{+})\,
         e(u_{-})\,e(u_{+})\,h(u_{-})} \right)
\nonumber \\
\ear
\bear
&& a_{10}^{4}(u,v)=-\left( \frac{\left( \frac{f_{1}(u_{+})}{h(u_{+})} -
          \frac{f(u_{+})\,g_{1}(u_{+})}{e(u_{+})\,h(u_{+})}
          \right) \,h_{1}(u_{-})\,i(u_{+})\,
        i_{1}(u_{-})}{{j(u_{-})}^2\,j(u_{+})} \right)  +
   \frac{e_{1}(u_{+})\,f_{1}(u_{-})}
    {j(u_{-})\,j(u_{+})}
\nonumber \\
&&   - \frac{f_{1}(u_{-})\,g(u_{+})\, g_{1}(u_{+})}{e(u_{+})\,j(u_{-})\,j(u_{+})} - \left( \frac{i(u_{-})\,i_{1}(u_{+})}{j(u_{-})\,j(u_{+})} +
        \frac{ \left|M_{3,4}^{(+)}(2\,u)\right| }{ \left|M_{3,3}^{(+)}(2\,u)\right| } \right)
\nonumber \\
&& \times  \frac{\left( e(u_{+})\,f_{1}(u_{+}) -
        f(u_{+})\,g_{1}(u_{+}) \right) \,
      \left( j(u_{-})\,q(u_{-}) - i(u_{-})\,i_{1}(u_{-}) \right) }{e(u_{+})\,h(u_{-})\,h(u_{+})\,j(u_{-})}
\ear
\bear
b_{1}^{1}(u,v)=\frac{a(-u_{-})\,h(u_{+})}{a(u_{+})\,h(-u_{-})}
\ear
\bear
&& b_{1}^{2}(u,v)= \left( e(u_{+})\,q(u_{+}) -
       r(u_{+})\,f(u_{+}) \right)
\nonumber \\
&& \times  \frac{\left( -\left( \frac{\left( -\left( g_{1}(u_{-})\,
                 i(u_{-}) \right)  + j(u_{-})\,m(u_{-})
              \right) \,\left( i(u_{-})\,m(u_{-}) - g(u_{-})\,q(u_{-}) \right)
              }{e(u_{-})\,i(u_{-})\,
            \left( -\left( i(u_{-})\,i_{1}(u_{-}) \right)  +
              j(u_{-})\,q(u_{-}) \right) } \right)  +
       \frac{-\left( g(u_{-})\,m(u_{-}) \right)  +
          i(u_{-})\,q(u_{-})}{e(u_{-})\,i(u_{-})} \right)  }{ e(u_{+})\,e(u_{+})}
\nonumber \\
\ear
\bear
&& b_{1}^{3}(u,v)= \left( h_{1}(u_{-})\,l_{1}(u_{-}) -
       f_{1}(u_{-})\,r_{1}(u_{-}) \right)
\nonumber \\
&& \times   \frac{\left( \frac{l_{1}(u_{+})}{h(u_{+})} -
       \frac{m(u_{+})\,\left( a(u_{+})\,m(u_{+}) -
            b(u_{+})\,c(u_{+}) \right) }{h(u_{+})\,
          \left( -\left( b(u_{+})\,b(u_{+}) \right)  + a(u_{+})\,l(u_{+})
            \right) } - \frac{c(u_{+})\,
          \left( -\left( b(u_{+})\,m(u_{+}) \right)  +
            l(u_{+})\,c(u_{+}) \right) }{h(u_{+})\,
          \left( -\left( b(u_{+})\,b(u_{+}) \right)  + a(u_{+})\,l(u_{+})
            \right) } \right)}{  h(u_{-})\,h_{1}(u_{-})}
\ear
\bear
b_{1}^{4}(u,v)=\frac{e_{1}(u_{-})\,
     \left( \frac{e_{1}(u_{+})}{j(u_{+})} -
       \frac{g(u_{+})\,\left( -\left( g_{1}(u_{+})\,q(u_{+})
               \right)  + f_{1}(u_{+})\,r(u_{+}) \right) }{
          j(u_{+})\,\left( -\left( e(u_{+})\,q(u_{+}) \right)  +
            f(u_{+})\,r(u_{+}) \right) } -
       \frac{\left( e(u_{+})\,f_{1}(u_{+}) -
            f(u_{+})\,g_{1}(u_{+}) \right) \,
          r_{1}(u_{+})}{j(u_{+})\,
          \left( e(u_{+})\,q(u_{+}) - f(u_{+})\,r(u_{+}) \right) } \right) }
     {j(u_{-})}
\ear
\bear
c_{22}^{1}(u,v)=-\left( \frac{b(u_{+})}{a(u_{+})} \right)
\ear
\bear
c_{21}^{1}(u,v)=-\left( \frac{b(2\,u)\,b(u_{+})}{a(2\,u)\,a(u_{+})} \right)  -
   \frac{b(u_{-})\,e(u_{+})}{a(u_{+})\,e(u_{-})}
\ear
\bear
c_{12}^{1}(u,v)=-\left( \frac{b(2\,v)\,b(u_{+})}{a(2\,v)\,a(u_{+})} \right)  +
   \frac{b(u_{-})\,e(u_{+})}{a(u_{+})\,e(u_{-})}
\ear
\bear
c_{11}^{1}(u,v)=\frac{b(2\,u)\,\left( -\left( \frac{b(2\,v)\,b(u_{+})}
           {a(2\,v)\,a(u_{+})} \right)  +
        \frac{b(u_{-})\,e(u_{+})}{a(u_{+})\,e(u_{-})} \right) }{a(2\,u)} +
   \frac{b(u_{+})\,e(u_{-})}{a(u_{+})\,e(u_{-})} -
   \frac{b(u_{-})\,b(2\,v)\,e(u_{+})}
    {a(2\,v)\,a(u_{+})\,e(u_{-})}
\ear
\bear
c_{23}^{2}(u,v)=-\left( \frac{f(u_{+})}{e(u_{+})} \right)
\ear
\bear
c_{22}^{2}(u,v)=\frac{\frac{b(u_{+})\,f(u_{-})\,b(u_{+})}{a(u_{+})} - f(u_{-})\,l(u_{+}) -
     \frac{f(u_{+})\,h(u_{-})\,\left( -\left( c(2\,u)\,b(2\,u) \right)  +
          a(2\,u)\,m(2\,u) \right) }{-\left( b(2\,u)\,b(2\,u) \right)  +
        a(2\,u)\,l(2\,u)}}{e(u_{+})\,h(u_{-})}
\ear
\bear
&& c_{21}^{2}(u,v)=\left(-\left( \frac{c(2\,u)\,f(u_{+})\,h(u_{-})}{a(2\,u)} \right)  +
     \left( \frac{b(2\,u)\,b(u_{+})}{a(2\,u)\,a(u_{+})} +
        \frac{b(u_{-})\,e(u_{+})}{a(u_{+})\,e(u_{-})} \right)
        \,f(u_{-})\,b(u_{+}) \right.
\nonumber \\
&& \left. - \frac{b(2\,u)\,f(u_{-})\,l(u_{+})}{a(2\,u)} -
     c(u_{-})\,f(u_{+})\right) \, / \,(\,e(u_{+})\,h(u_{-})\,)
\ear
\bear
c_{13}^{2}(u,v)=-\left( \frac{b(2\,v)\,f(u_{+})}{a(2\,v)\,e(u_{+})} \right)  +
   \frac{f(u_{-})\,h(u_{+})}{e(u_{+})\,h(u_{-})}
\ear
\bear
&& c_{12}^{2}(u,v)=-\left( \frac{\left( -\left( \frac{b(2\,v)\,b(u_{+})}
             {a(2\,v)\,a(u_{+})} \right)  +
          \frac{b(u_{-})\,e(u_{+})}{a(u_{+})\,e(u_{-})} \right) \,f(u_{-})\,
        b(u_{+})}{e(u_{+})\,h(u_{-})} \right)  +
   \frac{f(u_{+})\,l(u_{-})}{e(u_{+})\,h(u_{-})}
\nonumber \\
&&   +   \frac{f(u_{-})\,h(u_{+})\,\left|M_{2,3}^{(+)}(2\,u)\right| }{e(u_{+})\,h(u_{-})\, \left|M_{2,2}^{(+)}(2\,u)\right| } -
   \frac{b(2\,v)\,\left( f(u_{-})\,l(u_{+}) +
        \frac{f(u_{+})\,h(u_{-})\, \left|M_{2,3}^{(+)}(2\,u)\right| }{ \left|M_{2,2}^{(+)}(2\,u)\right| } \right) }
      {a(2\,v)\,e(u_{+})\,h(u_{-})}
\ear
\bear
&& c_{11}^{2}(u,v)=\left( \, \frac{c(2\,u)\,f(u_{-})\,h(u_{+})}{a(2\,u)} -
     \left( \frac{b(2\,u)\,\left( -\left( \frac{b(2\,v)\,b(u_{+})}
                {a(2\,v)\,a(u_{+})} \right)  +
             \frac{b(u_{-})\,e(u_{+})}{a(u_{+})\,e(u_{-})} \right) }{a(2\,u)}
         + \frac{b(u_{+})\,e(u_{-})}{a(u_{+})\,e(u_{-})}  \right. \right.
\nonumber \\
&& \left. -  \frac{b(u_{-})\,b(2\,v)\,e(u_{+})}
         {a(2\,v)\,a(u_{+})\,e(u_{-})} \right) \,f(u_{-})\,b(u_{+}) +
     \frac{b(2\,u)\,f(u_{+})\,l(u_{-})}{a(2\,u)} +
     c(u_{+})\,f(u_{-}) 
\nonumber \\
&&  \left. -
     \frac{b(2\,v)\,\left( \frac{c(2\,u)\,f(u_{+})\,h(u_{-})}{a(2\,u)} +
          \frac{b(2\,u)\,f(u_{-})\,l(u_{+})}{a(2\,u)} +
          c(u_{-})\,f(u_{+}) \right) }{a(2\,v)}\right)\, / \, \left(\,e(u_{+})\,
     h(u_{-})\,\right)
\ear
\bear
c_{24}^{3}(u,v)=-\left( \frac{i(u_{+})}{h(u_{+})} \right)
\ear
\bear
c_{23}^{3}(u,v)= -\, \frac{i(u_{+})\,\left|M_{3,4}^{(+)}(2\,u)\right|}{h(u_{+})\,\left|M_{3,3}^{(+)}(2\,u)\right| }   -
   \frac{i(u_{-})\,q(u_{+})}{h(u_{+})\,j(u_{-})} +
   \frac{f(u_{+})\,i(u_{-})\,r(u_{+})}{e(u_{+})\,h(u_{+})\,j(u_{-})}
\ear
\bear
&& c_{22}^{3}(u,v)=\frac{b(u_{+})\,g(u_{-})\,c(u_{+})}
    {a(u_{+})\,h(u_{+})\,j(u_{-})} -
   \frac{g(u_{-})\,m(u_{+}) +
      \frac{i(u_{+})\,j(u_{-})\,
         \left|M_{2,4}^{(+)}(2\,u)\right| }{\left|M_{2,2}^{(+)}(2\,u)\right|} +
      \frac{i(u_{-})\,\left|M_{2,3}^{(+)}(2\,u)\right| \,q(u_{+})}{\left|M_{2,2}^{(+)}(2\,u)\right|}}{h(u_{+})\,j(u_{-})}
\nonumber \\
&&    -\,  \frac{i(u_{-})\,\left( \frac{b(u_{+})\,f(u_{-})\,b(u_{+})}{a(u_{+})} -
        f(u_{-})\,l(u_{+}) - \frac{f(u_{+})\,h(u_{-})\,
           \left( -\left( c(2\,u)\,b(2\,u) \right)  + a(2\,u)\,m(2\,u) \right)
             }{-\left( b(2\,u)\,b(2\,u) \right)  + a(2\,u)\,l(2\,u)} \right)
       r(u_{+})}{e(u_{+})\,h(u_{-})\,h(u_{+})\,j(u_{-})}
\ear
\bear
&& c_{21}^{3}(u,v)=\frac{\left( \frac{b(2\,u)\,b(u_{+})}{a(2\,u)\,a(u_{+})} +
        \frac{b(u_{-})\,e(u_{+})}{a(u_{+})\,e(u_{-})} \right)
        \,g(u_{-})\,c(u_{+})}{h(u_{+})\,j(u_{-})}
\nonumber \\
&&  -\,  \frac{d(u_{-})\,g(u_{+}) +
      \frac{d(2\,u)\,i(u_{+})\,j(u_{-})}{a(2\,u)} +
      \frac{b(2\,u)\,g(u_{-})\,m(u_{+})}{a(2\,u)} +
      \frac{c(2\,u)\,i(u_{-})\,q(u_{+})}{a(2\,u)}}{h(u_{+})\,j(u_{-})}
\nonumber \\
&&    -\,  \left( i(u_{-})\,r(u_{+})\,\left( -\left( \frac{c(2\,u)\,f(u_{+})\,h(u_{-})}
           {a(2\,u)} \right)  +
        \left( \frac{b(2\,u)\,b(u_{+})}{a(2\,u)\,a(u_{+})} +
           \frac{b(u_{-})\,e(u_{+})}{a(u_{+})\,e(u_{-})}
           \right) \,f(u_{-})\,b(u_{+}) \right. \right.
\nonumber \\
&&  \left. \left.  -    \frac{b(2\,u)\,f(u_{-})\,l(u_{+})}{a(2\,u)} -
        c(u_{-})\,f(u_{+}) \right) \right)\, / \,\left(\, e(u_{+})\,h(u_{-})\,
      h(u_{+})\,j(u_{-})\,\right)
\ear
\bear
c_{14}^{3}(u,v)=-\left( \frac{b(2\,v)\,i(u_{+})}{a(2\,v)\,h(u_{+})} \right)  +
   \frac{i(u_{-})\,j(u_{+})}{h(u_{+})\,j(u_{-})}
\ear
\bear
&& c_{13}^{3}(u,v)=\frac{i(u_{-})\,j(u_{+})\, \left|M_{3,4}^{(+)}(2\,u)\right| }{h(u_{+})\,
      j(u_{-})\,\left|M_{3,3}^{(+)}(2\,u)\right| } +
   \frac{i(u_{+})\,q(u_{-})}{h(u_{+})\,j(u_{-})} -
   \frac{b(2\,v)\,\left( \frac{i(u_{+})\, \left|M_{3,4}^{(+)}(2\,u)\right| }{
           h(u_{+})\,\left|M_{3,3}^{(+)}(2\,u)\right| } +
        \frac{i(u_{-})\,q(u_{+})}{h(u_{+})\,j(u_{-})} \right) }{a(2\,v)}
\nonumber \\
&&  -  \frac{\left( -\left( \frac{b(2\,v)\,f(u_{+})}{a(2\,v)\,e(u_{+})} \right)
            + \frac{f(u_{-})\,h(u_{+})}{e(u_{+})\,h(u_{-})} \right) \,
      i(u_{-})\,r(u_{+})}{h(u_{+})\,j(u_{-})}
\ear
\bear
&& c_{12}^{3}(u,v)=\frac{g(u_{+})\,m(u_{-})}{h(u_{+})\,j(u_{-})} +
   \frac{i(u_{-})\,j(u_{+})\,\left|M_{2,4}^{(+)}(2\,u)\right| }{h(u_{+})\,j(u_{-})\,
      \left|M_{2,2}^{(+)}(2\,u)\right| }
\nonumber \\
&&    -  \frac{\left( -\left( \frac{b(2\,v)\,b(u_{+})}{a(2\,v)\,a(u_{+})} \right)
            + \frac{b(u_{-})\,e(u_{+})}{a(u_{+})\,e(u_{-})} \right) \,
      g(u_{-})\,c(u_{+})}{h(u_{+})\,j(u_{-})} +
   \frac{i(u_{+})\,\left|M_{2,3}^{(+)}(2\,u)\right| \,q(u_{-})}{h(u_{+})\,j(u_{-})\, \left|M_{2,2}^{(+)}(2\,u)\right| }
\nonumber \\
&&   -  \frac{b(2\,v)\,\left( \frac{g(u_{-})\,m(u_{+})}
         {h(u_{+})\,j(u_{-})} +
        \frac{i(u_{+})\,\left|M_{2,4}^{(+)}(2\,u)\right| }{h(u_{+})\,
           \left|M_{2,2}^{(+)}(2\,u)\right|} + \frac{i(u_{-})\, \left|M_{2,3}^{(+)}(2\,u)\right|
             \,q(u_{+})}{h(u_{+})\,j(u_{-})\, \left|M_{2,2}^{(+)}(2\,u)\right|
             } \right) }{a(2\,v)}
\nonumber \\
&&     -  \left(i(u_{-})\,\left( -\left( \frac{\left( -\left( \frac{b(2\,v)\,
                    b(u_{+})}{a(2\,v)\,a(u_{+})} \right)  +
               \frac{b(u_{-})\,e(u_{+})}{a(u_{+})\,e(u_{-})} \right) \,
             f(u_{-})\,b(u_{+})}{e(u_{+})\,h(u_{-})} \right)  +
        \frac{f(u_{+})\,l(u_{-})}{e(u_{+})\,h(u_{-})} \right. \right.
\nonumber \\
&&  +  \frac{f(u_{-})\,h(u_{+})\,
           \left( -\left( c(2\,u)\,b(2\,u) \right)  + a(2\,u)\,m(2\,u) \right)
             }{e(u_{+})\,h(u_{-})\,
           \left( -\left( b(2\,u)\,b(2\,u) \right)  + a(2\,u)\,l(2\,u) \right)
             }
\nonumber \\
&&  \left. \left.   - \frac{b(2\,v)\,
           \left( f(u_{-})\,l(u_{+}) +
             \frac{f(u_{+})\,h(u_{-})\,
                \left( -\left( c(2\,u)\,b(2\,u) \right)  +
                  a(2\,u)\,m(2\,u) \right) }{-\left( b(2\,u)\,b(2\,u) \right)
                    + a(2\,u)\,l(2\,u)} \right) }{a(2\,v)\,e(u_{+})\,h(u_{-})}
        \right) \,r(u_{+})\right)\, / \,\left(\, h(u_{+})\,j(u_{-})\, \right)
\nonumber \\
\ear
\bear
&& c_{11}^{3}(u,v)=\frac{d(u_{+})\,g(u_{-})}{h(u_{+})\,j(u_{-})} +
   \frac{d(2\,u)\,i(u_{-})\,j(u_{+})}{a(2\,u)\,h(u_{+})\,j(u_{-})} +
   \frac{b(2\,u)\,g(u_{+})\,m(u_{-})}
    {a(2\,u)\,h(u_{+})\,j(u_{-})}
\nonumber \\
&&  - \frac{\left( \frac{b(2\,u)\,\left( -\left( \frac{b(2\,v)\,b(u_{+})}
                {a(2\,v)\,a(u_{+})} \right)  +
             \frac{b(u_{-})\,e(u_{+})}{a(u_{+})\,e(u_{-})} \right) }{a(2\,u)}
         + \frac{b(u_{+})\,e(u_{-})}{a(u_{+})\,e(u_{-})} -
        \frac{b(u_{-})\,b(2\,v)\,e(u_{+})}
         {a(2\,v)\,a(u_{+})\,e(u_{-})} \right) \,g(u_{-})\,
      c(u_{+})}{h(u_{+})\,j(u_{-})} 
\nonumber \\
&&     +   \frac{c(2\,u)\,i(u_{+})\,q(u_{-})}{a(2\,u)\,h(u_{+})\,j(u_{-})} -
   \frac{b(2\,v)\,\left( \frac{d(2\,u)\,i(u_{+})}{a(2\,u)\,h(u_{+})} +
        \frac{d(u_{-})\,g(u_{+})}{h(u_{+})\,j(u_{-})} +
        \frac{b(2\,u)\,g(u_{-})\,m(u_{+})}
         {a(2\,u)\,h(u_{+})\,j(u_{-})} +
        \frac{c(2\,u)\,i(u_{-})\,q(u_{+})}{a(2\,u)\,h(u_{+})\,j(u_{-})}
        \right) }{a(2\,v)} 
\nonumber \\
&&  - \left( i(u_{-})\,r(u_{+})\,
      \left( \frac{c(2\,u)\,f(u_{-})\,h(u_{+})}{a(2\,u)} -
        \left( \frac{b(2\,u)\,\left( -\left( \frac{b(2\,v)\,b(u_{+})}
                   {a(2\,v)\,a(u_{+})} \right)  +
                \frac{b(u_{-})\,e(u_{+})}{a(u_{+})\,e(u_{-})} \right) }{a(
              2\,u)} \right. \right. \right.
\nonumber \\
&&  \left.    + \frac{b(u_{+})\,e(u_{-})}
            {a(u_{+})\,e(u_{-})} -
           \frac{b(u_{-})\,b(2\,v)\,e(u_{+})}
            {a(2\,v)\,a(u_{+})\,e(u_{-})} \right) \,f(u_{-})\,b(u_{+}) +
        \frac{b(2\,u)\,f(u_{+})\,l(u_{-})}{a(2\,u)} +
        c(u_{+})\,f(u_{-})
\nonumber \\
&&  \left. \left.  -    \frac{b(2\,v)\,\left( \frac{c(2\,u)\,f(u_{+})\,h(u_{-})}{a(2\,u)} +
             \frac{b(2\,u)\,f(u_{-})\,l(u_{+})}{a(2\,u)} +
             c(u_{-})\,f(u_{+}) \right) }{a(2\,v)} \right) \right)\, / \,\left(\,e(u_{+})\,h(u_{-})\,h(u_{+})\,j(u_{-})\,\right)
\nonumber \\
\ear
\bear
d_{2}^{1}(u,v)=-\left( \frac{f(u_{+})}{e(u_{+})} \right)
\ear
\bear
d_{1}^{1}(u,v)=-\left( \frac{h(u_{+})\,
        \left( -\left( f(u_{-})\,c(u_{-}) \right)  +
          h(u_{-})\,f(u_{-}) \right) }{e(
         u_{+})\,\left( h(u_{-})\,l(u_{-}) - f(u_{-})\,r(u_{-}) \right) }
      \right)  - \frac{b(2\,u)\,f(u_{+})}
    {a(2\,u)\,e(u_{+})}
\ear
\bear
d_{3}^{2}(u,v)=-\left( \frac{a(u_{+})\,m(u_{+}) -
       b(u_{+})\,c(u_{+})}{-\left( b(u_{+})\,b(u_{+}) \right)
           + a(u_{+})\,l(u_{+})} \right)
\ear
\bear
&& d_{2}^{2}(u,v)=\left[-\left( \frac{\left|M_{2,3}^{(+)}(2\,u)\right| \,
          \left( a(u_{+})\,m(u_{+}) -
            b(u_{+})\,c(u_{+}) \right) }{\left|M_{2,2}^{(+)}(2\,u)\right|} \right)\right.
\nonumber \\
&&  -   \frac{a(u_{+})\,\left( -\left( g_{1}(u_{-})\,i(u_{-})
               \right)  + j(u_{-})\,m(u_{-}) \right) \,
          q(u_{+})}{-\left( i(u_{-})\,
             i_{1}(u_{-}) \right)  + j(u_{-})\,q(u_{-})} +
       \frac{b(u_{+})\,r(u_{-})\,f(u_{+})}{h(u_{-})}
\nonumber \\
&& \left.  +  \frac{\left( \frac{a(u_{+})\,r(u_{+})\,
               \left( -\left( g_{1}(u_{-})\,i(u_{-}) \right)  +
                 j(u_{-})\,m(u_{-}) \right) }{-\left(
                  i(u_{-})\,i_{1}(u_{-}) \right)  +
               j(u_{-})\,q(u_{-})} -
            \frac{b(u_{+})\,e(u_{+})\,r(u_{-})}{h(u_{-})}
            \right) \,f(u_{+})}{e(u_{+})}\right]
\nonumber \\
&&  / \,\left(\, a(u_{+})\,l(u_{+}) - b(u_{+})\,b(u_{+}) \, \right)
\ear
\bear
&& d_{1}^{2}(u,v)=\left(\frac{c(2\,u)\,\left( -\left( a(u_{+})\,m(u_{+})
             \right)  + b(u_{+})\,c(u_{+}) \right) }{a(2\,u)} \right.
\nonumber \\
&&  - \frac{a(u_{+})\,i_{1}(u_{+})\,
        \left( -\left( d(u_{-})\,i(u_{-}) \right)  +
          g(u_{-})\,j(u_{-}) \right) }{-\left( i(u_{-})\,
           i_{1}(u_{-}) \right)  + j(u_{-})\,q(u_{-})}
\nonumber \\
&&  -    \frac{a(u_{+})\,b(2\,u)\,\left( -\left( g_{1}(u_{-})\,
             i(u_{-}) \right)  + j(u_{-})\,m(u_{-}) \right)
         q(u_{+})}{a(2\,u)\,
        \left( -\left( i(u_{-})\,i_{1}(u_{-}) \right)  +
          j(u_{-})\,q(u_{-}) \right) }
\nonumber \\
&&   +
     \frac{b(u_{+})\,h(u_{+})\,
        \left( -\left( l(u_{-})\,c(u_{-}) \right)  +
          r(u_{-})\,f(u_{-}) \right) }{-\left( h(u_{-})\,
           l(u_{-}) \right)  + f(u_{-})\,r(u_{-})} + \left( j(u_{-})\,m(u_{-}) - g_{1}(u_{-})\,i(u_{-}) \right)
\nonumber \\
&& \left. \times   \frac{a(u_{+})\,r(u_{+})\,
              \left( \frac{h(u_{+})\,
             \left( -\left( f(u_{-})\,c(u_{-}) \right)  +
               h(u_{-})\,f(u_{-}) \right) }{e(u_{+})\,
	       \left( h(u_{-})\,l(u_{-}) - f(u_{-})\,r(u_{-})
               \right) } + \frac{b(2\,u)\,f(u_{+})}
           {a(2\,u)\,e(u_{+})} \right) }{-\left( i(u_{-})\,
           i_{1}(u_{-}) \right)  + j(u_{-})\,q(u_{-})} \right)
\nonumber \\
&&    / \, \left(\, -\left(  b(u_{+})\,b(u_{+}) \right)  + a(u_{+})\,l(u_{+}) \, \right)
\ear
\bear
d_{4}^{3}(u,v)=-\left( \frac{e(u_{+})\,f_{1}(u_{+}) -
       f(u_{+})\,g_{1}(u_{+})}{e(u_{+})\,q(u_{+}) -
       f(u_{+})\,r(u_{+})} \right)
\ear
\bear
&& d_{3}^{3}(u,v)=\left( \frac{\left( -\left( e(u_{+})\,f_{1}(u_{+}) \right)  +
          f(u_{+})\,g_{1}(u_{+}) \right) \,
        \left|M_{3,4}^{(+)}(2\,u)\right| }{ \left|M_{3,3}^{(+)}(2\,u)\right| } \right.
\nonumber \\
&&  +  \left( e(u_{+})\,f_{1}(u_{-})\,h(u_{+})\,
        \left( \frac{l_{1}(u_{+})}{h(u_{+})} -
          \frac{m(u_{+})\,\left( a(u_{+})\,m(u_{+}) -
               b(u_{+})\,c(u_{+}) \right) }{h(u_{+})\,
             \left( -\left( b(u_{+})\,b(u_{+}) \right)  +
               a(u_{+})\,l(u_{+}) \right) } \right. \right.
\nonumber \\
&&  \left. \left.   - \frac{c(u_{+})\,\left( -\left( b(u_{+})\,
                  m(u_{+}) \right)  +
               l(u_{+})\,c(u_{+}) \right) }{h(u_{+})\,
             \left( -\left( b(u_{+})\,b(u_{+}) \right)  +
               a(u_{+})\,l(u_{+}) \right) } \right) \,
        \left( h_{1}(u_{-})\,l_{1}(u_{-}) -
          f_{1}(u_{-})\,r_{1}(u_{-}) \right) \right)
\nonumber \\
&& \left.   / \,\left(
        h_{1}(u_{-})\,
        \left( f_{1}(u_{-})\,r_{1}(u_{-}) - h_{1}(u_{-})\,l_{1}(u_{-}) \right) \, \right) \,\, \right)\,/\,
    \left(\,e(u_{+})\,q(u_{+}) - f(u_{+})\,r(u_{+}) \, \right)
\ear
\bear
&& d_{2}^{3}(u,v)=-\left( \frac{e(u_{+})\,f_{1}(u_{-})\,
          l_{1}(u_{+})\, \left|M_{2,3}^{(+)}(2\,u)\right|
          }{h_{1}(u_{-})\,\left|M_{2,2}^{(+)}(2\,u)\right|} + \frac{\left( e(u_{+})\,f_{1}(u_{+}) -
            f(u_{+})\,g_{1}(u_{+}) \right) \, \left|M_{2,4}^{(+)}(2\,u)\right| }{\left|M_{2,2}^{(+)}(2\,u)\right|} \right.
\nonumber \\
&&  -  \frac{f(u_{+})\,\left( i_{1}(u_{-})\,
             m(u_{-}) - g_{1}(u_{-})\,q(u_{-})
            \right) \,q(u_{+})}{i(u_{-})\,
           i_{1}(u_{-}) - j(u_{-})\,q(u_{-})} +
       \frac{n(u_{-})\,e(u_{+})\,f_{1}(u_{+})}
        {h_{1}(u_{-})} 
\nonumber \\
&&  +   \frac{\left( -\left( \frac{n(u_{-})\,e(u_{+})\,
                 g_{1}(u_{+})}{h_{1}(u_{-})} \right)
                + \frac{f(u_{+})\,r(u_{+})\,
               \left( i_{1}(u_{-})\,m(u_{-}) -
                 g_{1}(u_{-})\,q(u_{-}) \right) }{i(u_{-})\,
                i_{1}(u_{-}) - j(u_{-})\,q(u_{-})} \right) \,
          f(u_{+})}{e(u_{+})}
\nonumber \\
&&   - \, \left( e(u_{+})\,f_{1}(u_{-})\,m(u_{+})\,
          \left( \frac{\left|M_{2,3}^{(+)}(2\,u)\right| \,
               \left( a(u_{+})\,m(u_{+}) -
                 b(u_{+})\,c(u_{+}) \right) }{\left|M_{2,2}^{(+)}(2\,u)\right|} \right. \right.
\nonumber \\
&&  +  \frac{a(u_{+})\,\left( -\left( g_{1}(u_{-})\,
                    i(u_{-}) \right)  + j(u_{-})\,m(u_{-})
                 \right) \,q(u_{+})}{-\left( i(u_{-})\,
                  i_{1}(u_{-}) \right)  + j(u_{-})\,q(u_{-})} -
            \frac{b(u_{+})\,r(u_{-})\,f(u_{+})}{h(u_{-})}
\nonumber \\
&& \left. \left.   -   \frac{\left( \frac{a(u_{+})\,r(u_{+})\,
                    \left( -\left( g_{1}(u_{-})\,i(u_{-}) \right)
                          + j(u_{-})\,m(u_{-}) \right) }
                    {-\left( i(u_{-})\,i_{1}(u_{-}) \right)  +
                    j(u_{-})\,q(u_{-})} -
                 \frac{b(u_{+})\,e(u_{+})\,r(u_{-})}
                  {h(u_{-})} \right) \,f(u_{+})}{
                e(u_{+})} \right) \right)
\nonumber \\
&&  / \, \left(h_{1}(u_{-})\,\left( a(u_{+})\,l(u_{+}) - b(u_{+})\,b(u_{+}) \, \right) \,\, \right)
      + c(u_{+})\,e(u_{+})\,f_{1}(u_{-})
\nonumber \\
&&  \times \left( \frac{\left|M_{2,3}^{(+)}(2\,u)\right| \,
               \left( b(u_{+})\,m(u_{+}) -
                 l(u_{+})\,c(u_{+}) \right) }{\left|M_{2,2}^{(+)}(2\,u)\right|}\right.
\nonumber \\
&&  +
            \frac{b(u_{+})\,\left( -\left( g_{1}(u_{-})\,
                    i(u_{-}) \right)  + j(u_{-})\,m(u_{-})
                 \right) \,q(u_{+})}{-\left( i(u_{-})\,
                  i_{1}(u_{-}) \right)  + j(u_{-})\,q(u_{-})} -
            \frac{l(u_{+})\,r(u_{-})\,f(u_{+})}{h(u_{-})}
\nonumber \\
&& \left.  - \frac{\left( \frac{r(u_{+})\,b(u_{+})\,
                    \left( -\left( g_{1}(u_{-})\,i(u_{-}) \right)
                          + j(u_{-})\,m(u_{-}) \right) }
                    {-\left( i(u_{-})\,i_{1}(u_{-}) \right)  +
                    j(u_{-})\,q(u_{-})} -
                 \frac{e(u_{+})\,l(u_{+})\,r(u_{-})}
                  {h(u_{-})} \right) \,f(u_{+})}{
                e(u_{+})} \right)
\nonumber \\
&& \left.  / \,\left( h_{1}(
           u_{-})\,\left( a(u_{+})\,l(u_{+}) - b(u_{+})\,b(u_{+})  \right) \, \right) \,\, \right)\, / \, \left( \, e(u_{+})\,q(u_{+}) -
       f(u_{+})\,r(u_{+}) \, \right)
\ear
\bear
&& d_{1}^{3}(u,v)=\left(\frac{d(2\,u)\,\left( -\left( e(u_{+})\,f_{1}(u_{+})
             \right)  + f(u_{+})\,g_{1}(u_{+}) \right) }{a(2\,u)} \right.
\nonumber \\
&&  + \frac{f(u_{+})\,i_{1}(u_{+})\,
        \left( g(u_{-})\,i_{1}(u_{-}) -
          d(u_{-})\,q(u_{-}) \right) }{i(u_{-})\,
         i_{1}(u_{-}) - j(u_{-})\,q(u_{-})}
\nonumber \\
&&  +  \frac{a(-u_{-})\,e(u_{+})\,f_{1}(u_{-})\,h(u_{+})\,
        \left( \frac{c(2\,u)\,h(u_{-})\,h(u_{+})}{a(2\,u)} +
          c(u_{-})\,c(u_{+}) +
          \frac{b(2\,u)\,f(u_{-})\,r(u_{+})}{a(2\,u)} \right) }{a(u_{+})\,
        h(-u_{-})\,\left( -\left( h_{1}(u_{-})\,
             l_{1}(u_{-}) \right)  +
          f_{1}(u_{-})\,r_{1}(u_{-}) \right) }
\nonumber \\
&&  + \frac{b(2\,u)\,\left( \frac{f(u_{+})\,
             \left( i_{1}(u_{-})\,m(u_{-}) -
               g_{1}(u_{-})\,q(u_{-}) \right) \,
             q(u_{+})}{i(u_{-})\,i_{1}(u_{-}) -
             j(u_{-})\,q(u_{-})} -
          \frac{e(u_{+})\,\left( f_{1}(u_{-})\,
                r_{1}(u_{-}) -
               n(u_{-})\,l_{1}(u_{-}) \right) \,
             f_{1}(u_{+})}{-\left( h_{1}(u_{-})\,
                l_{1}(u_{-}) \right)  +
             f_{1}(u_{-})\,r_{1}(u_{-})} \right) }
        {a(2\,u)}
\nonumber \\
&&   - \left( \frac{f(u_{+})\,r(u_{+})\,
           \left( i_{1}(u_{-})\,m(u_{-}) -
             g_{1}(u_{-})\,q(u_{-}) \right) }{i(u_{-})\,
            i_{1}(u_{-}) - j(u_{-})\,q(u_{-})} \right.
\nonumber \\
&& \left.  -  \frac{e(u_{+})\,g_{1}(u_{+})\,
           \left( f_{1}(u_{-})\,r_{1}(u_{-}) -
             n(u_{-})\,l_{1}(u_{-}) \right) }
           {-\left( h_{1}(u_{-})\,l_{1}(u_{-})
              \right)  + f_{1}(u_{-})\,r_{1}(u_{-})}
        \right)
\nonumber \\
&& \left. \times \left( \frac{h(u_{+})\,
           \left( -\left( f(u_{-})\,c(u_{-}) \right)  +
             h(u_{-})\,f(u_{-}) \right) }{e(
            u_{+})\,\left( h(u_{-})\,l(u_{-}) - f(u_{-})\,r(u_{-}) \right) } +
         \frac{b(2\,u)\,f(u_{+})}
         {a(2\,u)\,e(u_{+})} \right) \right)
\nonumber \\
&&  /\,\left(\,e(u_{+})\,q(u_{+}) -  f(u_{+})\,r(u_{+})\, \right)
\ear
\bear
&& a_{2}^{4}(u,v)=-\left( \frac{b(2\,v)\,n(u_{+})\,d(u_{-})}
      {a(2\,v)\,j(u_{-})\,j(u_{+})} \right)  +
   \frac{a(u_{-})\,n(u_{-})\,d(u_{+})\,e(u_{+})}
    {a(u_{+})\,e(u_{-})\,j(u_{-})\,j(u_{+})}
\nonumber \\
&&  +  \left( \frac{d(2\,u)}{a(2\,u)} +
      \frac{d(u_{-})\,d(u_{+})}{j(u_{-})\,j(u_{+})} +
      \frac{b(2\,u)\,g(u_{-})\,g_{1}(u_{+})}
       {a(2\,u)\,j(u_{-})\,j(u_{+})} +
      \frac{c(2\,u)\,i(u_{-})\,i_{1}(u_{+})}
       {a(2\,u)\,j(u_{-})\,j(u_{+})} \right)
\nonumber \\
&&  \times  \left( \frac{e(u_{+})\,b(-u_{-})}{a(u_{+})\,e(-u_{-})} +
      \frac{b(2\,v)\,b(u_{+})}{a(2\,v)\,a(u_{+})} \right)
\nonumber \\
&&  +   \frac{f_{1}(u_{-})\,g(u_{+})\,
      \left( a(u_{+})\,b(u_{-})\,e(u_{-})\,r(u_{+})- a(u_{-})\,b(u_{+})\,e(u_{+})\,r(u_{-}) \right) }{a(u_{+})\,e(u_{-})\,
      e(u_{+})\,h(u_{-})\,j(u_{-})\,j(u_{+})}
\nonumber \\
&&   - \left( \frac{g(u_{-})\,g_{1}(u_{+})}{j(u_{-})\,j(u_{+})} +
      \frac{i(u_{-})\,i_{1}(u_{+})\,\left|M_{2,3}^{(+)}(2\,u)\right| }
         {j(u_{-})\,j(u_{+})\,\left|M_{2,2}^{(+)}(2\,u)\right| } +
      \frac{ \left|M_{2,4}^{(+)}(2\,u)\right| }{ \left|M_{2,2}^{(+)}(2\,u)\right| } \right)
\nonumber \\
&& \times   \left( \frac{e(u_{+})\,b(-u_{-})\,
         \left( \frac{b(2\,u)}{a(2\,u)} +
           \frac{b(u_{-})\,b(u_{+})}{e(u_{-})\,e(u_{+})} \right) }{a(u_{+})\,
         e(-u_{-})} + \frac{b(2\,v)\,
         \left( \frac{b(2\,u)\,b(u_{+})}{a(2\,u)\,a(u_{+})} +
           \frac{b(u_{-})\,\left( {b(u_{+})}^2 - a(u_{+})\,l(u_{+}) \right) }
            {a(u_{+})\,e(u_{-})\,e(u_{+})} \right) }{a(2\,v)} \right.
\nonumber \\
&&  \left. +   \frac{a(u_{-})\,b(u_{+})\,
         \left( h(u_{-})\,l(u_{-}) - f(u_{-})\,r(u_{-}) \right) }{a(u_{+})\,
         {e(u_{-})}^2\,h(u_{-})} +
      \frac{b(u_{-})\,f(u_{-})\,r(u_{+})}{e(u_{-})\,e(u_{+})\,h(u_{-})}
      \right) + h_{1}(u_{-})\,i(u_{+})\,
\nonumber \\
&&  \times  \frac{ c(u_{-})\,g_{1}(u_{+}) -\left( \frac{a(u_{-})\,c(u_{+})\,e(u_{+})\,
             g_{1}(u_{-})}{a(u_{+})\,e(u_{-})} \right) -  \frac{f(u_{+})\,i_{1}(u_{-})\,
           \left( a(u_{+})\,b(u_{-})\,e(u_{-})\,r(u_{+}) - a(u_{-})\,b(u_{+})\,e(u_{+})\,r(u_{-}) \right) }{a(u_{+})\,
           e(u_{-})\,e(u_{+})\,h(u_{-})} }{h(u_{+})\,{j(u_{-})}^2\,
      j(u_{+})}
\nonumber \\
&&    - \left( \frac{i(u_{-})\,i_{1}(u_{+})}
       {j(u_{-})\,j(u_{+})} +  \frac{ \left|M_{3,4}^{(+)}(2\,u)\right| }{ \left|M_{3,3}^{(+)}(2\,u)\right| } \right)
     \left( \frac{c(u_{-})\,g_{1}(u_{+})\,i(u_{-})}
       {h(u_{-})\,h(u_{+})\,j(u_{-})} -  \frac{b(2\,v)\,c(u_{-})\,m(u_{+})}
       {a(2\,v)\,h(u_{-})\,h(u_{+})} \right.
\nonumber \\
&&   +  \frac{a(u_{-})\,c(u_{+})\,e(u_{+})\,
         \left( j(u_{-})\,m(u_{-}) - g_{1}(u_{-})\,i(u_{-}) \right) }{a(u_{+})\,e(u_{-})\,
         h(u_{-})\,h(u_{+})\,j(u_{-})}
\nonumber \\
&&    +   \frac{f(u_{+})\,\left( j(u_{-})\,q(u_{-}) - i(u_{-})\,i_{1}(u_{-}) \right) \,
         \left( a(u_{+})\,b(u_{-})\,e(u_{-})\,r(u_{+}) - a(u_{-})\,b(u_{+})\,e(u_{+})\,r(u_{-}) \right) }{a(u_{+})\,
         e(u_{-})\,e(u_{+})\,{h(u_{-})}^2\,h(u_{+})\,j(u_{-})}
\nonumber \\
&&    - \left( \frac{e(u_{+})\,b(-u_{-})\,
            \left( \frac{b(2\,u)}{a(2\,u)} +
              \frac{b(u_{-})\,b(u_{+})}{e(u_{-})\,e(u_{+})} \right) }{
            a(u_{+})\,e(-u_{-})} +
         \frac{b(2\,v)\,\left( \frac{b(2\,u)\,b(u_{+})}{a(2\,u)\,a(u_{+})} +
              \frac{b(u_{-})\,\left( {b(u_{+})}^2 - a(u_{+})\,l(u_{+}) \right)
                   }{a(u_{+})\,e(u_{-})\,e(u_{+})} \right) }{a(2\,v)} \right.
\nonumber \\
&& \left.  +     \frac{a(u_{-})\,b(u_{+})\,
            \left( h(u_{-})\,l(u_{-}) - f(u_{-})\,r(u_{-}) \right) }{
            a(u_{+})\,{e(u_{-})}^2\,h(u_{-})} +
         \frac{b(u_{-})\,f(u_{-})\,r(u_{+})}{e(u_{-})\,e(u_{+})\,h(u_{-})}
         \right)
\nonumber \\
&&   \times  \left( \frac{-\left( c(2\,u)\,b(2\,u) \right)  +
            a(2\,u)\,m(2\,u)}{-\left( b(2\,u)\,b(2\,u) \right)  +
            a(2\,u)\,l(2\,u)} + \frac{f(u_{-})\,r(u_{+})}{h(u_{-})\,h(u_{+})}
         \right)
\nonumber \\
&&  \left. + \frac{\left( e(u_{+})\,b(-u_{-}) +
           \frac{b(2\,v)\,e(-u_{-})\,b(u_{+})}{a(2\,v)} \right) \,
         \left( \frac{c(2\,u)}{a(2\,u)} +
           \frac{c(u_{-})\,c(u_{+})}{h(u_{-})\,h(u_{+})} +
           \frac{b(2\,u)\,f(u_{-})\,r(u_{+})}{a(2\,u)\,h(u_{-})\,h(u_{+})}
           \right) }{a(u_{+})\,e(-u_{-})} \right)
\ear
\bear
&& a_{8}^{4}(u,v)=-\left( \frac{a(-u_{-})\,f(u_{+})\,
        \left( \frac{d(2\,u)}{a(2\,u)} +
          \frac{d(u_{-})\,d(u_{+})}{j(u_{-})\,j(u_{+})} +
          \frac{b(2\,u)\,g(u_{-})\,g_{1}(u_{+})}
           {a(2\,u)\,j(u_{-})\,j(u_{+})} +
          \frac{c(2\,u)\,i(u_{-})\,i_{1}(u_{+})}
           {a(2\,u)\,j(u_{-})\,j(u_{+})} \right) }{a(u_{+})\,e(-u_{-})}
      \right)
\nonumber \\
&&  + \frac{f_{1}(u_{-})\,g(u_{+})\,
      \left( b(u_{+})\,b(u_{-})\,r(u_{-})\,f(u_{+}) - a(u_{+})\,e(u_{-})\,c(u_{-})\,q(u_{+}) \right) }
      {a(u_{+})\,e(u_{-})\,e(u_{+})\,h(u_{-})\,j(u_{-})\,j(u_{+})}
\nonumber \\
&&   - \frac{n(u_{-})\,d(u_{+})\,b(u_{-})\,
      f(u_{+})}{a(u_{+})\,e(u_{-})\,j(u_{-})\,j(u_{+})}
      -  \left( \frac{g(u_{-})\,g_{1}(u_{+})}{j(u_{-})\,j(u_{+})} +
      \frac{i(u_{-})\,i_{1}(u_{+})\,\left|M_{2,3}^{(+)}(2\,u)\right| }
         {j(u_{-})\,j(u_{+})\,\left|M_{2,2}^{(+)}(2\,u)\right| } +
      \frac{\left|M_{2,4}^{(+)}(2\,u)\right|}{\left|M_{2,2}^{(+)}(2\,u)\right|} \right)
\nonumber \\
&& \times  \left( -\left( \frac{a(-u_{-})\,f(u_{+})\,
           \left( \frac{b(2\,u)}{a(2\,u)} +
             \frac{b(u_{-})\,b(u_{+})}{e(u_{-})\,e(u_{+})} \right) }{
           a(u_{+})\,e(-u_{-})} \right)  +
      \frac{q(u_{+})\,
         \left( -\left( f(u_{-})\,c(u_{-}) \right)  +
           h(u_{-})\,f(u_{-}) \right) }{e(u_{-})\,e(u_{+})\,
         h(u_{-})} \right.
\nonumber \\
&& \left.  - \frac{b(u_{+})\,b(u_{-})\,
         \left( h(u_{-})\,l(u_{-}) - f(u_{-})\,r(u_{-}) \right) \,
         f(u_{+})}{a(u_{+})\,{e(u_{-})}^2\,e(u_{+})\,
         h(u_{-})} \right)  - h_{1}(u_{-})\,i(u_{+})
\nonumber \\
&&  \times \frac{ \frac{d(u_{-})\,f_{1}(u_{+})}
         {h(u_{+})\,j(u_{-})} -
        \frac{c(u_{+})\,g_{1}(u_{-})\,b(u_{-})\,
           f(u_{+})}{a(u_{+})\,e(u_{-})\,h(u_{+})\,j(u_{-})}
         + \frac{f(u_{+})\,i(u_{-})\,
           \left( b(u_{+})\,b(u_{-})\,r(u_{-})\,f(u_{+}) - a(u_{+})\,e(u_{-})\,c(u_{-})\, q(u_{+}) \right) }
           {a(u_{+})\,e(u_{-})\,e(u_{+})\,h(u_{-})\,h(u_{+})\,j(u_{-})} }{j(u_{-})\,j(u_{+})}
\nonumber \\
&&   - \left( \frac{i(u_{-})\,i_{1}(u_{+})}{j(u_{-})\,j(u_{+})} +
      \frac{ \left|M_{3,4}^{(+)}(2\,u)\right| }{ \left|M_{3,3}^{(+)}(2\,u)\right| } \right) \,
    \left(   \frac{\left( -\left( d(u_{-})\,i(u_{-}) \right)  +
           g(u_{-})\,j(u_{-}) \right) \,
         f_{1}(u_{+})}{h(u_{-})\,h(u_{+})\,j(u_{-})}   \right.
\nonumber \\
&&    -\left( \frac{a(-u_{-})\,f(u_{+})\,
           \left( \frac{c(2\,u)}{a(2\,u)} +
             \frac{c(u_{-})\,c(u_{+})}{h(u_{-})\,h(u_{+})} +
             \frac{b(2\,u)\,f(u_{-})\,r(u_{+})}{a(2\,u)\,h(u_{-})\,h(u_{+})}
             \right) }{a(u_{+})\,e(-u_{-})} \right)
\nonumber \\
&&   -  \frac{c(u_{+})\,b(u_{-})\,
         \left( -\left( g_{1}(u_{-})\,i(u_{-}) \right)  +
           j(u_{-})\,m(u_{-}) \right) \,
         f(u_{+})}{a(u_{+})\,e(u_{-})\,h(u_{-})\,h(u_{+})\,
         j(u_{-})} +
\nonumber \\
&&   \frac{f(u_{+})\,
         \left( j(u_{-})\,q(u_{-}) - i(u_{-})\,i_{1}(u_{-}) \right) \,
         \left( b(u_{+})\,b(u_{-})\,r(u_{-})\,f(u_{+}) - a(u_{+})\,e(u_{-})\,c(u_{-})\,q(u_{+}) \right) }
         {a(u_{+})\,e(u_{-})\,e(u_{+})\,{h(u_{-})}^2\,h(u_{+})\,j(u_{-})}
\nonumber \\
&&   -  \left( \frac{-\left( c(2\,u)\,b(2\,u) \right)  + a(2\,u)\,m(2\,u)}
          {-\left( b(2\,u)\,b(2\,u) \right)  + a(2\,u)\,l(2\,u)} +
         \frac{f(u_{-})\,r(u_{+})}{h(u_{-})\,h(u_{+})} \right)
\nonumber \\
&&  \times  \left( -\left( \frac{a(-u_{-})\,f(u_{+})\,
              \left( \frac{b(2\,u)}{a(2\,u)} +
                \frac{b(u_{-})\,b(u_{+})}{e(u_{-})\,e(u_{+})} \right) }{
              a(u_{+})\,e(-u_{-})} \right)  +
         \frac{q(u_{+})\,
            \left( h(u_{-})\,f(u_{-}) - f(u_{-})\,c(u_{-}) \right) }{e(u_{-})\,
            e(u_{+})\,h(u_{-})} \right.
\nonumber \\
&&  \left. \left. -
         \frac{b(u_{+})\,b(u_{-})\,
            \left( h(u_{-})\,l(u_{-}) - f(u_{-})\,r(u_{-}) \right) \,
            f(u_{+})}{a(u_{+})\,{e(u_{-})}^2\,e(u_{+})\,
            h(u_{-})} \right)  \right)
\ear

\addcontentsline{toc}{section}{Appendix D}
\section*{\bf Appendix D: Relations for arbitrary $S$ }
\setcounter{equation}{0}
\renewcommand{\theequation}{D.\arabic{equation}}

In this  Appendix we present certain expressions concerning the unwanted terms of the one-particle
problem as well as the construction of the two-particle vector for arbitrary S.

The commutation rules used in the solution of one-particle eigenvalue problem
come from the entries [1,2], [2,3], $\dots$, [2S,2S+1], [2,2+2S+1], 
[3,3+2S+1], $\dots$, [2S+1,2(2S+1)] of the boundary Yang-Baxter equation 
(\ref{intertwRELcodificada}). 
To cancel the unwanted
terms we need to know how to compute the 
ratio $\frac{q_{i}^{(2)}(\lambda,\lambda_{1})}{q_{i}^{(1)}(\lambda,\lambda_{1})}$  which is not expected
to have a dependence on the i-th index. This means that this ratio  can be calculated collecting the
simplest unwanted contributions which turns out to be those coming  from the commutation rules  
between the fields $\widetilde{A}_{2S}(\lambda)$,
$\widetilde{A}_{2S+1}(\lambda)$ and $B_{12}(\lambda_1)$. Considering the
help of mathematical induction we find that the function
$q_{2S}^{(2)}(\lambda,\lambda_{1})$  is
\bear
&& q_{2S}^{(2)}(\lambda,\lambda_{1}) = -\omega_{2S}^{(+)}(\lambda) \frac{R_{1,2S}^{2S+1,2}(\lambda+\lambda_{1})}{R_{1,2S}^{2S,1}(\lambda+\lambda_{1})} + \omega_{2S+1}^{(+)}(\lambda)
\nonumber \\
&\times& \left( \frac{R_{1,2S}^{2S+1,2}(\lambda+\lambda_{1})}{R_{1,2S}^{2S,1}(\lambda+\lambda_{1})}\frac{\left| M_{2S,2S+1}^{(+)}(2\lambda)
\right|}{\left| M_{2S,2S}^{(+)}(2\lambda) \right|} - \frac{R_{1,2S}^{2S+1,2}(\lambda-\lambda_{1})}{R_{1,2S+1}^{2S+1,1}(\lambda-\lambda_{1})} \frac{\left|
\begin{array}{cc}
R_{1, 2S}^{2S, 1}(\lambda+\lambda_{1}) & R_{1, 2S}^{2S+1, 2}(\lambda+\lambda_{1}) \\
R_{2, 2S+1}^{2S, 1}(\lambda+\lambda_{1}) & R_{2, 2S+1}^{2S+1, 2}(\lambda+\lambda_{1})
\end{array} \right|}{R_{1, 2S}^{2S, 1}(\lambda+\lambda_{1}) R_{1, 2S+1}^{2S+1, 1}(\lambda+\lambda_{1})} \right)
\nonumber\\
\label{q1}
\ear
while 
$q_{2S}^{(1)}(\lambda,\lambda_{1})$  is given by
\bear
&& q_{2S}^{(1)}(\lambda,\lambda_{1}) = \omega_{2S}^{(+)}(\lambda) \left( \frac{R_{1,2S}^{2S+1,2}(\lambda-\lambda_{1})R_{1,2S+1}^{2S+1,1}(\lambda+\lambda_{1})}{R_{1,2S+1}^{2S+1,1}(\lambda-\lambda_{1}) R_{1,2S}^{2S,1}(\lambda+\lambda_{1})} - \frac{R_{1,2S}^{2S+1,2}(\lambda+\lambda_{1})}{R_{1,2S}^{2S,1}(\lambda+\lambda_{1})}\frac{\left| M_{1,2}^{(+)}(2\lambda_{1})
\right|}{\left| M_{1,1}^{(+)}(2\lambda_{1}) \right|} \right)
\nonumber \\
&+& \omega_{2S+1}^{(+)}(\lambda) \left[ - \frac{\left| M_{2S,2S+1}^{(+)}(2\lambda) \right|}{\left| M_{2S,2S}^{(+)}(2\lambda) \right|} \left( \frac{R_{1,2S}^{2S+1,2}(\lambda-\lambda_{1})R_{1,2S+1}^{2S+1,1}(\lambda+\lambda_{1})}{R_{1,2S+1}^{2S+1,1}(\lambda-\lambda_{1}) R_{1,2S}^{2S,1}(\lambda+\lambda_{1})} - \frac{R_{1,2S}^{2S+1,2}(\lambda+\lambda_{1})}{R_{1,2S}^{2S,1}(\lambda+\lambda_{1})}\frac{\left| M_{1,2}^{(+)}(2\lambda_{1})
\right|}{\left| M_{1,1}^{(+)}(2\lambda_{1}) \right|} \right) \right.
\nonumber \\
&-& \frac{\left| M_{1 ~ 2}^{(+)}(2\lambda_{1})\right|}{\left| M_{1 ~ 1}^{(+)}(2\lambda_{1}) \right|} \frac{R_{1,2S}^{2S+1,2}(\lambda-\lambda_{1})}{R_{1,2S+1}^{2S+1,1}(\lambda-\lambda_{1})} \frac{\left|
\begin{array}{cc}
R_{1, 2S}^{2S, 1}(\lambda+\lambda_{1}) & R_{1, 2S}^{2S+1, 2}(\lambda+\lambda_{1}) \\
R_{2, 2S+1}^{2S, 1}(\lambda+\lambda_{1}) & R_{2, 2S+1}^{2S+1, 2}(\lambda+\lambda_{1})
\end{array} \right|}{R_{1, 2S}^{2S, 1}(\lambda+\lambda_{1}) R_{1, 2S+1}^{2S+1, 1}(\lambda+\lambda_{1})}
\nonumber\\
&+&\left. \frac{R_{1,2S}^{2S,1}(\lambda-\lambda_{1}) R_{1,2S}^{2S+1,2}(\lambda+\lambda_{1}) R_{2,2S+1}^{2S+1,2}(\lambda-\lambda_{1})-R_{1,2S}^{2S+1,2}(\lambda-\lambda_{1}) R_{2,2S+1}^{2S,1}(\lambda+\lambda_{1}) R_{1,2S}^{2S+1,2}(\lambda-\lambda_{1})}{R_{1,2S+1}^{2S+1,1}(\lambda-\lambda_{1}) R_{1,2S}^{2S,1}(\lambda+\lambda_{1}) R_{1,2S+1}^{2S+1,1}(\lambda-\lambda_{1})} \right]
\nonumber\\
\label{q2}
\ear

Taking into account the explicit  expressions for the Boltzmann  weights  and for the functions $\omega_{j}^{\pm}(\lambda)$
one can verify that the ratio 
$\frac{q_{2S}^{(2)}(\lambda,\lambda_{1})}{q_{2S}^{(1)}(\lambda,\lambda_{1})}$ satisfy Eq.(\ref{RATIOT}).

We close this Appendix discussing  the construction of the two-particle state.  The appropriate commutation
relation is derived by combining the entries  
[1,3] and [1,3+2S] of 
Eq.(\ref{intertwRELcodificada}). After some algebra we find that commutation relation between the
creation operators $B_{12}(u)$ and $B_{12}(v)$ is
\bear
&&B_{12}(u)B_{12}(v)+\frac{R_{1,2}^{3,2}(u_{+})}{R_{1,2}^{2,1}(u_{+})}B_{13}(u)A_{2}(v) + \sum_{j=4}^{2S+1}\frac{R_{1,2}^{j,j-1}(u_{+})}{R_{1,2}^{2,1}(u_{+})} B_{1j}(u)C_{j-1 2}(v)
\nonumber\\
&-&\frac{R_{1,2}^{3,2}(u_{-})}{R_{1,3}^{3,1}(u_{-})}\left(\frac{R_{1,3}^{3,1}(u_{+})}{R_{1,2}^{2,1}(u_{+})}B_{13}(u)A_{1}(v)+\sum_{j=4}^{2S+1}\frac{R_{1,3}^{j,j-2}(u_{+})}{R_{1,2}^{2,1}(u_{+})} B_{1j}(u)C_{j-2 1}(v)\right)  =
\nonumber \\
&=&Z_S(u,v)
\left[ B_{12}(v)B_{12}(u)+\frac{R_{3,2}^{1,2}(u_{+})}{R_{2,1}^{1,2}(u_{+})} B_{13}(v)A_{2}(u) 
+ \sum_{j=4}^{2S+1}\frac{R_{j,j-1}^{1,2}(u_{+})}{R_{2,1}^{1,2}(u_{+})} B_{1j}(v)C_{j-1 2}(u) \right.
\nonumber \\
&+&  \left. \frac{\left|
\begin{array}{cc}
R_{1, 2}^{3, 2}(u_{-}) & R_{1, 3}^{3, 1}(u_{-}) \\
R_{3, 2}^{1, 2}(u_{-}) & R_{3, 3}^{1, 1}(u_{-})
\end{array} \right|}{\left|
\begin{array}{cc}
R_{1, 2}^{3, 2}(u_{-}) & R_{1, 3}^{3, 1}(u_{-}) \\
R_{2, 2}^{2, 2}(u_{-}) & R_{2, 3}^{2, 1}(u_{-})
\end{array} \right|} 
\left \{ 
\frac{R_{3,1}^{1,3}(u_+)}{R_{2,1}^{1,2}(u_+)} B_{13}(v)A_{1}(u)+
\sum_{j=4}^{2S+1} \frac{R_{j,j-2}^{1,3}(u_{+})}
{R_{2,1}^{1,2}(u_{+})} B_{1j}(v)C_{j-2 1}(u) \right \} \right]
\nonumber\\
\label{D2}
\ear
where function $Z_S(u,v)$ has been defined in Eq.(\ref{ZZU}).

The above relation allows us to define the following vector
\bear
\phi(u,v)& =& B_{12}(u)B_{12}(v)+\frac{R_{1,2}^{3,2}(u_{+})}{R_{1,2}^{2,1}(u_{+})}B_{13}(u)A_{2}(v) 
+ \sum_{j=4}^{2S+1}\frac{R_{1,2}^{j,j-1}(u_{+})}{R_{1,2}^{2,1}(u_{+})} B_{1j}(u)C_{j-1 2}(v)
\nonumber\\
&-&\frac{R_{1,2}^{3,2}(u_{-})}{R_{1,3}^{3,1}(u_{-})}\left(\frac{R_{1,3}^{3,1}(u_{+})}{R_{1,2}^{2,1}(u_{+})}B_{13}(u)A_{1}(v)+\sum_{j=4}^{2S+1}\frac{R_{1,3}^{j,j-2}(u_{+})}{R_{1,2}^{2,1}(u_{+})} B_{1j}(u)C_{j-2 1}(v)\right)  
\label{VECC}
\ear
which 
is symmetric under the exchange of the variables $u$ and $v$, thanks to certain identities between the Boltzmann
weights. More precisely, we have
\EQ
\phi(u,v)= Z_S(u,v) \phi(v,u)
\EN

The two-particle state is now obtained by acting the vector (\ref{VECC}) on the pseudovacuum 
$\ket{\bar{0}_{S}}$ leading us to
\bear
\ket{\bar{\psi}_{2}(\lambda_{1},\lambda_{2})} &=& \biggl( B_{12}(\lambda_{1})B_{12}(\lambda_{2})+\frac{R_{1,2}^{3,2}(\lambda_{1}+\lambda_{2})}{R_{1,2}^{2,1}(\lambda_{1}+\lambda_{2})}B_{13}(\lambda_{1})A_{2}(\lambda_{2})
\nonumber\\
&-&\frac{R_{1,2}^{3,2}(\lambda_{1}-\lambda_{2})}{R_{1,3}^{3,1}(\lambda_{1}-\lambda_{2})} \frac{R_{1,3}^{3,1}(\lambda_{1}+\lambda_{2})}{R_{1,2}^{2,1}(\lambda_{1}+\lambda_{2})}B_{13}(\lambda_{1})A_{1}(\lambda_{2}) \biggr) \ket{\bar{0}_{S}}
\ear

Finally, taking into account Eq.(\ref{Acombination})
we then recover the expression (\ref{TWOS}) exhibited in section \ref{spinSsol}.

\newpage

\begin{table}[h]
\begin{center}
\begin{tabular}{|c|c|c|}
\hline Manifold & $\varepsilon=\epsilon_{+}/\epsilon_{-}$ & $\widetilde{K}_{S}^{(-)}(\lambda)$ \\
\hline\hline
I              &  $+$  & Upper      \\
\cline{1-2} II &  $-$  & Triangular \\
\hline I       &  $-$  & Lower      \\
\cline{1-2} II &  $+$  & Triangular \\
\hline
\end{tabular}
\end{center}
\caption{The triangular property dependence of $\widetilde{K}_S^{(-)}(\lambda)$ on the
ratio $\varepsilon=\epsilon_{+}/\epsilon_{-}$. } \label{tab1}
\end{table}

\end{document}